\documentclass[dutch,english]{book}

\usepackage{graphicx}
\usepackage{amsmath}
\usepackage{amscd}
\usepackage{amsfonts}
\usepackage{amsthm}
\usepackage{fancybox}
\usepackage{psfrag}
\usepackage{babel}
\usepackage[a4paper=true,breaklinks=true]{hyperref}
\usepackage{axodraw}
\usepackage{epsfig,multicol}

 \usepackage{verbatim}

\def\ID{\mathbb{D}}

\newcommand{\pp}{{=\!\!\!|}}

\newcommand{\Li}{\operatorname{Li}}
\newcommand{\diag}{\operatorname{diag}}

\newcommand{\rref}[1]{(\ref{#1})}

\newcommand{\DOUBLEFIGURE}[5][ht]{\@dblfloat{figure}[#1]\centerline{%
        \parbox{.45\textwidth}{\centerline{\epsfig{file=#2}}}~~~~
        \parbox{.45\textwidth}{\centerline{\epsfig{file=#3}}}}
        \centerline{\parbox[t]{.45\textwidth}{\caption{#4}}~~~~
        \parbox[t]{.45\textwidth}{\caption{#5}}}\end@dblfloat}

\newcommand{\hepth}[1]{\href{http://xxx.lanl.gov/abs/hep-th/#1}{\tt hep-th/#1}}

\newcommand\spires[1]{\href{http://www-spires.slac.stanford.edu/spires/find/hep/www?j=#1}}

\newcommand\npb[3]{\spires{NUPHA\%2CB#1\%2C#3}{{\it Nucl.\ Phys.\ }{\bf B #1} (#2) #3}}
\newcommand\prsla[3]{{\it Proc.\ Roy.\ Soc.\ Lond.\ }{\bf A #1} (#2) #3}
\newcommand\jhep[3]{{\spires{JHEPA\%2C#1\%2C#3}{{\it J. High Energy Phys.\ }{\bf #1} (#2) #3}}}
\newcommand\prl[3]   {\spires{PRLTA\%2C#1\%2C#3}{{\it Phys.\ Rev.\ Lett.\ }{\bf #1} (#2) #3}}
\newcommand\plb[3]   {\spires{PHLTA\%2CB#1\%2C#3}{{\it Phys.\ Lett.\ }{\bf B #1} (#2) #3}}
\newcommand\mpla[3]  {\spires{MPLAE\%2CA#1\%2C#3}{{\it Mod.\ Phys.\ Lett.\ }{\bf A #1} (#2) #3}}
\newcommand\cmp[3]   {\spires{CMPHA\%2C#1\%2C#3}{{\it Commun.\ Math.\ Phys.\ }{\bf #1} (#2) #3}}
\newcommand\prd[3]   {\spires{PHRVA\%2CD#1\%2C#3}{{\it Phys.\ Rev.\ }{\bf D #1} (#2) #3}}
\newcommand\cqg[3]   {\spires{CQGRD\%2C#1\%2C#3}{{\it Class.\ and Quant.\ Grav.\ }{\bf #1} (#2) #3}}

\newcommand\ptp[3]   {\spires{PTPKA\%2C#1\%2C#3}{{\it Prog.\ Theor.\ Phys.\ }{\bf #1} (#2) #3}}

\newcommand\ijmp[3] {\spires{IMPAE\%2C#1\%2C#3}{{\it Int.\ J.\ Mod.\ Phys.\ }{\bf #1} (#2) #3}}
\newcommand\fort[3] {\spires{FPYKA\%2C#1\%2C#3}{{\it Fortsch.\ Phys.\ }{\bf #1} (#2) #3}}

\catcode`@=11
\@addtoreset{equation}{section}

\def\lhead{\@ifnextchar[{\@xlhead}{\@ylhead}}
\def\@xlhead[#1]#2{\gdef\@elhead{#1}\gdef\@olhead{#2}}
\def\@ylhead#1{\gdef\@elhead{#1}\gdef\@olhead{#1}}

\def\chead{\@ifnextchar[{\@xchead}{\@ychead}}
\def\@xchead[#1]#2{\gdef\@echead{#1}\gdef\@ochead{#2}}
\def\@ychead#1{\gdef\@echead{#1}\gdef\@ochead{#1}}

\def\rhead{\@ifnextchar[{\@xrhead}{\@yrhead}}
\def\@xrhead[#1]#2{\gdef\@erhead{#1}\gdef\@orhead{#2}}
\def\@yrhead#1{\gdef\@erhead{#1}\gdef\@orhead{#1}}

\def\lfoot{\@ifnextchar[{\@xlfoot}{\@ylfoot}}
\def\@xlfoot[#1]#2{\gdef\@elfoot{#1}\gdef\@olfoot{#2}}
\def\@ylfoot#1{\gdef\@elfoot{#1}\gdef\@olfoot{#1}}

\def\cfoot{\@ifnextchar[{\@xcfoot}{\@ycfoot}}
\def\@xcfoot[#1]#2{\gdef\@ecfoot{#1}\gdef\@ocfoot{#2}}
\def\@ycfoot#1{\gdef\@ecfoot{#1}\gdef\@ocfoot{#1}}

\def\rfoot{\@ifnextchar[{\@xrfoot}{\@yrfoot}}
\def\@xrfoot[#1]#2{\gdef\@erfoot{#1}\gdef\@orfoot{#2}}
\def\@yrfoot#1{\gdef\@erfoot{#1}\gdef\@orfoot{#1}}

\newdimen\headrulewidth
\newdimen\footrulewidth
\newdimen\plainheadrulewidth
\newdimen\plainfootrulewidth
\newdimen\headwidth
\newif\if@fancyplain \@fancyplainfalse
\def\fancyplain#1#2{\if@fancyplain#1\else#2\fi}

\footrulewidth\z@
\plainheadrulewidth\z@
\plainfootrulewidth\z@

\def\@fancyhead#1#2#3#4#5{#1\hbox to\headwidth{\vbox{\hbox
{\rlap{\parbox[b]{\headwidth}{\raggedright#2\strut}}\hfill
\parbox[b]{\headwidth}{\centering#3\strut}\hfill
\llap{\parbox[b]{\headwidth}{\raggedleft#4\strut}}}\headrule}}#5}

\def\@fancyfoot#1#2#3#4#5{#1\hbox to\headwidth{\vbox{\footrule
\hbox{\rlap{\parbox[t]{\headwidth}{\raggedright#2\strut}}\hfill
\parbox[t]{\headwidth}{\centering#3\strut}\hfill
\llap{\parbox[t]{\headwidth}{\raggedleft#4\strut}}}}}#5}

\def\headrule{{\if@fancyplain\headrulewidth\plainheadrulewidth\fi
\hrule\@height\headrulewidth\@width\headwidth \vskip-\headrulewidth}}

\def\footrule{{\if@fancyplain\footrulewidth\plainfootrulewidth\fi
\vskip-0.3\normalbaselineskip\vskip-\footrulewidth
\hrule\@width\headwidth\@height\footrulewidth\vskip0.3\normalbaselineskip}}

\def\ps@fancy{
\let\@mkboth\markboth
\@ifundefined{chapter}{\def\sectionmark##1{\markboth
{\ifnum \c@secnumdepth>\z@
 \thesection\hskip 1em\relax \fi ##1}{}}
\def\subsectionmark##1{\markright {\ifnum \c@secnumdepth >\@ne
 \thesubsection\hskip 1em\relax \fi ##1}}}
{\def\chaptermark##1{\markboth {\ifnum \c@secnumdepth>\m@ne
 \@chapapp\ \thechapter. \ \fi ##1}{}}
\def\sectionmark##1{\markright{\ifnum \c@secnumdepth >\z@
 \thesection. \ \fi ##1}}}
\def\@oddhead{\@fancyhead\relax\@olhead\@ochead\@orhead\hss}
\def\@oddfoot{\@fancyfoot\relax\@olfoot\@ocfoot\@orfoot\hss}
\def\@evenhead{\@fancyhead\hss\@elhead\@echead\@erhead\relax}
\def\@evenfoot{\@fancyfoot\hss\@elfoot\@ecfoot\@erfoot\relax}

\headwidth\textwidth}
\def\ps@fancyplain{\ps@fancy \let\ps@plain\ps@plain@fancy}
\def\ps@plain@fancy{\@fancyplaintrue\ps@fancy}

\catcode`\@=12

\begin{document}


\baselineskip13pt

\pagestyle{fancy}

\frontmatter

\begin{titlepage}

\begin{center}
\large{Vrije Universiteit Brussel\\
Faculteit Wetenschappen\\
Theoretische Natuurkunde\\}
\vspace{4cm}
\begin{picture}(354,0)
\put(0,0){\line(1,0){354}} 
\end{picture}  \\
\Large{Higher Derivative Corrections to the Abelian Born-Infeld Action using Superspace Methods}\\
\vspace{-8pt}
\begin{picture}(354,0)
\put(0,0){\line(1,0){354}} 
\end{picture}  \\
\vspace{12pt}
\large{\textsc{Stijn Nevens}}\\
\vspace{12pt}
\end{center}
\vspace{4cm}
Promotor: Prof. Dr. A. Sevrin
\vspace{-2.1em}
\begin{flushright}
Proefschrift ingediend\\
met het oog op het\\
behalen van de graad van\\
Doctor in de Wetenschappen\\
\end{flushright}
\vspace{1.6cm}
\begin{center}
2006
\end{center}
\end{titlepage}

 \newpage
\thispagestyle{empty}
\mbox{}

\newpage
\thispagestyle{empty}
\mbox{}

\vspace*{7cm}
\begin{flushright}
{\it Guts, guts, got no guts and stitches don't help at all.} \\
{\scriptsize \sc John Cale (Guts)}
\end{flushright}

\newpage
 \thispagestyle{empty}
\mbox{}

 \chapter*{Acknowledgements}

I would like to start by thanking my supervisor, Alexander Sevrin, for numerous reasons. First of all for his courses that opened up the broad world of quantum field theory, general relativity and ultimately string theory to me. Next for his everlasting support and enthusiasm during the last five years. And lastly for proofreading the work presented here. Further thanks goes to the members of my jury: Professors Ben Craps, Robert Roosen, Walter Troost and Stefan Vandoren.

This manuscript would contain a lot more mistakes against physics and the english language had it not been thoroughly read by Alexander Wijns whom I also would like to thank for numerous suggestions and discussions regarding this thesis. 

For collaborations I thank Paul Koerber, Alexander Sevrin, Walter Troost and Alexander Wijns while for discussions in the office, KK, coffee room and car, Ben Craps, Arjan Keurentjes, Paul Koerber and Alexander Wijns deserve being mentioned. I would also like to thank all the people from Leuven and the ULB for good times during and after our numerous seminars together and all the people I met at various conferences all over Europe. Especially Andres for sharing my interest in exploring the culinary aspirations of the places and introducing me to the black gold.

I would like to mention 
my high-shool physics teacher Ferdinand Van Schuylenbergh for sharpening my latent interest in physics by his enthusiastic teaching.

Of course I should not forget the people not related to physics but important to me anyhow, you know who you are (or so I hope). Making a list would take up too much space and inevitably lead to errors and thus I will refrain from doing so. Making an exception I would like to express my  gratitude to my parents Agnes and Jan, my brothers Daan and Dries and my girlfriend Katleen $\heartsuit$.

The work presented in this thesis would not have been possible without the financial support from various institutions. During the first four years I received a Ph.D. grant (aspirant) from the ``FWO-Vlaanderen ''. Furthermore our group was supported in part by the Free University of Brussels (Vrije Universiteit Brussel), in part by the Belgian Federal Science Policy Office through the  Interuniversity Attraction Pole P5/27, in part by the European Commission FP6 RTN programme MRTN-CT-2004-005104 and in part by the ``FWO-Vlaanderen'' through project G.0428.06.

\newpage
\thispagestyle{empty}

\parskip 6pt plus 1pt minus 1pt

\lhead[\fancyplain{}{}]{\fancyplain{}{}}

\rhead[\fancyplain{}{}]{\fancyplain{}{}}

\chead{\fancyplain{}{\bf Contents}}

\cfoot{\fancyplain{}{}}

\tableofcontents

\newpage
\thispagestyle{empty}

\mainmatter

\parskip 5pt plus 1pt minus 1pt

\lhead[\fancyplain{}{\bf\thepage}]{\fancyplain{}{}}

\chead[\fancyplain{}{\bf\leftmark}]{\fancyplain{}{\bf\rightmark}}

\rhead[\fancyplain{}{}]{\fancyplain{}{\bf\thepage}}

\cfoot{\fancyplain{\bf\thepage}{}}


 \chapter{Introduction}

\begin{quote}{\it ``Man's mind, once streched by a new idea, never regains its original dimensions.''} 
\vspace*{-0,6cm}
\begin{flushright}{\scriptsize \sc Oliver Wendell Holmes}\end{flushright}
\end{quote}

All the efforts done in theoretical physics today can grosso modo be split into two categories. The first is finding the fundamental underlying principles that govern physical phenomena and the second is trying to explain what we observe in nature starting from these principles. Of course these two domains rely heavily on each other and the line between them is often a very fuzzy one.  The reason to make this distinction is that the quest for  a theory of everything is primordially an effort to find a single underlying principle. From it everything else can, in principle, be explained. We say in principle since finding such a theory does not necessarily mean that we will be able to explain/calculate/predict/\ldots  everything since some phenomena will presumably remain too complicated.

At the moment there are two complementary theories which combined give an excellent description
of nature: the Standard Model and General Relativity. The Standard Model is a spontaneously broken $SU(3) \times SU(2) \times U(1)$ gauge theory and provides an outstanding description of the electroweak  and the strong force, both on small and large length scales. A drawback is that the model depends on twenty-six free parameters which must be determined experimentally. Gravity is the only force not incorporated in the Standard Model. Although its long distance behavior is very adequately described by General Relativity, it is not compatible with quantum mechanics. Not only does General Relativity yield a non-renormalizabele quantum field theory, the existence of black holes also gives rise to fundamental problems and paradoxes.  

From these observations it is apparent that more work is needed towards a single underlying theory. In the ideal case we need a theory that unifies gravity with the other forces and correctly predicts the twenty-six free parameters in the Standard Model.  A promising candidate is String Theory in which one assumes that elementary particles are small vibrating strings rather than point-particles. In the low energy regime, the resulting model is essentially a supersymmetric version of general relativity coupled to a supersymmetric gauge theory. Although it has many pleasing features it also is spawned with various connected problems. Maybe the most apparent problem is that there is not one but five consistent string theories. All of these are only well defined in ten dimensions, a second problem.

Efforts to solve the second problem, by compactifying the excess dimensions, naturally opened up a the door to the discovery of D-branes which revolutionized string theory. A D$p$-brane is a $p$-dimensional dynamical object, having a $p+1$-dimensional worldvolume, defined by the fact that open strings can end on it. Via duality transformations, in which D-branes play a crucial role, it became clear that all five string theories are limiting cases of a more fundamental theory dubbed M-theory. 

A tantalizing aspect of D-branes is their intimate relation with gauge theories. The $p+1$-dimensional worldvolume is described by a $p+1$-dimensional field theory containing $9-p$ massless scalar fields, describing the transversal position of the D$p$-brane in the ten dimensional ambient space, and a $U(1)$ gaugefield induced by the open strings, for which the groundstates are $U(1)$ gauge particles, ending on it.

The effective action of a single brane in the slowly varying field limit is the Born-Infeld action which in leading order is nothing more than the action for ordinary Maxwell theory. This is the result not including derivatives on the fieldstrength of the $U(1)$ gauge field. The first paper to study derivative correction was \cite{abelianbi4derivative} where it was shown that the term containing two derivatives vanishes and it was not until recently \cite{wyllard} that the four derivative corrections where calculated. 

\section*{Roadmap}

Chapter \ref{What_is_string_theory} serves a compact introduction to string theory in general but with a strong focus on the elements needed in the following chapters.  Starting with the point particle we quickly pass to the bosonic string which we use to introduce some important notions of string theory. Not totally satisfied with some features 
of bosonic strings, we turn to superstrings. Along the way we introduce the concept of superspace which will be relevant for chapter 3. We then move on to describe non perturbative aspects of string theory  as we introduce, passing via $T$-duality, D-branes. Lastly we concentrate on various methods to calculate the Born-Infeld action which is, as already stated, the effective action for  D-branes.

In the next chapter we study two-dimensional supersymmetric non-linear sigma-models with boundaries. We derive the most general family of boundary conditions in the non-supersymmetric case. Next we show that no further conditions arise when passing to the $N=1$ model and we present a manifest $N=1$ off-shell formulation. The analysis is greatly simplified compared to previous studies and there is no need to introduce non-local superspaces nor to go (partially) on-shell. Whether or not torsion is present does not modify
the discussion. Subsequently, we determine under which conditions a second supersymmetry exists. As for the case without boundaries, two covariantly constant complex structures are needed. However, because of the presence of the boundary, one gets expressed in terms of the other one and the remainder of the geometric data.
Finally we recast some of our results in $N=2$ superspace.

In chapter \ref{Beta-Function_Calculations_in_Boundary_Superspace} we calculate the beta-functions for an open string sigma-model in the presence of a $U(1)$ background. Passing to $N=2$ boundary superspace (a special case of the general setup developed in chapter 3) in which the background is fully characterized by a scalar potential, significantly facilitates the calculation. Performing the calculation through three loops yields the effective action for a single D-brane in trivial bulk background fields through four derivatives on the fieldstrengths and to all orders in $ \alpha'$. Finally, the
calculation presented there shows that demanding ultra-violet finiteness of the non-linear sigma-model can be reformulated as the requirement that the background is a deformed stable holomorphic $U(1)$ bundle.

We end with our conclusions in chapter \ref{Closing_Remarks}. Our conventions can be found in appendix A alongside a brief summary in dutch written down in appendix B. 

The papers I co-authered are \cite{susyboundary}, \cite{towards} and \cite{betastijnalexalexwalter}.


 \chapter{What is String Theory?} \label{What_is_string_theory}

We will not try to answer the question raised by Joseph Polchinski in 1994 at Les Houches \cite{Whatisstring} if only because at this point the question is still open and the main topic of the very active research domain that is string theory. The present chapter merely serves to set the stage for the later chapters by introducing the reader to the basic elements of string theory. We will also introduce more advanced material and techniques needed in the remainder of this thesis.  The treatment presented here does not claim to be complete nor to be fully original.  For a more detailed overview we refer the reader to the following excellent books \cite{bookGSW}, \cite{bookpolchinski}, \cite{cvj} and \cite{bookzwiebach}.

\section{Perturbative String Theory}

Before turning to string theory, a theory of one-dimensional objects, we briefly review the theory of point-particles, zero-dimensional objects, thereby introducing some core-elements in string theory in a more familiar setting.

\subsection{The Point-Particle}

Consider the action describing a relativistic point-particle with mass $m$ moving in a $D$-dimensional Minkowski-space with metric $\eta_{\mu\nu}=\diag \, (-,+,\ldots,+)$,
\begin{equation}
\label{puntactie1} \mathcal{S}=-m \int d \tau
\left(-\frac{dx^\mu}{d\tau}\frac{dx^\nu}{d\tau}\eta_{\mu\nu}\right)^{1/2}.
\end{equation}
This is exactly the length of the particle's world-line parametrized by $\tau$. Upon looking at this action two things draw our attention. The first is that this action is incapable of handling massless particles and the second is the presence of the square root residing in the action.

Both problems are easily resolved by introducing an auxiliary variable $g^{\tau\tau}(\tau)$, which can be interpreted as a metric\footnote{In this simple case $g^{\tau\tau} = (g_{\tau\tau})^{-1}$} on the world-line and a new action,
\begin{equation}
\label{puntactie2}\mathcal{S}= -\frac{1}{2}\int d
\tau
 \sqrt{-g_{\tau\tau}} \left[g^{\tau\tau} \left(\frac{dx^\mu}{d\tau}\right)^2+m^2\right],
\end{equation}
which is equivalent to the previous one upon eliminating $g_{\tau\tau}$ using it's equation of motion,
\begin{equation}
\label{bewegingsvgl e} \dot{x}^2 - g_{\tau\tau} m^2 = 0.
\end{equation}
This action exhibits the following two symmetries:
\begin{itemize}
\item Poincar\'e invariance,
\begin{align} 
    x^\mu &\rightarrow \Lambda^{\mu}_{\ \nu}
    x^\nu + a^\mu,
 \end{align}
with $\Lambda^{\mu}_{\ \nu}$ a Lorentz transformation and $a^\mu$ a translation.
\item Reparametrisation invariance,
\begin{align} \begin{split}
    \tau & \rightarrow \tau', \\
    g_{\tau\tau} (\tau) & \rightarrow
    g_{\tau\tau} '(\tau') =
    \frac{\partial\tau}{\partial\tau'} \frac{\partial\tau}{\partial\tau'}
    g_{\tau\tau} (\tau), \\
     x^\mu(\tau) & \rightarrow  x'^\mu(\tau') =
    x^\mu(\tau).
\end{split} \end{align}
\end{itemize}
The first symmetry tells us that the action is Poincar\'e invariant which is built in by construction. The second symmetry only reflects the fact that the action should not depend on the particular choice of world-line-parametrisation. 

The notation chosen is very suggestive of seeing the action \rref{puntactie2} as a one-dimensional theory containing $D$ free scalar fields $x^\mu(\tau)$ and reinterpret the repara\-metrisation invariance as an invariance under general coordinate transformations in one dimension. The Poincar\'e invariance then has to be seen as an internal symmetry of the theory. Quantizing this theory would just lead to ordinary quantum mechanics for a point-particle so nothing new is to be learned here. However it is a procedure precisely like the one described in this paragraph that is at the heart of string theory.

\subsection{Bosonic Strings} \label{bosonicstringssection}

We are now ready to tackle strings. Just as a point-particle sweeps out a world-line when moving in space-time, a string sweeps out a world-sheet. Motivated by what we have learned from the relativistic point-particle we introduce two parameters $\tau = \sigma^0$ and $\sigma = \sigma^1$ to parametrize this world-sheet. The former can be seen as a time coordinate and the latter as a space coordinate. Just as before we embed the string in a $D$-dimensional Minkowski space having coordinates $X^\mu$. Pushing the analogy even further we take the action proportional to the surface of the world-sheet \cite{nambu,goto} leading to,
\begin{equation} \label{Nabuagotoactie}
\mathcal{S}_{NG}=-\frac{1}{2\pi\alpha'}\int d^2 \sigma
 \left(-\det\frac{dX^\mu}{d\sigma^a}\frac{dX^\nu}{d\sigma^b}\eta_{\mu\nu}\right)^{1/2},
\end{equation}
called the Nambu-Goto action.  The parameter $\alpha'$ appearing in the action is related to the fundamental string length as follows, $l_s=\sqrt{\alpha'}$, and is needed to make the action dimensionless. It is the whole pre-factor $T = \frac{1}{2 \pi\alpha'}$ that can be seen as the tension of the string that is the analogue of the mass $m$ for the point-particle.

As in the case of the point-particle we can conveniently reformulate this action, eliminating the awkward square root, by introducing an auxiliary field $h_{ab}(\tau,\sigma)$, which can be seen as a metric on the world-sheet. The resulting action is called the Polyakov action \cite{bdh,dz,polyakov1,polyakov2},
\begin{equation} \label{polyakovactie1}
 \mathcal{S}_{P}= - \frac{1}{4\pi\alpha'}\int d^2 \sigma
\sqrt{-\det h_{ab}} \ h^{ab}\partial_a X^\mu\partial_b
X^\nu\eta_{\mu\nu}.
\end{equation}
Checking the equivalence of both actions is easy using the equations of motion for $h_{ab}$ ,
\begin{align}
    \label{energiemoment} T_{ab} = \frac{1}{2}
    \partial_a X^\mu \partial_b X_\mu - \frac{1}{4}
    h_{ab} h^{cd} \partial_c X^\mu
    \partial_d X_\mu &= 0,
\end{align}
which just says that the two-dimensional energy-momentum tensor, $T_{ab}$, should vanish.

The Polyakov action exhibits the following symmetries,
\begin{itemize}
\item Poincar\'e invariance,
\begin{align} \begin{split}
    X^\mu(\tau,\sigma) &\rightarrow \Lambda^{\mu}_{\ \nu}
    X^\nu(\tau,\sigma) + a^\mu, \\
    h_{ab}(\tau,\sigma) &\rightarrow
    h_{ab}(\tau,\sigma),
\end{split} \end{align}
with $\Lambda^{\mu}_{\ \nu}$ a Lorentz transformation and $a^\mu$ a translation.
\item Reparametrisation invariance,
\begin{align} \begin{split}
    \sigma^a & \rightarrow \sigma'^a, \\
    X^\mu(\sigma^0,\sigma^1) & \rightarrow  X'^\mu(\sigma'^0,\sigma'^1) =
    X^\mu(\sigma^0,\sigma^1), \\
    h_{ab}(\sigma^0,\sigma^1) & \rightarrow
    h'_{ab}(\sigma'^0,\sigma'^1) =
    \frac{\partial\sigma^c}{\partial\sigma'^a}
    \frac{\partial\sigma^d}{\partial\sigma'^b}
    h_{cd}(\sigma^0,\sigma^1).
\end{split} \end{align} 
\item Weyl rescaling of the world-sheet metric,
\begin{align} \begin{split}
    X^\mu(\tau,\sigma) & \rightarrow X^\mu(\tau,\sigma), \\
    h_{ab}(\tau,\sigma) & \rightarrow
    h'_{ab}(\tau,\sigma) = \exp (2\omega(\tau,\sigma)) \
    h_{ab}(\tau,\sigma).
\end{split} \end{align}
\end{itemize} 
One can show that it is always possible to go, at least locally, to the so called conformal gauge $h_{ab}= \eta_{ab} = \text{diag} \, [-1,1]$ by choosing a suitable reparametrisation and Weyl rescaling. When one tries to generalize the analysis in this section  in a natural way  for $n$-dimensional objects one sees that the possibility to gauge away the $h_{ab}$ is only possible for $n \leq 1$. This, next to the fact that higher dimensional objects give rise to non-renormalizable field theories, is crucial to string theory.
After imposing the conformal gauge there is some residual symmetry left, called conformal symmetry, which is generated by the infinite-dimensional Virasoro algebra. Its generators are nothing more than the Fourier modes, $L_m$, of the energy-momentum tensor.  So the theory we are left with is a so called Conformal Field Theory (CFT) and is highly symmetric. When going to the quantum theory the Virasoro algebra gets an anomaly, called central charge, and the resulting algebra is called the central extension of the Virasoro algebra. 

In conformal gauge the Polyakov action \rref{polyakovactie1} reduces to,
\begin{align}
\mathcal{S}_{P}&=-\frac{1}{4\pi\alpha'}\int d^2 \sigma \ 
\eta^{ab} \partial_a X^\mu\partial_b
X_\mu, \\
\label{polyakov-actie in conforme ijk}  &=\frac{1}{4\pi\alpha'}\int
d^2 \sigma
\left[\left(\partial_{\tau}X\right)^2-\left(\partial_{\sigma}X\right)^2\right].
\end{align}
The variation of \rref{polyakov-actie in conforme ijk}  gives,
\begin{equation} \label{variatiepolyakovactie}
\delta\mathcal{S}_P = \frac{1}{2\pi\alpha'}\int d^2 \sigma \delta
X^\mu \left(\partial_\sigma^2-\partial_\tau^2\right)X_\mu -
\frac{1}{2\pi\alpha'} \int d \tau
\, \partial_{\sigma}X_\mu \delta X^\mu \big|^{\sigma=\bar{\sigma}}_{\sigma=0},
\end{equation}
where the second term is a boundary term which is to be evaluated at the two endpoints of the string, $0$ and $\bar{\sigma}$. 

Now is a good moment to point out that we have two different kind of strings depending on wether the endpoints of the string are joined or not:
\begin{itemize}
\item {\bf Open strings} As the name suggests this type of string has two endpoints which are not joined. In this case the world-sheet has boundaries and one can use a conformal transformation to map the world-sheet to a disk. We will choose $\bar{\sigma}=\pi$ as is traditional in the literature. 
\item {\bf Closed strings} In this case the string is closed and thus has no endpoints. Upon choosing the usual convention  $\bar{\sigma}=2 \pi$ we impose the following periodicity condition,
\begin{equation}\label{geslotensnaren}
X^\mu(\sigma) = X^\mu(\sigma+2\pi).
\end{equation}
In this case the world-sheet has no boundaries and one can pick a conformal transformation to map it to a sphere.
\end{itemize}
We cannot avoid that the two ends of an open string join thus forming a closed string. So having open strings necessarily means we have closed strings too.

The equation of motion for the Polyakov action in conformal gauge \rref{polyakov-actie in conforme ijk} is simply the free wave equation in two dimensions,
\begin{equation}
\label{bewegingsvergelijking}\ \Box X^\mu =
\left(\partial_\sigma^2-\partial_\tau^2\right)X^\mu =0.
\end{equation}
For closed strings the boundary term in \rref{variatiepolyakovactie} is automatically zero due to the periodicity conditions \rref{geslotensnaren}. Requiring that the boundary term vanishes for open strings we have the following options for each direction $X^\mu$:
\begin{itemize}
\item Neumann boundary conditions,
\begin{align}\label{restricties1}
\partial_\sigma X^\mu = 0 \quad \text{at} \quad \sigma=0,\pi.
\end{align}
This describes freely moving endpoints. A consequence of this condition is that no momentum can flow off the endpoints  of the string.
\item Dirichlet boundary conditions,
\begin{align}\label{restricties2}
 \delta X^\mu = 0 
\quad \text{at} \quad \sigma=0,\pi.
\end{align}
This condition tells us that we no longer consider the endpoints of the string to be dynamical but rather see them as having a fixed position. This implies we have a preferred position and hence we break Lorentz invariance.  It also implies momentum can flow off the endpoints of the string which may seem strange because momentum should be conserved. A simple example can shed light on this problem. Think about an ordinary (rubber) string with one end attached to a sufficiently massive wall. We can consider the wall, and hence also the endpoint, to be fixed even if we transfer momentum to it. So in this case we just need some massive object (wall) to attach the string endpoints to and both problems are resolved. We can transfer momentum to the wall without moving it (a lot) and we naturally break Lorentz invariance because we just put an object in space-time. With hindsight it seems very naive to exclude this possibility but in the early days of string theory it was discarded precisely because there was no such object known. Now we know these massive objects do exist, they are called D-branes \cite{Dai:1989ua,polchinskidbranesrrcharges}.
\end{itemize}
In the following we will consider Neumann boundary conditions and let the Dirichlet boundary conditions and D-branes come and haunt us later.

The general solution to the free wave equation in two dimensions can be written as
\begin{equation}
X^\mu(\tau,\sigma)=X_L^\mu(\tau + \sigma)+X_R^\mu(\tau - \sigma).
\end{equation}
with  $X^\mu_L$ and $X^\mu_R$ two arbitrary functions which describe left and right moving modes respectively.

In the case of closed strings the most general solution looks like, 
\begin{equation}
\begin{split}\label{AlgemeneoplossinggeslotensnarenLR}
  &X_L^\mu(\tau+\sigma)=\frac{x^\mu}{2} + \frac{\hat{x}^\mu}{2} +\frac{\alpha'}{2}p^\mu(\tau+\sigma) + i\sqrt{\frac{\alpha'}{2}} \sum_{n \neq 0}\frac{\alpha_n^\mu}{n} e^{-in(\tau+\sigma)}, \\
& X_R^\mu(\tau-\sigma)  =\frac{x^\mu}{2} - \frac{\hat{x}^\mu}{2} +\frac{\alpha'}{2}p^\mu(\tau-\sigma) + i\sqrt{\frac{\alpha'}{2}}\sum_{n \neq 0}\frac{\widetilde{\alpha}_n^\mu}{n} e^{-in(\tau-\sigma)},
\end{split}
\end{equation}
resulting in,
\begin{align} \label{Algemeneoplossinggeslotenensnaren}
X^\mu(\tau,\sigma)=x^\mu +\alpha'p^\mu\tau +
i\sqrt{\frac{\alpha'}{2}}\sum_{n \neq
0}\frac{1}{n}\left(\alpha_n^\mu
e^{-in(\tau+\sigma)}+\widetilde{\alpha}_n^\mu
e^{-in(\tau-\sigma)}\right).
\end{align}
With $\alpha^\mu_n$ and $\widetilde{\alpha}_n^\mu$ Fourier coefficients, $x^\mu$ the mass center postition at $\tau=0$ and $p^\mu$ the centre of mass momentum. So we see the string moves in space-time and has left and right moving oscillations which are independent. Reality of $X^\mu$ implies  $\alpha_{-n}^\mu = \left(\alpha_{n}^\mu \right)^\dag$ and $\widetilde{\alpha}_{-n}^\mu = \left(\widetilde{\alpha}_{n}^\mu \right)^\dag$. 
The coefficients obey the following Poisson brackets,
\begin{align}
    \left\{ \alpha_n^\mu,\alpha_m^\nu \right\}  =
    \left\{ \widetilde{\alpha}_n^\mu,\widetilde{\alpha}_m^\nu
    \right\} = -i m \delta_{m+n} \eta^{\mu\nu}, \
    \left\{ \alpha_n^\mu,\widetilde{\alpha}_m^\nu \right\} = 0, \
    \left\{ x^\mu , p^\nu \right\}  = \eta^{\mu\nu},
\end{align}
which we recognize as those for a two-dimensional harmonic oscillator and for a particle in mechanics.

In the case of open strings the most general solution forces left and right moving modes,

\begin{equation}
\begin{split}
 \label{AlgemeneoplossingopensnarenLR}
  &X_L^\mu(\tau+\sigma)=\frac{x^\mu}{2} + \frac{\hat{x}^\mu}{2} +\alpha'p^\mu(\tau+\sigma) + i\sqrt{\frac{\alpha'}{2}}\sum_{n \neq 0}\frac{\alpha_n^\mu}{n} e^{-in(\tau+\sigma)},\\
 &X_R^\mu(\tau-\sigma)=\frac{x^\mu}{2} - \frac{\hat{x}^\mu}{2} +\alpha'p^\mu(\tau-\sigma) + i\sqrt{\frac{\alpha'}{2}}\sum_{n \neq 0}\frac{\alpha_n^\mu}{n} e^{-in(\tau-\sigma)},
\end{split}
\end{equation}
 to combine into standing waves,
\begin{align}\label{Algemeneoplossingopensnaren}
X^\mu(\tau,\sigma)=x^\mu +2\alpha'p^\mu\tau +
2i\sqrt{\frac{\alpha'}{2}}\sum_{n \neq
0}\frac{\alpha_n^\mu}{n}e^{-in\tau}\cos(n\sigma).
\end{align}
And everything obeys the the same reality conditions and Poison brackets as before.


We can quantize the theory in the canonical way by making the following replacement $  \{ \ , \ \} \rightarrow \frac{1}{i} [ \ , \ ]$ and seeing the $X^\mu$ as operators. As with the harmonic oscillator\footnote{In the following we will only consider $\alpha_m^\mu$, adding the tilde is straightforward.}
 we take the $\alpha_m^\nu$ to be creation operators for $m<0$ and annihilation operators for
$m>0$. The vacuum state $\left| 0;k \right\rangle$ is defined to be annihilated by all annihilation operators and to be an eigenstate of $p^\mu$ with eigenvalue $k^\mu$.
We build the Fock-space by successive application of raising operators. Upon doing this we encounter negative norm states, $\alpha_m^0 \left| 0;k \right\rangle$ for $m<0$, since $\left[ \alpha_m^0,\alpha_{-m}^0\right] = -m$. Looking back we see that we gauged away the $h_{\alpha\beta}$, however we still have to impose its equations of motion, in other words we have to impose the vanishing of the energy-momentum tensor. Just like in the Gubta-Bleuler treatment of electrodynamics we impose the weaker condition that its positive frequency components annihilate the vacuum \cite{GiudiceVecchiaVirasoro}. One finds that the spectrum can only be ghost-free if $D = 26$ \cite{d26b,d26gt}, this is called the critical dimension. So we see that the generators of the infinite-dimensional Virasoro algebra, that represent a symmetry of the Polyakov action in conformal gauge, are needed to remove the ghosts from the physical spectrum \cite{Virasoro_crit_dim}.

All in all we find the following spectrum. The ground state of the open string, $\left| 0;k \right\rangle$, is a tachyon\footnote{A state with negative mass squared which signifies an instability of the vacuum.} and thus indicates that the open string vacuum is unstable. This problem has been resolved in recent years \cite{senDRABDB,senUTP,pieter_jan_phd} and it turns out that the existence of the tachyon is a perturbative consequence of the instability of the space-filling D25-brane and that the condensation of the tachyon should correspond to the decay of this brane. The perturbatively stable vacuum should then correspond to the closed string vacuum without the presence of open strings. 

The first exited state is $ \left| \zeta;k \right\rangle = \zeta \cdot \alpha_{-1} \left| 0;k \right\rangle$ with $\zeta$ a polarization vector. This state obeys the following conditions, $\zeta \cdot k  = 0$ and $k^2=0$. Moreover it has the following symmetry, $\zeta \to \zeta + k \lambda$, with $\lambda$ an arbitrary function of the momentum. All this hints towards the interpretation of this state as a massless $U(1)$ gauge boson (Maxwell field). The first condition is then the Lorentz gauge condition, the second tells us it has mass zero and the symmetry can be seen as the residual gauge symmetry after imposing the Lorentz gauge. By introducing $N$ extra degrees of freedom, known as Chan-Paton factors, on the string's endpoints we can upgrade this to a $U(N)$ group. 
In the language of D-branes this corresponds to having the endpoints stuck to $N$ coinciding D-branes. 


The ground state of the closed string  $\left| 0;k \right\rangle$, is again a tachyon but in this case it is not known if a true vacuum exitst and what it should look like \cite{HMS_Closed_String_Tachyon}. Recently some calculations  \cite{YB_Closed_String_Tachyon} were made to support the claim that what happens is analogous to the case of open strings and so one would expect no closed strings in the true vacuum. Without gravity excitations (see the next paragraph) spacetime ceases to be dynamical and it would seem that, for all intents and purposes, it has dissappeared. 

Due to the level matching condition, which imposes the number of left and right moving oscillators to be equal\footnote{It arises as follows. The operator $P_{\sigma}$ that generates $\sigma$-translations is proportional to the number of left moving oscillators minus the number of right moving ones. As for closed strings there is no physical significance to where on the string we are, the physics should be invariant under such translations and hence we should impose $P_{\sigma}=0$ as an operator condition on our physical states, resulting in the level matching condition. },  the first excited state is given by $\left| \xi ;k \right\rangle = \xi_{\mu\nu}\alpha_{-1}^\mu \widetilde{\alpha}_{-1}^\nu \left| 0;k \right\rangle $ with $\xi$ a polarization tensor. It obeys the following conditions, $ \xi_{\mu\nu} k^{\mu} = \xi_{\mu\nu} k^{\nu} = 0$ and $k^2=0$. From the first condition we see that only states of transverse polarization occur while the second tells us that this state is massless.  
This state can be split in three parts. The traceless symmetric part of $\left| \xi ;k \right\rangle$ is the graviton, $G_{\mu\nu}$, the antisymmetric part, $B_{\mu\nu}$, is called the Kalb-Ramond field, and the trace is the dilaton $\Phi$. The presence of the graviton is most probably the `Raison d'\^etre' of string theory. 

A remark is in order. The state described in the previous paragraph also possesses  the following symmetry, $\xi_{\mu\nu} \to \xi_{\mu\nu} + k_{\mu} \iota_{\nu} + \kappa_{\mu} k_{\nu} $, where $\iota$ and $\kappa$ are arbitrary vectors orthogonal to the momentum $k$. For the antisymmetric part this results in $\xi_{[\mu\nu]} \to \xi_{[\mu\nu]} + k_{\mu} \lambda_{\nu} - \lambda_{\mu} k_{\nu} $ with $\lambda$ an arbitrary vector orthogonal to the momentum $k$. This is nothing more as the Fourier transform of the standard gauge transformation for a 2-form,
\begin{equation}
B_{\mu\nu} \to B_{\mu\nu} + \partial_{\mu} \Lambda_{\nu} - \partial_{\nu} \Lambda_{\mu}.
\end{equation}



A natural question that arises is what happens if we let our string move in a background containing lots of strings. It turns out that if we have a coherent state of massless closed string states the action we have to consider is given by,
\begin{align}
 \mathcal{S}_{\text{bulk}}= &- \frac{1}{4\pi\alpha'} \int d^2 \sigma \sqrt{-\det h_{ab}}
 \left[ \left(h^{ab}G_{\mu\nu}(X) + \epsilon^{ab}B_{\mu\nu}(X)\right)\partial_a X^\mu\partial_b X^\nu \right] \nonumber \\  \label{polyakovactiecurvedbulk}  
  &- \frac{1}{4\pi} \int d^2 \sigma \sqrt{-\det h_{ab}} \ \Phi(X) R,
\end{align}
where the term containing the gravity field is the one obtained from the Polyakov action \rref{polyakovactie1} by changing from a flat metric to a curved one. Then we have a term describing the coupling of the string to the Kalb-Ramond field. Just as a point particle couples in natural way to a Maxwell field the string couples in a natural way to this field and the associated charge carried by the string can be visualized as Maxwell current on the string. The last term contains the two-dimensional Ricci scalar $R$ and the dilaton $\Phi$. We will postpone its interpretation till after the next paragraph. An action like this in which the kinetic term has a field-dependent coefficient is called a non-linear sigma-model.

For strings moving in a background carrying a coherent state of massless open strings we have to add the following boundary term,
\begin{align}
 \mathcal{S}_{\text{boundary}} & =   \int d^2 \sigma \ \partial_\sigma ( A_\mu \partial_\tau X^\mu) \nonumber \\  \label{polyakovactiecurvedboundary}
 & =  \int d\tau \left( A_\mu \partial_\tau X^\mu|_{\sigma=\pi}- A_\mu \partial_\tau X^\mu|_{\sigma=0}\right),
 \end{align}
which describes the coupling of string endpoints to the Maxwell field. 
This is precisely the term we have to add to the bulk action if we want the coupling to the Kalb-Ramond field to be gauge invariant for open strings. Closely related is the fact that string charge can not stop flowing at the endpoint of a string and, in a modern language, flows into the D-brane where the open string ends on. This is possible since the electric field lines on the D-brane carry string charge and thus couple to the Kalb-Ramond field just as strings do. To conclude we point out that the boundary term can easily be absorbed in the bulk action if we replace the Kalb-Ramond field $B_{\mu\nu} $ by $\mathcal{F}_{\mu\nu} = B_{\mu\nu}  + 2 \pi \alpha' F_{\mu\nu}$. 



To explain the role of the dilaton we first notice that,
\begin{equation}
\chi = \frac{1}{4\pi} \int d^2 \sigma \sqrt{-\det h_{ab}} \  R = 2 - 2g -b,
\end{equation}
is only function of the genus $g$ and the number of boundaries $b$ of the two-dimensional manifold that is the string world-sheet.
This topological invariant is called the Euler characteristic of a manifold. We then turn to the path integral,
\begin{equation}
{\cal Z}= \int {\cal D}X {\cal D}h \ e^{-  \mathcal{S}_{\text{bulk,e}}},
\end{equation}
where $\mathcal{S}_{\text{bulk,e}}$ is the Wick rotated version of the bulk action. From this it is easy to see that amplitudes will be weighted by a factor 
$e^{-\left\langle  \Phi \right\rangle\chi}$ with $\left\langle  \Phi \right\rangle$ the vacuum expectation value of the dilaton.
Adding a handle to the world-sheet, which corresponds to the emission and absorption of an closed string, increases the genus by one and hence adds a factor $e^{2 \left\langle  \Phi \right\rangle}$.  From this we can see that the process of emitting a closed string is proportional to $e^{ \left\langle  \Phi \right\rangle}$. Similarly adding a strip to the world-sheet, which corresponds to the emission and absorption of an open string, adds one boundary and hence adds a factor  $e^{ \left\langle  \Phi \right\rangle}$. Emitting an open string is therefore a process proportional to $e^{ \left\langle  \Phi \right\rangle / 2}$. Now it is fairly easy to see that the string coupling constants obey,
\begin{equation}
g_{\text{s,closed}} \sim g_{\text{s,open}}^2 \sim g_{\text{s}} \equiv  e^{ \left\langle  \Phi \right\rangle}.
\end{equation}
So we see that the string coupling is not a free parameter but has a dynamical value set by the vacuum expectation value of the dilaton.

As was pointed out in the beginning of this section Weyl invariance is crucial to string theory. Indeed, the action \rref{polyakovactiecurvedbulk} only defines a consistent string theory if the two-dimensional quantum field theory it defines is Weyl invariant. Weyl invariance at the quantum level is equivalent with the vanishing of the renormalization group beta-funtions. 
Up to first order in $\alpha'$ this gives,
\begin{eqnarray}\begin{split}
\beta_{\mu\nu}^G &= \alpha' \left( R_{\mu\nu}  + 2 \nabla_\mu \nabla_\nu \Phi - \frac{1}{4} H_{\mu\kappa\sigma} H_\nu^{\phantom{\nu}\kappa\sigma} \right) + \mathcal{O}(\alpha'^{2}) = 0, \\ 
\beta_{\mu\nu}^B &= \alpha' \left( - \frac{1}{2} \nabla^\kappa H_{\kappa\mu\nu} + \nabla^\kappa \Phi H_{\kappa\mu\nu} \right)  + \mathcal{O}(\alpha'^{2}) = 0, \\
\beta^\Phi &= \alpha' \left({D-26\over6\alpha^\prime} - \frac{1}{2}\nabla^2\Phi+\nabla_\kappa\Phi\nabla^\kappa \Phi -\frac{1}{24}H_{\kappa\mu\nu}H^{\kappa\mu\nu} \right)  + \mathcal{O}(\alpha'^{2}) = 0,
\end{split}
\end{eqnarray}
where $H_{\mu\nu\kappa}\equiv\partial_{\mu} B_{\nu\kappa}+\partial_{\nu} B_{\kappa\mu}+\partial_{\kappa} B_{\mu\nu}$. 

These equations tell us in which backgrounds we can consistently define string theory and must coincide with the equations of motion for these background fields if they are to have any physical interpretation. In the absence of $B$ and $\Phi$ the vanishing of the first equation is nothing more than the Einstein equations\footnote{Calculating $\beta_{\mu\nu}^G$ up to order $\alpha'^2$ we get the following equation, $R_{\mu\nu} + \frac{\alpha'}{2}R_{\mu\kappa\lambda\tau}R_{\nu}^{\phantom{\nu}\kappa\lambda\tau} = 0$,  where the second term is a stringy correction to general relativity which vanishes if $\alpha' \to 0$.} in vacuum, $R_{\mu\nu} =0$, which agrees with the assertion made in the previous sentence. An even better test is that these equations can be shown to be the Euler-Lagrange equations comming from the following action,
\begin{eqnarray} \label{supergravityorderalphaprime}
\mathcal{S} = \frac{1}{2\kappa^2_0} \int d^{26}x \ \sqrt{-G} e^{-2\Phi} \left[R+4\nabla_\mu \Phi\nabla^\mu\Phi
-\frac{1}{12}H_{\mu\nu\lambda} H^{\mu\nu\lambda}\right],
\end{eqnarray}
where we have implemented the requirement that $D=26$ as can now be seen from the vanishing of $\beta^\Phi$.  This is called the low energy effective action.

If we include the boundary term \rref{polyakovactiecurvedboundary}  and subsequently require the beta-functions for $A^\mu$ to vanish we get the effective action for the Maxwell field which is nothing more than the D-brane effective action. We will describe this in a more detailed way in the last section of this chapter, thereby illustrating this method.

\subsection{Superstrings}

Even after witnessing all the marvels of bosonic string theory there are some loose ends. Maybe the most important being that in sharp contrast to the wealth of gauge particles we found, not a single fermion was present in the spectrum. The second is the troublesome vacuum for both open en closed strings. To solve these problems we add
world-sheet Majorana fermions, $\psi^\mu$, with Dirac action to the bosonic string action\footnote{With respect to \rref{polyakov-actie in conforme ijk}  we absorbed a factor $(\alpha')^{-1/2}$ into $X^\mu$ and removed an overall factor $(2 \pi)^{-1}$.} \cite{RNSSuperstring1,RNSSuperstring2}. The resulting action in conformal gauge is,
\begin{align}  \label{supersymmetrypolyakov1} 
- \int d^2 \sigma  \left[ \frac{1}{2}  \eta^{ab} \partial_a X^\mu \partial_b X_\mu - i \bar{\psi}^\mu \rho^a \partial_a \psi_\mu \right ].
\end{align}
We choose our Dirac matrices to be, $\rho^0 = \sigma^2$ and $\rho^1 =i \sigma^1$ where $\sigma^1$ and $\sigma^2$ are the Pauli matrices. Hence, $\{\rho^a,\rho^{b}\}=-2\eta^{ab}$. $\psi$ can be written as,
 \begin{align} 
\psi =  \left[\begin{matrix} \psi_- \\ \psi_+ \end{matrix}\right] 
\end{align}
where the components are two Majorana Weyl fermions of opposite chirality, $\rho \psi_\pm = \mp \psi_\pm$, with $\rho=\rho^0\rho^1$. We will raise and lower spinor indices as follows,
$\psi_\mp=\pm\psi^\pm$. For Majorana spinors $\bar{\psi} =  \psi^T \rho^0 $.

For the bosonic string theory the infinite Virasoro algebra as a symmetry algebra was crucial to remove the ghosts. These problems persist here and it can be shown that the spectrum is equally sick for the fermionic sector. To cure this we need an enlarged symmetry algebra. Luckily the action is invariant under the following global transformation of the fields,
\begin{align} \label{supersymmetrytransformpolyakov} 
\begin{split}
\delta X^\mu & = \bar{\varepsilon} \psi^\mu, \\
\delta \psi  &= -\frac{i}{2} \rho^a \partial_a X^\mu \varepsilon,
\end{split}
\end{align}
with $\varepsilon$ a constant Majorana spinor. This symmetry which mixes bosons and fermions is known under the name supersymmetry. By virtue of this the theory has an enhanced symmetry called superconformal symmetry which is generated by the super-Virasoro algebra.  Just like in the bosonic case we should have started from a theory that was reparametrisation invariant, this forces the supersymmetry to become local. Next to that we have to add a zweibein and a Rarita-Schwinger field as auxiliary fields. It can then be shown that, upon gauge fixing, the super-Virasoro conditions needed to remove the ghosts emerge as residual gauge conditions for this full theory.

A basic property of supersymmetry transformations is that the commutator of two of them gives a translation,
\begin{align} \label{supersymmetrycommutator} 
[\delta_1,\delta_2] X^\mu = g^a \partial_a X^\mu \quad \text{and} \quad [\delta_1,\delta_2] \Psi^\mu = g^a \partial_a \Psi^\mu,
\end{align}
where $g^a$ is a constant vector. 
The second relation is only valid using the equations of motion for the fermions. In other words the supersymmetry algebra closes only on-shell.

The equations of motion for $X^\mu$ are just the same as in the previous section and the whole analysis preformed there can be repeated in this case.  The equations of motion for the fermions are nothing more than the two dimensional Dirac equation $\rho^a \partial_a \psi^\mu = 0$. With our conventions this can be show to split into a pair of decoupled equations,
\begin{align} \label{eomfermions}
\begin{split}
( \partial_ \tau + \partial_ \sigma ) \psi_-^\mu = 0, \\
( \partial_ \tau - \partial_ \sigma ) \psi_+^\mu =0.
\end{split}
\end{align}
And so we see  $\psi_-$ and $\psi_+$ describe right and left moving  fermionic modes respectively. We can rewrite the action \rref{supersymmetrypolyakov1} in a form that makes this decoupling manifest by introducing light-cone coordinates, $\sigma^==\tau - \sigma$ and $\sigma^\pp \, =\sigma+\tau$ and derivatives $\partial_\pp \, =\frac 1 2 ( \partial_ \tau + \partial_ \sigma )$ and $\partial_==\frac 1 2 ( \partial_ \tau - \partial_ \sigma)$,
\begin{align} \label{polyakov-actie in conforme ijk susyversion} 
\mathcal{S} & = \int d^2 \sigma \eta_{\mu\nu} \left[ 2 \partial_\pp  \,X^\mu \partial_= X^\nu + 2i \psi_-^\mu \partial_\pp \, \psi_-^\nu  + 2i \psi_+^\mu \partial_= \psi_+^\nu \right].
\end{align}
From this it is fairly easy to read of the equations of motion. It also makes it apparent that we can put for instance $ \psi_+^\mu$ to zero and discuss a theory with right moving fermions only. This can indeed be done and we then get a theory with a left moving side coming from bosonic string theory and an right moving side from superstring theory, however this theory has to few left moving degrees of freedom to be consistent. This can be solved by adding extra left moving fields carrying gauge field indices and leads to the two so called heterotic string theories\footnote{Alternatively, these theories can be constructed by combining the right moving sector from the Type II superstring (discussed later) with the left moving sector from the bosonic string followed by compactifying the 16 extra left moving bosons on particular 16-dimensional tori leading to Type HO and Type HE.}: Type HO and Type HE \cite{Heteroticdiscovery1} and \cite{Heteroticdiscovery2}. Both have critical dimension $D=10$. 

In varying the action \rref{polyakov-actie in conforme ijk susyversion} we get the following boundary term for the fermions,
\begin{align} \label{polyakov-actie in conforme ijk susyversion boundaryterm} 
-i \int d \tau \eta_{\mu\nu}\left( \psi_-^\mu \delta \psi_-^\nu  - \psi_+^\mu \delta \psi_+^\nu \right) \big|^{\sigma=\bar{\sigma}}_{\sigma=0}. 
\end{align}
In the case of open strings this is zero if we impose $\psi_+^\mu=\pm \psi_-^\mu$ at each end. The overall relative sign is a matter of convention and we can set $\psi_+^\mu(\tau,0)=\psi_-^\mu(\tau,0)$ without loss of generality. 
Now for the other end we have two possibilities:
\begin{itemize}
\item Ramond (R) boundary conditions
\begin{align} \label{ramond} 
\psi_+^\mu(\tau,\pi)=\psi_-^\mu(\tau,\pi).
\end{align}
We then get the following mode expansion,
\begin{align} \label{ramondsolution}
\psi_\pm^\mu(\tau,\sigma) = \frac{1}{\sqrt{2}}\sum_{n \in Z} d_n^\mu
e^{-in(\tau\pm\sigma)}.
\end{align}
\item Neveu-Schwarz (NS) boundary conditions
\begin{align} \label{neveuschwarz} 
\psi_+(\tau,\pi)=-\psi_-(\tau,\pi).
\end{align}
We then get the a mode expansion over the half integers,
\begin{align} \label{neveuschwarzsolution}
\psi_\pm^\mu(\tau,\sigma) = \frac{1}{\sqrt{2}}\sum_{r \in Z + 1/2} b_r^\mu
e^{-ir(\tau\pm\sigma)}.
\end{align}
\end{itemize}

For closed strings we can put the boundary term to zero by requiring periodicity or antiperiodicity for each component of $\psi$ separately. Demanding periodicity or antiperiodicity leads to the same mode expansion as for Ramond respectively Neveu-Schwarz boundary conditions. Since we can choose NS or R for the left and right sector separately we get four different sectors for the closed string: NS-NS, NS-R, R-NS and R-R. The spectrum for the closed string is the product of the spectrum for the left en right moving side.

We then proceed by quantizing this theory much in the same way as we did for the bosonic case, replacing the Poison brackets by commutators or anticommutators depending on wether the variables involved are bosonic or fermionic and interpreting the fields as operators. We take $\alpha_m^\nu, d_n^\mu$ and $b_r^\mu$ to be creation operators for $m,r<0$ and annihilation operators for $m,r>0$. Having two distinct string states depending on wether we have R or NS boundary conditions means we have two sectors which we analyse separately. We then impose the super-Virasoro constraints to eliminate the ghosts from the spectrum and find it to be ghostfree in both cases if $D=10$. 

As we will see the spectrum again has a tachyon, but in this case we can truncate the spectrum using the so called GSO projection \cite{GSODiscovery1,GSODiscovery2} to eliminate it. The GSO projection is needed to make the theory modular invariant\footnote{See \cite{bookpolchinski} for more on modular invariance and it's relation to the GSO projection.} which in turn is needed to make the theory consistent. The GSO projection tells us we should only keep states with a fixed eigenvalue $f$ under the operator $(-1)^F$ where $F$ is the fermion number of this state. With our conventions we have to take $f=+1$ in the NS sector and $f=\pm1$ in the R sector.  

In the NS-sector we define the vacuum state $\left| 0;k;\text{NS} \right\rangle$ to be annihilated by $\alpha_m^\nu$ and $b_r^\mu$ for $m,r>0$. This state is a tachyon and we assign odd fermion number to it in order to have it removed by the GSO projection.  The first exited state is $ \left| \zeta;k \right\rangle = \zeta \cdot b_{-1/2} \left| 0;k;\text{NS} \right\rangle$ with $\zeta$ a polarization vector. All properties are just the same as for the first excited state for open strings in the bosonic case. And we identify this as a massless vector boson $A^\mu$. This has $8$ physical degrees of freedom.

In the R-sector we define the vacuum state $\left| 0;k;\text{R} \right\rangle$ to be annihilated by $\alpha_m^\nu$ and $d_n^\mu$ for $m>0$. This state has 32-fold degeneracy, since the $d_0^\mu$ take groundstates into groundstates. After imposing the super-Virasoro constraints we find the groundstate to be a Majorana spinor having 16 components obeying the Dirac equation. In ten dimensions this can be split into two Majorana-Weyl spinors of different chirality having 8 components. The chirality operator can be  shown to coincide with $(-1)^F$ for the vacuum and so we are free to keep either of the two depending on which GSO projection we choose. Both are equivalent as making a spacetime reflection on a single axis flips the sign of the chirality. 

All in all we find the ground state for the open superstring to be a vector multiplet of $D = 10$, $N=1$ spacetime supersymmetry. We will not show this but point out that at least one of the necessary conditions, namely the presence of an equal amount of bosons and fermions, is obeyed and this by virtue of the GSO projection which halved the amount of fermions in the R sector.

Before turning to theories with open 
 strings we will discuss theories containing only closed strings. As we saw above the spectrum is the product of the spectrum for the left en right moving side. We thus expect the NS-NS and R-R sector to contain bosons while the fermions reside in the NS-R and R-NS sector. 

It turns out we have two different theories depending on wether we choose the same or opposite GSO-projection (chirality) for the left and the right movers in the Ramond sector. Choosing opposite GSO-projection leads to a non-chiral theory called Type IIA while 
choosing the same leads to a chiral theory called Type IIB. Both live in 10 dimensions.

Analyzing Type IIA first we get the same massless states as for the closed bosonic string in the NS-NS sector: $G_{\mu\nu}$, $B_{\mu\nu}$ and $\Phi$. The R-R sector contains a vector $C_{[1]}$ and a three-form $C_{[3]}$. Finally, the 
NS-R and R-NS sector is made up of two Majorana-Weyl spinors and two gravitini (spin $3/2$ particles) of opposite chirality. These states combine to form a multiplet of $N=2$ non-chiral supergravity in ten dimensions. 

For Type IIB the NS-NS sector has the same contents as before: : $G_{\mu\nu}$, $B_{\mu\nu}$ and $\Phi$.  The R-R sector contains scalar $C_{[0]}$ a two-form $C_{[2]}$ and a four-form $C_{[4]}^+$ with self dual field strength. And the NS-R and R-NS sector is again made up of two Majorana-Weyl spinors and two gravitini but this time of equal chirality. These states combine to form a multiplet of $N=2$ chiral supergravity in ten dimensions.

As was argued before theories with open strings also contain closed strings and open strings have $N=1$ supersymmetry. This implies that the closed strings we need can not be the Type II theories as they have to much supersymmetry to couple in a consistent way. To resolve this problem we notice that since Type IIB has the same chirality on both sides we can define a parity operator $\Omega$ which exchanges left and right moving modes. 
Obviously $\Omega^2 = 1$ and we can use it to construct a projection operator, $U=\frac{1}{2}(\Omega+1)$ which we can use to project out the states which are not invariant under $\Omega$. Doing so we get an unoriented closed string theory having $N=1$ supersymmetry called Type I closed unoriented string theory. Although this theory by itself is not consistent it is precisely the one that couples to open strings (which is already unoriented). As it turns out there is a further inconsistency which can be resolved by adding extra degrees of freedom, known as Chan-Paton factors, to the strings endpoints. In the case of unoriented strings this enhances the gauge group to $Sp(n)$ or $SO(n)$ but only the choice $SO(32)$ leads to a consistent theory called Type I.

\subsection{Superstrings in Superspace} \label{generalsuperspacesection}

To gain more insight in the supersymmetry transformation \rref{supersymmetrytransformpolyakov} it is useful to recast the superstring-action \rref{supersymmetrypolyakov1} in $N=(1,1)$ superspace \cite{FairlieMartinSuspa,MontonenSuspa,superspaceGates1983nr}.  To do so we supplement the two world-sheet coordinates $\sigma^a$ with two Majorana-Weyl fermionic coordinates, $\theta^+$ and $\theta^-$. A general function in superspace or superfield can be written as,
\begin{equation}
\Sigma(\sigma,\theta)= \Sigma_0(\sigma) - i \theta^+ \Sigma_+(\sigma)  - i \theta^- \Sigma_-(\sigma)  + i \theta^- \theta^+ \Sigma_2(\sigma) .
\label{superfield}
\end{equation}
This equation is the complete power series expansion in powers of $\theta^\pm$ since for these anticommuting variables $(\theta^\pm)^2=0$ holds. The transformation properties of the components of course depend on the transformation properties of the superfield $\Sigma(\sigma,\theta)$ which equal those of the first component. Needless to say, the product of two superfields is again a superfield.

The real coordinate superfield is given by,
\begin{equation}
\Phi^\mu = X^\mu - i \theta^+ \psi_+^\mu - i \theta^- \psi_-^\mu + i \theta^- \theta^+ F^\mu,
\label{realcoordinatsuperfield}
\end{equation}
and unites $X^\mu$ and $\psi^\mu$ together with a new field $F^\mu$ which as we will see is an auxiliary field.

Supersymmetry on superspace is represented by the following supersymmetry generators or supercharges,
\begin{equation}
Q_+ = \frac{\partial}{\partial \theta^+} + i \theta^+ \partial_{\pp} \, , \quad \quad
Q_- = \frac{\partial}{\partial \theta^-} + i \theta^- \partial_{=} ,
\end{equation}
which satisfy,
\begin{equation} \label{whatisrelationssupercharges}
Q_+^2 =  i \partial_{\pp} \, , \quad \quad Q_-^2 = i \partial_{=}, \quad \quad \left\{ Q_+ ,Q_- \right\} = 0 .
\end{equation}
We define the transformation of a superfield as follows,
\begin{equation}  \label{whatissusyn=11suspa}
\delta(\varepsilon^+) \Sigma = \left[ \varepsilon^+ Q_+,\Sigma \right], \quad \quad
\delta(\varepsilon^-) \Sigma = \left[ \varepsilon^- Q_-,\Sigma \right],
\end{equation}
with $\varepsilon^\pm$ constant anticommuting spinors. Using the relations \rref{whatisrelationssupercharges} it is easy to check that commutators of two of the same kind of these $\left[ \delta(\varepsilon^\pm_1), \delta(\varepsilon^\pm_2) \right]$ give a translation on the worldsheet and the commutators of two of a different kind give zero, $\left[ \delta(\varepsilon^\pm_1), \delta(\varepsilon^\mp_2) \right] = 0$. 

Performing the $N=(1,1)$ supersymmetry transformation $\varepsilon^+ Q_+ + \varepsilon^- Q_-$  on the superfield $\Phi^\mu$ one can easily read of the transformations for each component,
\begin{equation}
\begin{split} \label{whatissusytransformationcomponents}
\delta X^\mu & = -i \varepsilon^+ \psi_+^\mu - i \varepsilon^- \psi_-^\mu, \\
\delta \psi_+^\mu & = \varepsilon^+ \partial_{\pp} \, X^\mu + \varepsilon^- F^\mu, \\
\delta \psi_-^\mu & = \varepsilon^- \partial_{=} X^\mu - \varepsilon^+ F^\mu, \\
\delta F^\mu & = + i \varepsilon^+ \partial_{\pp} \, \psi_-^\mu - i \varepsilon^- \partial_{=} \psi_+^\mu,
\end{split}
\end{equation}
which perfectly agrees with \rref{supersymmetrytransformpolyakov} if we set $F^\mu=0$, corresponding to the absence of the auxiliary field in our initial formulation, and impose the Dirac equations, \rref{eomfermions}. By virtue of the auxiliary field we have achieved closure of the supersymmetry algebra without using the equations of motion, in other words, it closes off-shell.

Next we introduce covariant derivatives,
\begin{equation}
D_+ = \frac{\partial}{\partial \theta^+} - i \theta^+ \partial_{\pp} \, , \quad \quad
D_- = \frac{\partial}{\partial \theta^-} - i \theta^- \partial_{=} .
\end{equation}
They obey the following relations,
\begin{equation} \label{whatisrelationssupercovariantderivatives}
\begin{split}
D_+^2 =  -i \partial_{\pp} \, , \quad \quad D_-^2 = -i \partial_{=}, \quad \quad \left\{ D_+ ,D_- \right\} = 0, \\
 \left\{Q_\pm ,D_+ \right\} = 0 ,  \quad \quad   \left\{Q_\pm ,D_- \right\} = 0 . \quad \quad \quad
\end{split}
\end{equation}
The second line implies the covariant derivatives commute with the supersymmetry transformations, thus justifying their name.

To build an action we also need a notion of integration over superspace. The correct definition turns out to be the Berezin integration,
\begin{equation}
\int d\theta^+(A+\theta^+B) = B , \quad \quad \int d\theta^-(C+\theta^-D) = D,
\end{equation}
which follows automatic if we demand integration to be a linear operation invariant under a translation, $\theta^\pm \to \theta^\pm +  \theta^\pm_t$. So we see that integration over anticommuting variables acts as a derivative,
\begin{equation}
\int d\theta^+ \Sigma = \frac{\partial}{\partial \theta^+} \Sigma =\left[ D_+ \Sigma \right], \quad \quad \int d\theta^- \Sigma = \frac{\partial}{\partial \theta^-} \Sigma =\left[ D_- \Sigma \right], 
\end{equation}
where the square brackets denote putting $\theta^+=0$ and $\theta^-=0$ respectively. The last equalities are easy to see from the form of $D_\pm$.

Building actions invariant under supersymmetry is straightforward since by construction any scalar action of the following form is invariant,
\begin{equation} \label{scalaractionwhatis}
\int d^2\sigma d\theta^+  d\theta^- \Sigma = \int d^2\sigma \left[ D_+ D_- \Sigma \right],
\end{equation}
where the square brackets now denote $\theta^+= \theta^-=0$. This can explicitly be checked using integration by parts.

An action of special interest is the following,
\begin{equation}
\begin{split}
{\cal S}_{\text{bulk}}&= 2 \int d^2\sigma d\theta^+ d\theta^-  \left(  D_+ \Phi^\mu D_- \Phi_\mu \right)\\
& = 2 \int d^2\sigma \left[ D_+ D_- \left( D_+ \Phi^\mu D_- \Phi_\mu \right) \right].
\end{split}
\label{bulksperspacewhatisstringohnebgfields}
\end{equation}
Performing the integration over $\theta^\pm$ explicitly we get,
\begin{equation}
\begin{split} \label{bulkwhatisstringohnebgfieldscomponents}
{\cal S}_{\text{bulk}}& = \int d^2 \sigma \eta_{\mu\nu} \left[ 2 \partial_\pp  \,X^\mu \partial_= X^\nu + 2i \psi_-^\mu \partial_\pp \, \psi_-^\nu  + 2i \psi_+^\mu \partial_= \psi_+^\nu
+2 F^\mu F^\nu  \right].
\end{split}
\end{equation}
Of course this action is still invariant under the supersymmetry transformations \rref{whatissusytransformationcomponents}. It is also easy to see the field equations say $F^\mu =0$, so we can simply set $F^\mu$ to zero and forget it. Upon doing so we find our action from the previous section \rref{polyakov-actie in conforme ijk susyversion} but with more insight in how supersymmetry comes about.

\subsection{Superstrings in Background Fields} \label{Superstrings_in_background_fields}

Just as in the bosonic case one can ask the question what happens if background fields are present. We will first consider the  superstring moving in a NS-NS background containing  only a gravitational field $G_{\mu\nu}$ and a Kalb-Ramond field $B_{\mu\nu}$ since these are the only closed string background fields we will need in the remainder of this thesis. To do so the superspace technology introduced in the  previous section pays of as we can easily generalize the action \rref{bulksperspacewhatisstringohnebgfields} to nontrivial backgrounds. The superspace $N=(1,1)$ non-linear sigma-model describing the coupling to the  fields mentioned above reads,
\begin{equation}
{\cal S}_{\text{bulk}}= 2\, \int d^2\sigma \left[ D_+ D_- \left((G_{\mu\nu}+B_{\mu\nu}) D_+ \Phi^\mu D_- \Phi^\nu\right)\right],
\label{bulksupwhatissuperstringsbackgroundfields}
\end{equation}
where the term containing $G_{\mu\nu}$ is the obvious generalisation of the action \rref{bulksperspacewhatisstringohnebgfields} to a curved background and the term containing $B_{\mu\nu}$ describes the coupling to the Kalb-Ramond field. This action is by construction invariant under the supersymmetry transformations \rref{whatissusyn=11suspa}.

In the remainder of this thesis we will also need the corresponding Lagrangian in components. Working out the the integration over $\theta^\pm$ explicitly we get,
\begin{equation}
\begin{split}
{\cal L}_{\text{bulk}}& = 
2(G_{\mu\nu}+B_{\mu\nu}) \partial_{\pp} \, X^\mu \partial_= X^\nu
+ 2i G_{\mu\nu} \psi^\mu_+\nabla_=^{(+)} \psi ^\nu_+ \\
& + 2i G_{\mu\nu} \psi ^\mu_-\nabla_{\pp}^{(-)} \psi ^\nu_-
 +R^{(-)}_{\mu\nu\kappa\lambda} \psi _-^\mu \psi _-^\nu \psi _+^\kappa \psi _+^\lambda \\
&+2 (F^\mu-i \eta \Gamma ^{\ \mu}_{(-)\kappa\lambda} \psi ^\kappa_- \psi ^\lambda_+ )G_{\mu\nu}
(F^\nu-i \eta \Gamma ^{\ \nu}_{(-)\iota \rho} \psi ^\iota_- \psi ^\rho_+ ).\label{whatislag11}
\end{split}
\end{equation}
which is obviously a generalisation of \rref{bulkwhatisstringohnebgfieldscomponents} but as can be seen from its form it would have been difficult to derive it as an evident generalisation from there. Again this demonstrates the power of superspace techniques.

To interpret the Lagrangian given above we need some notation wich we will introduce here. The torsion is given by the curl of the Kalb-Ramond field,
\begin{equation}
T_{\mu\nu\kappa}=-\frac 3 2 B_{[\mu\nu,\kappa]}.
\end{equation}
We use this to introduce two connections,
\begin{equation}
\Gamma^{\ \mu}_{(\pm) \nu\kappa} = \big\{{}^\mu{}_{\nu\kappa}\big\}\pm T^\mu{}_{\nu\kappa},
\label{torsieconn}
\end{equation}
where $\big\{{}^\mu{}_{\nu\kappa}\big\}$ is the standard Christoffel connection. The
connections are used to define covariant derivatives,
\begin{equation}
\begin{split}
\nabla^{(\pm)}_\mu V^\nu&= \partial_\mu V^\nu+\Gamma^{\ \nu}_{(\pm) \kappa\mu}V^\kappa,\\
\nabla^{(\pm)}_\mu V_\nu&= \partial_\mu V_\nu-\Gamma^{\ \kappa}_{(\pm) \nu\mu}V_\kappa.
\end{split}
\end{equation}
We define curvature tensors by the following relation,
\begin{equation}
{[} \nabla_\mu^{(\pm)},\nabla_\nu^{(\pm)}{]}V^\kappa=\frac 1 2 V^\lambda R_{(\pm)\lambda\mu\nu}^{\ \kappa}\pm
T^\lambda{}_{\mu\nu}\nabla^{(\pm)}_\lambda V^\kappa,  \label{muntwhatis}
\end{equation}
and we get explicitly,
\begin{equation}
\begin{split}
R^{\ \mu}_{(\pm)\nu\kappa\lambda}&= \Gamma^{\ \mu}_{(\pm)\nu\lambda,\kappa}+ \Gamma ^{\ \mu}_{(\pm)\rho\kappa}\Gamma ^{\ \rho}_{(\pm)\nu\lambda}
- \left( \kappa\leftrightarrow \lambda\right), \\
R^{(\pm)}_{\mu\nu\kappa\lambda}&= \Gamma ^{(\pm)}_{\mu\nu\lambda,\kappa}+\Gamma ^{(\pm)}_{\rho\mu\lambda} \Gamma ^{\ \rho}_{(\pm)\nu\kappa}-\left(\kappa\leftrightarrow \lambda\right).
\end{split}
\end{equation}
The curvature tensors $R^{(\pm)}_{\mu\nu\kappa\lambda}$ are anti-symmetric in the first and the last two indices, they also
satisfy
\begin{equation}
R^+_{\mu\nu\kappa \lambda }=R^-_{\kappa\lambda\mu\nu}.
\end{equation} 

For strings moving in a background carrying a coherent state of massless open strings in the NS sector we have to add a boundary term containing the gauge field $A^\mu$. 
We need not write it explicitly as this can then be encoded as before by replacing  $B_{\mu\nu}$ by $\mathcal{F}_{\mu\nu} = B_{\mu\nu}  + 2 \pi \alpha' F_{\mu\nu}$ in \rref{bulksupwhatissuperstringsbackgroundfields} and \rref{whatislag11}. However, as we will discuss in great detail in the next chapter, actions like \rref{scalaractionwhatis} are not supersymmetric in the presence of boundaries and cannot used to describe open strings. This implies we can certainly not use them to describe how these couple to gauge field backgrounds. For this we again refer to the next chapter.

\section{Non Perturbative String Theory and Dualities}

In the previous section we have seen that there are five consistent string theories which all live in ten dimensions and not in the more familiar four dimensions we are used to and live in. A natural solution is then to reduce a $d$-dimensional subspace to a $d$-dimensional compact space. Ultimately, one would want this proces to appear dynamically so that $d$ and the precise form of the compact space are determined by the theory and not imposed by hand. We are not yet at this point. 


\subsection{Compactification and $T$-duality for Closed Strings}

The simplest case one can consider is to compactify one direction, say $X^{l}$, on a circle of radius $R$,
\begin{equation} \label{identificatie}
X^l \cong X^l + 2 \pi R.
\end{equation}
Since the action is a local object and making a compactification is a topological change this does not alter the equations of motion, hence we can suffice by studying the effects on the boundary conditions. For closed strings the periodicity conditions \rref{geslotensnaren} ensure that the boundary term in the variation of the action is automatically zero.  For the compact dimension we can relax this condition to,
\begin{equation} \label{open_snaar_in_compacte_dimensie}
X^{l}(\sigma + 2 \pi) = X^{l}(\sigma) + 2 \pi R w, \quad \quad w \in \bf{Z},
\end{equation}
which tells us we must allow the string to wind $w$ times around the circle. This has the following effect on the general solution in the compact dimension\footnote{In this $(\text{oscill. contr.})$ stands for the oscillator contributions as found in \rref{Algemeneoplossinggeslotenensnaren}.},
\begin{equation}
\begin{split}
 \label{AlgemeneoplossinggeslotensnarenLRcomp}
  &X_L^l(\sigma^+)=\frac{x^l}{2} + \frac{\hat{x}^l}{2} +\frac{\alpha'}{2}(p^{l} + \frac{wR}{\alpha'})(\tau+\sigma) + (\text{oscill. contr.}), \\
&X_R^l(\sigma^-)  =\frac{x^l}{2} - \frac{\hat{x}^l}{2} +\frac{\alpha'}{2}(p^{l} - \frac{wR}{\alpha'})(\tau-\sigma) + (\text{oscill. contr.}),
\end{split}
\end{equation}
resulting in,
\begin{equation}
\label{Algemeneoplossinggeslotenensnarencomp}
X^l(\tau,\sigma)=x^l + \alpha' p^{l} \tau + w R \sigma + (\text{oscill. contr.}).
\end{equation}

Now consider the operator $\exp(i p^{l} 2\pi R )$ which describes a translation over a distance $2 \pi R$ in the compact direction. Obviously this operator has to be equal to unity and we find that the momentum in the compact direction has to be quantized according to,
\begin{equation} \label{kwantisatie impuls}
p^{l} = \frac{n}{R} , \quad \quad  n \in \bf{Z}.
\end{equation}

Taking this into account, one finds the following operator for the mass,
\begin{equation} \label{massoperatorkk}
    M^2 =   \frac{n^2}{R^2} +
    \frac{w^2R^2}{\alpha'^2} +  (\text{oscill. contr.}),
\end{equation}
where  $M^2 = - p_{\tilde \mu} p^{\tilde \mu}$  with $\tilde \mu$ running over the non-compact directions, in other words, this is the mass as seen from the non-compact piece of space-time. The second factor is the energy necessary to wind the string around the circle. 

Upon taking the limit $R\rightarrow\infty$ we observe that all states with $w \neq 0$ get infinite mass and decouple while the discrete set of states with $w = 0$ and $n \in \bf{Z}$ turn into a continuous set of states. This is to be expected if we undo the compactification by letting the radius grow to infinity. Taking the inverse limit  $R\rightarrow 0 $ we see that all states with $n \neq 0$ get infinite mass and decouple. In ordinary field theory this is all we expect. Here however, the states with $n =0$ and $w \in \bf{Z}$ open up a new continuum since it costs little effort to wrap a small circle. So we see that upon letting the radius approach zero we again end up in the uncompactified theory. 

It is not hard to see that the mass \rref{massoperatorkk} is invariant under the following transformation,
\begin{equation} \label{tdualnwrdualr}
\begin{split}
R   \rightarrow & \, R' = \frac{\alpha'}{R}, \\
w  & \leftrightarrow n,
\end{split}
\end{equation}
which is known as $T$-duality and where $R'$ is called the dual radius. As a matter of fact, the whole theory is invariant under this transformation.  From this we observe that the space of inequivalent theories is the halfline $R \geq \sqrt{\alpha'}$.

If we define the action of $T$-duality on the oscillators to be given by, $\alpha_{n} \leftrightarrow \tilde \alpha_{n}$ this, combined with \rref{tdualnwrdualr}, results in the following transformation on the fields,
\begin{equation} \label{geslotensnarenniewevar}
X^l(\tau,\sigma)= X^l_L(\sigma^+) + X^{l}_R(\sigma^-)  \rightarrow  \hat X^l(\tau,\sigma) = X^l_L(\sigma^+) - X^{l}_R(\sigma^-).
\end{equation}

The analysis up to here was only valid for the bosonic string and the bosonic piece of the superstring. Deriving the way $T$-duality acts on the fermions is quite easy if we take in account that after $T$-dualizing the theory still has to be supersymmetric.  We can then easily read of the following rules from the supersymmetry transformations \rref{supersymmetrytransformpolyakov} or \rref{whatissusytransformationcomponents},
\begin{equation} \label{geslotensnarenniewevarfermions}
\begin{split}
 \psi_+^{l} & \rightarrow  \psi_+^l \\
 \psi_-^l & \rightarrow  - \psi_-^{l}
 \end{split}
\end{equation}

We observe that $T$-duality acts essentially as a spacetime parity operator on the right moving side of the worldsheet. A consequence of this is that it reverses the chirality of the right moving Ramond groundstate and hence reverses the relative chiralities of the left and right moving Ramond groundstates. This implies that $T$-duality changes Type IIA into Type IIB theory and vice-versa. 

\subsection{$T$-duality for Open Strings}

Open strings have no way to wind non-trivially around the circle and thus the boundary conditions are not altered. This implies that the general solution in the compact dimension is still given by \rref{AlgemeneoplossingopensnarenLR} and \rref{Algemeneoplossingopensnaren}. The mass as seen from the non-compact dimensions is now given by,
\begin{equation}
M^2 = \frac{n^2}{R^2} + (\text{oscill. contr.}).
\end{equation}

Performing the same analysis as for closed strings we see that upon taking the limit $R \rightarrow 0$ the states with nonzero momentum get infinite mass and decouple, but there is no new continuum due to the fact that there are no winding states. So just as in ordinary field theory we are left with a theory with one dimension less than we started with. 

As we already mentioned, theories with open strings have closed strings as well. This gives rise to the unsatisfactory observation that in the $R \rightarrow 0$ limit closed strings live in $D$ dimensions while open strings live in $D-1$. This seeming paradox can be resolved by going to the $T$-dual theory,
\begin{equation}
\label{Algemeneoplossingopenensnarencomp}
\begin{split}
 \hat{X}^l(\tau,\sigma) & = X^l_L(\sigma^+) - X^{l}_R(\sigma^-), \\
& =\hat{x}^l + 2 n  R' \sigma + (\text{oscill. contr.}).
\end{split}
\end{equation}
The Neumann boundary conditions on the original coordinate become Dirichlet boundary conditions on the dual coordinate,
\begin{align} \label{dirichletdualneumann}
0 = \partial_\sigma X^l = \partial_\tau \hat{X}^l \quad \text{at} \quad \sigma=0,\pi.
\end{align}

Due to the Dirichlet boundary conditions we observe that the endpoints of the open string do not move in the $\hat{X}^l$ direction but are confined to a fixed plane,
\begin{align} \label{opentfixedplane}
 \hat{X}^l(\pi)  =  \hat{X}^l(0) + 2\pi R' n.
\end{align}
This tells us that the open string warps the dual circle $n$ times which is possible due to the fixed endpoints. Just like in the closed string case the quantized momentum is turned into winding number.  
 The $(D-1)$-dimensional plane to which the endpoints are confined are nothing more than the worldvolume swept out by a $(D-2)$-dimensional object, called a D-brane \cite{Dai:1989ua}. We will threat them in the next section. 

After $T$-dualizing the mass becomes,
\begin{equation}
M^2 =     \frac{n^2 R'^2}{\alpha'^2} + (\text{oscill. contr.}).
\end{equation}
Now we see that in the $R' \rightarrow 0$ limit we again end up in the uncompactified theory but with the endpoints of the open string stuck to a $(D-1)$-dimensional hyperplane thus resolving our problem.

The observations made in the preceding paragraphs are in accordance with the intuitive idea that the interior of the open string is indistinguishable from the closed string and should still be vibrating in $D$ dimensions after $T$-duality. The only point where open and closed strings differ are the endpoints and so it are only these that should be restricted to a $(D-1)$-dimensional hyperplane.

\subsection{$T$-duality and Wilson Lines}

Now let us consider what happens if we turn on a  constant background gauge potential of the following form,
\begin{equation} \label{tdualwila}
A_{l}(X^{\mu}) = \frac{\vartheta}{2 \pi R} = -i \Lambda^{-1} \frac{\partial}{\partial X^{l}}  \Lambda,
\end{equation}
with,
\begin{equation}
\Lambda(X^{l}) = \exp \left(  i \frac{\vartheta}{2 \pi R} X^{l} \right).
\end{equation}
Locally a field of this form is pure gauge and so it can be gauged away. Globally this can not be done as can be observed from the fact that the following gauge invariant quantity, called Wilson line, 
\begin{equation}
W = \exp \left(  i \oint dX^{l}  A_{l} \right) = \exp  \left(  i \vartheta \right),
\end{equation}
is non-zero.

Now assume for the sake of argument that only the $\sigma = \pi$ endpoint is charged under \rref{tdualwila}. Then the state describing such a string picks up a phase exactly equal to $W$ under a transformation $X^l \to X^l + 2 \pi R.$ This in turn makes the open string momentum fractional as follows,
\begin{equation} \label{kwantisatie impuls fractional}
p^{l} = \frac{n}{R} + \frac{\vartheta}{2 \pi R}    , \quad \quad  n \in \bf{Z}. 
\end{equation}
Relation \rref{opentfixedplane} is then observed to be modified to, 
\begin{align} \label{opentfixedplanewilsonline}
 \hat{X}^l(\pi)  =  \hat{X}^l(0) + (2\pi n + \vartheta) R'.
\end{align}
From this we see that, up to an arbitrary additive constant, the endpoint is at position,
\begin{align} \label{opentwilsonlinerelationgaugefield}
 \hat{X}^l  =   \vartheta R' = 2 \pi \alpha' A_{l}
\end{align}

\subsection{D-branes}

After unveiling the existence of D-branes we now focus on them exclusively. In this section we consider open strings in the presence of a $p$-dimensional D-brane which from here on we will call a D$p$-brane. We can obtain this configuration by applying $T$-duality in $9-p$ dimensions. Such a configuration is thus described by  Neumann boundary conditions in $p+1$ directions and Dirichlet boundary conditions in the remaining $9-p$ directions,
\begin{equation}\label{mixedneumanndirichletdbranes}
\begin{split}
 \partial_\sigma X^\mu = 0 \quad &\text{for} \quad \mu \in \{0,\ldots,p\}, \\
 \partial_\tau X^{\tilde \mu} = 0 \quad &\text{for} \quad \tilde \mu \in \{p+1,\ldots,9\}.
\end{split}
\end{equation}

From \rref{dirichletdualneumann} we observe that $T$-duality interchanges Dirichlet and Neumann boundary conditions. Thus applying $T$-duality in a direction tangent to the D$p$-brane produces a D$(p-1)$-brane while doing so in a direction normal to the brane results in an D$(p+1)$-brane.

Making a change in boundary conditions as in \rref{mixedneumanndirichletdbranes} yields the a modification to the open string spectrum. It is split as follows,
\begin{equation}
 A_{\mu} \quad \text{for} \quad  \mu \in \{0,\ldots,p\}, \quad \quad \text{and} \quad \quad  \Phi_{\tilde \mu} \quad \text{for} \quad \tilde \mu \in \{p+1,\ldots,9\},
\label{}
\end{equation}
where the $A_{\mu}$ describe a gauge theory living on the brane while the $\Phi_{\tilde \mu}$ are collective
coordinates describing the position of the brane and hence the shape of its embedding in space-time. This was to be expected since for the directions tangential to the D-brane nothing has changed and so we should still have our original gauge fields. For the directions normal to the brane we get new fields $\Phi_{\tilde \mu}$ describing the transverse position of the D-brane as we could anticipate from the fact that the value of $A_{\tilde \mu}$ has the interpretation of the position of the D-brane in the $T$-dual picture, \rref{opentwilsonlinerelationgaugefield}. This is in accordance with the fact that D-branes are in fact dynamical objects. The presence of a D-brane breaks translational invariance in the $9-p$ transversal directions and the $\Phi_{\tilde \mu} $ are nothing more than the Goldstone bosons associated with the breaking of symmetry.

D-branes had been around for a while, they were discovered using the $T$-duality arguments used in the preceding paragraphs in a beautiful paper \cite{Dai:1989ua}. However their importance was only realized much later in \cite{polchinskidbranesrrcharges} where it was shown, solving a long-standing problem, that they in fact are the basic carriers of RR charge. Indeed a D$p$-brane couples electrically to a $(p+1)$-form as follows,
\begin{equation} \label{branecoupletoforms}
\int_{\mathcal{M}_{p+1}} C_{[p+1]},
\end{equation}
where the integral is over the $(p+1)$-dimensional worldvolume of the D-brane.  On the other hand, a D$(6-p)$-brane couples magnetically to the same $(p+1)$-form as follows,
\begin{equation}
\int_{\mathcal{M}_{7-p}} {}^{\star} C_{[p+1]} 
\end{equation}
where ${}^{\star} C_{[p+1]}$ is the $(7-p)$-form for which the fieldstrength is the Hodge dual of the fieldstrength for $C_{[p+1]}$.

In \cite{polchinskidbranesrrcharges} it was argued that one should add D$p$-branes and the corresponding open strings to type II theories (these in fact then cease to be type II since the presence of D-branes breaks half of the supersymetry). This results in a consistent string theory provided $p$ is even in the IIA theory or odd in the IIB theory, which is in accordance with the RR field content of both theories. Indeed, the $C_{[1]}$ and $C_{[3]}$ forms from Type IIA theory couple to D0-, D2-, D4- and D6-branes while the $C_{[0]}$, $C_{[2]}$ and $C_{[4]}^+$ forms from Type IIB couple D$(-1)$-, D1-, D3-, D5- and D7-branes.
In addition there is a D$8$-brane in Type IIA theory and a D$9$-brane in Type IIB theory, the forms they couple to do not correspond to propagating states and hence we do not see them in the spectrum. Although it appears that we have modified the type II theories by adding something new to it, D-branes are actually intrinsic to any nonperturbative formulation of type II theories.


To finish we remark that it is possible to get the Type I theory from Type IIB theory. To do so we have to introduce a space-filling orientifold plane O$9$. Orientifold planes arise when one makes the theory invariant under a combination of spacetime and worldsheet parity. The resulting theory is inconsistent but we can cure it by inserting 16 D9-branes. This turns out to be equivalent to adding 32-valued Chan-Paton factors to the string endpoints resulting in the familiar $SO(32)$ gauge-group from Type I theory.





\section{The D-brane Effective Action}

If we want to study the dynamics for a D-brane, it is very useful to derive an action for it.  A D$p$-brane sweeps out a $(p+1)$-dimensional worldvolume when moving in spacetime.  Just as for the point-particle and the string we introduce coordinates, $\xi^{a}$ for $a \in \{0, \ldots, p \}$, on the D-brane worldvolume. The natural choice is to take an action proportional to the worldvolume as follows,
\begin{equation} \label{Nabuagotoactieforbranes}
\mathcal{S}_{p}=- \tau_{p} \int d^{p+1}\xi   \left(- \det  G_{ab} \right)^{1/2},
\end{equation}
where $G_{ab}$ is the pullback of the metric to the brane,
\begin{equation}
 G_{ab} = \frac{\partial X^{\mu}}{\partial \xi^{a}} \frac{\partial X^{\nu}}{\partial \xi^{b}} G_{\mu \nu},
\end{equation}
and $\tau_{p} $ is the tension of the D${p}$-brane. It can be rewritten  as,
\begin{equation} \label{tensiotau}
\tau_{p} = e^{-\Phi} \, T_{p} = g_{\text{s}}^{-1} T_{p}, 
\end{equation}
where $T_{p}$ is the tension with all dependence on the string coupling constant $g_{\text{s}}$ removed. Its explicit form is not needed for our purposes and deriving it would lead us too far. The fact that $\tau_{p}$ is proportional to $e^{-\Phi}$ is due to the fact  we are working at open string tree level\footnote{Compare to the factor $e^{-2\Phi}$ in  \rref{supergravityorderalphaprime} where we work at closed string tree level.}.

If there would be just gravity we would be done now, however we expect the D-brane to couple to other background fields and we should also write the action for the gauge field $A_{a}$ living on the brane. In the limit of constant fieldstrength $F_{ab}$, the action for the fields coupling to a D$p$-brane is known to all orders in $\alpha'$, it is the so called Dirac-Born-Infeld action\footnote{Note that next to this term there is the Wess-Zumino term describing the coupling to the RR-forms, it is just a modification of the coupling \rref{branecoupletoforms} and reduces to it when $B$ and $F$ are zero.}
\cite{BornInfeld,dirac},
\begin{equation}
\mathcal{S}_{\text{DBI},p} = -T_p \int d^{p+1}\xi \, e^{-\Phi}    \left[- \det  (G_{ab}+B_{ab}
+2\pi\alpha' F_{ab}) \right]^{1/2}, \label{diracborninfeld}
\end{equation}
where $B_{ab}$ is the pullback of the Kalb-Ramond field to the brane. 

The remainder of this section and chapter \ref{Beta-Function_Calculations_in_Boundary_Superspace} will be dedicated to calculating this action and derivative corrections to it. To facilitate life slightly we shall use the following premises:
\begin{itemize}
\item From here on we will always work in Euclidean space and denote the indices $i,j,k,... \in  \{1, \ldots, 10 \}$, as is usual. This amounts to making a Wick rotation on the timelike coordinate. (Alternatively one can consider only magnetic field strengths and thus work with a $p$-dimensional spatial part of space-time.) 
\item We will work in flat space $G_{ij} = \delta_{ij}$ since the metric is easy to reinstate.
\item Notice that we again see the combination $\mathcal{F}_{ab} = B_{ab} +2\pi\alpha' F_{ab}$ appearing which is necessary to make the action invariant under gauge transformations  of the Kalb-Ramond field as we already pointed out below equation \rref{polyakovactiecurvedboundary}. This enables us to put $B=0$ since we can always find the dependence on $B$ by replacing $2\pi\alpha' F$ by $\mathcal{F}$.
\item The dilation field $\Phi$ will be fixed to a constant value and absorbed in the tension $\tau_{p}$ as in \rref{tensiotau}. 
\item Any derivative corrections to any other field as the gauge field or gauge fieldstrength won't be considered.
\item We will align the D$p$-brane worldvolume with the first $p+1$ coordinates as follows,
\begin{equation}
\xi^{a} = X^{a} \quad \quad \text{for}  \quad \quad a \in \{1, \ldots, p+1 \}.
\end{equation}
This is the so called static gauge.
\item We will only derive the action for the extremal case of a spacefilling D$9$-brane since the action for lower dimensional branes can always be derived from it by dimensional reduction using $T$-duality (see for instance \cite{Tseytlin:1999dj}). 
\item Lastly we will absorb the factor $2\pi\alpha' $ in the gauge field $A_{i}$ .
\end{itemize} 
Under these assumptions the action \rref{diracborninfeld} reduces to the Born-Infeld action\footnote{We will always write Born-Infeld action when referring to the Abelian Born-Infeld action. When discussing the non-Abelian case we will state so explicitly.} \cite{fradkintseytlin1,ACNY,Leigh:1989jq} which in Euclidean space reads,
\begin{equation}
\mathcal{S}_{\text{BI}} =  \tau_{9} \int d^{10} X  \left[ \det  (G_{ij} + F_{ij}) \right]^{1/2}, \label{borninfeldeuclidean}
\end{equation}

There are many different ways to obtain the D-brane effective action. We will work out two methods explicitly and subsequently discuss 
some other methods relevant to this thesis. The first possibility we will discuss uses $T$-duality to build the action piece by piece. This method is straightforward end easy to follow, a drawback is that it only works in the limit of slowly varying fieldstrength and so derivative corrections can not be found using it.  

In the second method we will discuss, we start by deriving the beta-functions for the gauge field, $\beta_{i}^{A}$, from the open string action coupled to a gauge backgroundfield \rref{polyakovactiecurvedboundary}. Just as in the discussion for closed strings we then demand vanishing of the beta-functions which is equivalent to asking Weyl invariance. If the resulting equations, which tell us in which background one can consistently define open string theory, are to have any meaning we have to interpret them as the e.o.m.~for the D-brane. Although this method is less transparent, it does allow one to calculate derivative corrections and we will do so, albeit using a superspace sigma-model, in chapter \ref{Beta-Function_Calculations_in_Boundary_Superspace}.

\subsection{Using Tilted D-branes}

We start by considering a D2-brane extended in the $X^{1}$ and $X^{2}$ directions with a constant gauge field $F_{12}$ on it. We can always choose a gauge such that,
\begin{equation}
A_{1}=0 \quad \text{and} \quad A_{2}=X^{1}F_{12}.
\end{equation}
Now we $T$-dualize along the $X^{2}$ direction and use the relation between the potential and the dual coordinate \rref{opentwilsonlinerelationgaugefield}.  We then find the following relation\footnote{Remember that we rescaled the gauge field and hence the fieldstrength.},
\begin{equation}
\hat{X}^{2} =  X^{1}F_{12},
\end{equation}
which tells us that the resulting D1-brane is tilted at an angle $\arctan(F_{12})$ towards the $X^{2}$ axis. This gives rise to a geometrical factor in the D1-brane worldvolume action, 
\begin{equation} \label{D1worldvolumeaction}
\mathcal{S} \sim  
\int dX^{1} \left[1+ (F_{12})^{2} \right]^{1/2} =  \int d^{1} X  \left[
 \det  \left\{ \left(
\begin{array}{cc}
1  &  0    \\
 0 &   1  \\ 
\end{array}
\right) 
+
 \left(
\begin{array}{cc}
0 & F_{12}    \\
F_{21} & 0  \\ 
\end{array}
\right) 
 \right\} \right]^{1/2},
\end{equation}

We can always perform a set of rotations in order to bring $F$ to block-diagonal form, thereby effectively reducing the problem to a product of factors like  \rref{D1worldvolumeaction}
After doing so we find the following action,
\begin{equation}
\mathcal{S}_{\text{BI}} \sim \int d^{10} X  \left[ \det  (G_{ij} + F_{ij}) \right]^{1/2}. \label{borninfeldfromtiltedbranes}
\end{equation}
This precisely matches the Born-Infeld action  \rref{borninfeldeuclidean} from the previous section. From the derivation it should be clear that derivative corrections are out of the picture here.

\subsection{Using Beta-Function Calculations}  \label{WhatisUsingBetaFuntionCalculations}

The method to calculate the Born-Infeld action we present in this section was first used in \cite{ACNY} without any reference to D-branes. It was later redone and extended in \cite{Leigh:1989jq} in the context of D-branes shortly after these where discovered.

As a starting point we consider an open string in flat space coupled on the boundary to an electromagnetic field. 
To facilitate life we will make a Wick-rotation on the worldsheet $\tau$, followed by a conformal mapping of the worldsheet to the upper half plane. The boundary is then located at $ \sigma =0$.  From \rref{polyakov-actie in conforme ijk} and \rref{polyakovactiecurvedboundary} we then read of the following action\footnote{We also removed an overall factor $1/ 2 \pi \alpha'$.},
 \begin{equation} \label{whatisbetaaction1}
{\cal S} = \frac{1}{2} \int d^2 \sigma \, \partial_{a}  X_{i}  \partial^{a} X^{i} + i \int d\tau \, A_{i}   \partial_{\tau} X^{i},
\end{equation}
After this rewriting we can just verbatim copy the derivation in \cite{ACNY}, we choose not to do so and restrict ourselves to giving the key elements of the derivation. 

From the point of view of the two dimensional field theory on the worldsheet, $A^{i}$ is a coupling and as such will receive loop-corrections resulting in a bare coupling as follows,
\begin{equation}
A_{i}^{\text{bare}} = A_{i} + A_{i}^{\text{c}},
\end{equation}
where $A_{i}^{\text{c}}$ is the counterterm. Using minimal subtraction one finds the following one loop result, 
\begin{equation}
A_{i}^{\text{c}}(\Lambda) = \frac{1}{2 \pi} \ln \Lambda  \left( ( G - F^{2} )^{-1} \right)^{jk} \partial_{j} F_{ik},
\end{equation}
where we used a sharp short distance cutoff $\Lambda$ to regulate the theory. 

The beta-function can then be obtained by differentiating with respect to the cutoff $\Lambda$,
\begin{equation}
\beta^{A}_{i} = \Lambda \frac{\partial \, }{\partial \Lambda} A_{i}^{\text{c}} = \frac{1}{2 \pi} \left( ( G - F^{2} )^{-1} \right)^{jk} \partial_{j} F_{ik}.
\end{equation}
Requiring the beta-function to vanish amounts to the following equations of motion,
\begin{equation} \label{whatisstringeombifrombeta}
 \left( ( G - F^{2} )^{-1} \right)^{jk} \partial_{j} F_{ik} = 0,
\end{equation}
which are precisely the Euler-Lagrange equations derived from the Born-Infeld action  \rref{borninfeldeuclidean}.

This method to calculate the D-brane effective action can be extended to include derivative corrections, by including higher loop contributions to the counterterm $A_{i}^{\text{c}}$. Here is the catch, the model used in \cite{ACNY} and \cite{Leigh:1989jq} makes it very difficult to do this. 

Finding a possible solution to this problem is the main theme of this thesis. Motivated by the fact that superspace calculations usually are less difficult to perform than their partner calculations in ordinary space we set out to derive a suitable superspace model (chapter \ref{Non_Linear_Sigma-Models_with_Boundaries}) which we subsequently use to redo and extend the calculation in this subsection (chapter \ref{Beta-Function_Calculations_in_Boundary_Superspace}).

\subsection{Stable Holomorphic Vector Bundles} \label{Stable_Holomorphic_Vector_Bundles}

As we will show in this section, the notion of a stable holomorphic vector bundle is very important in the study of the D-brane effective actions. We start by considering ordinary Maxwell theory whose action is given by,
\begin{equation} \label{yangmillseuclidean}
\mathcal{S}_{\text{Maxwell}} = \int d^{10} X \, F_{ij} F^{ij},
\end{equation}
for which the equations of motion are,
\begin{equation} \label{yangmillseom}
\partial^{i} F_{ij} = 0.
\end{equation}
Making an expansion and keeping only the lowest order terms in $\alpha'$, \rref{borninfeldeuclidean} and \rref{whatisstringeombifrombeta} reduce to \rref{yangmillseuclidean} and \rref{yangmillseom} respectivly. So we observe that in lowest order in $\alpha'$ the D-brane dynamics is described by Maxwell theory.

We now want to see under which conditions the presence of a D-brane preserves some of the space-time supersymmetry. These are the so called BPS conditions. 
To do so we introduce complex coordinates
as follows,
\begin{equation}
\begin{split}
Z^\alpha &= \frac{1}{\sqrt 2}\left(X^{2\alpha -1}+iX^{2\alpha }\right) \\
\bar Z^{\bar\alpha} &= \frac{1}{\sqrt 2} \left(X^{2\alpha -1}-iX^{2\alpha }\right)
\label{cc}
\end{split} \quad \quad  \quad  \text{for} \quad  \quad \quad \alpha \in \left\{1, \ldots ,5 \right\}.
\end{equation}
Using this language it is easy to write down the BPS conditions. The gauge bundle should be a holomorphic vector bundle,
\begin{equation} \label{whatisholom}
F_{\alpha \beta} = F_{\bar \alpha \bar \beta} = 0,
\end{equation}
satisfying the Donaldson-Uhlenbeck-Yau stability condition or DUY condition for short,
\begin{equation} \label{whatsisDUY}
F_{\alpha \bar \alpha} \equiv G^{\alpha \bar \beta}F_{\alpha \bar \beta}=0.
\end{equation}
These two conditions solve the equations of motion as can be seen form,
\begin{equation}  \label{yangmillseominholvectorb}
0 = \partial^{i} F_{ij} = \partial_{\beta}F_{\bar{\beta}\alpha}+ \partial_{\bar{\beta}}F_{\beta\alpha} =  2\partial_{\bar{\beta}}F_{\beta\alpha} - \partial_{\alpha}F_{\beta\bar{\beta}},
\end{equation}
where we used the Bianchi identity in the last equality.

Equations \rref{whatisholom} and \rref{whatsisDUY} only represent the BPS conditions in the limit $\alpha' \to 0$ where the D-brane dynamics is described by Maxwell theory as we already remarked and we expect them to get corrections in the full theory. If we now consider the Born-Infeld action \rref{borninfeldeuclidean} the equations of motion \rref{whatisstringeombifrombeta} can be rewritten as,
\begin{equation}
\sum_{n=0}^{\infty}(F^{2n})^{jk} \partial_{j}F_{ki}=0.
\end{equation}
Imposing holomorphicity \rref{whatisholom} and using the Bianchi identities this reduces to,
\begin{equation} \label{whatiseombiholo}
0=\partial_{\bar \beta} \big(\sum_{n=0}^{\infty}\frac{1}{2n+1}(F^{2n+1})_{\alpha \bar \alpha}\big) = \partial_{\bar \beta} (\text{arcth} F)_{\alpha \bar \alpha} 
\end{equation}
So that the stability condition gets modified to,
\begin{equation}\label{whatsisDUYdeformed}
(\text{arcth} F)_{\alpha \bar \alpha} = 0,
\end{equation}
which together with the holomorphicity conditions \rref{whatisholom} defines the BPS conditions for the full theory in the case of slowly varying fieldstrengths.

The ideas in this subsection were used in \cite{liespaulalex} to derive the Born-Infeld action. Starting from Maxwell theory its deformations were studied by adding arbitrary powers of the fieldstrength to the Maxwell action. Demanding that a deformation of  \rref{whatisholom} and \rref{whatsisDUY} still solves the equations of motion, it was then shown that the deformation is uniquely determined and is precisely the Born-Infeld action. The holomorphicity condition  \rref{whatisholom} remains unchanged while the DUY stability condition gets deformed to \rref{whatsisDUYdeformed}. 

\subsection{Higher Derivative Corrections} \label{Higher_Derivative_Corrections}

Up till now we only discussed results not including derivatives on the fieldstrength. The first paper to study derivative correction was \cite{abelianbi4derivative} where it was shown that the term containing two derivatives vanishes. In \cite{wyllard} the four derivative corrections were calculated using the boundary state formalism (see also \cite{koerber:thesis}). In this section we will give an overview of the relevant results.

The action derived in \cite{wyllard}, including the four derivative corrections, reads,
\begin{equation} \label{whatiswyllardaction}
\begin{split}
{\cal S} =  \tau_9 \int d^{10}X\; \sqrt{h^{+}}\bigg[ 1 &+
\frac{1}{96} \Big( \frac 1 2 h_+^{ij}h_+^{kl}
S_{jk} S_{li} \\
&- h_+^{k_2 i_1}h_+^{i_2 k_1}h_+^{l_2j_1}h_+^{j_2 l_1}
S_{i_1 i_2 j_1j_2}S_{k_1k_2 l_1 l_2}\Big)\bigg],
\end{split}
\end{equation}
where we introduced the following,
\begin{eqnarray} \label{whatisopenstringmetric1}
h^\pm_{ij} \equiv G_{ij} \pm F_{ij} \quad \quad  \text{and} \quad \quad h^\pm \equiv \det (h^\pm_{ij} ),
\end{eqnarray}
and its inverse $h_\pm^{ij}$,
\begin{eqnarray} \label{whatisopenstringmetric2}
h^{ik}_+\,h^+_{kj}=h^{ik}_-\,h^-_{kj}= \delta ^i_j.
\end{eqnarray}
Lastly, we defined,
\begin{eqnarray}
S_{ijkl}=\partial_i \partial_j F_{kl} +  h_+^{mn} \partial_i F_{km}\; \partial_j F_{ln} -  h_+^{mn} \partial_i F_{lm}\; \partial_j F_{kn} 
\end{eqnarray}
and
\begin{equation}
S_{ij} = h_+^{kl} S_{klij}.
\end{equation}

The equations of motion following from \rref{whatiswyllardaction} were derived in \cite{koerber:thesis} and it was subsequently shown that they are satisfied provided the holomorphicity conditions \rref{whatisholom} hold and we modify the DUY stability condition to,
\begin{eqnarray} \label{modifiedduystabilityinwhatis}
G^{ \alpha \bar \beta } \left(\text{arcth}\, F\right)_{ \alpha \bar
\beta } + \frac{1}{96} S_{ij\alpha \bar \beta} S_{kl\gamma \bar
\delta}\; h_+^{jk}h_+^{li}\left( h_{+}^{\alpha \bar \delta} h_{+}^{\gamma \bar \beta} - h_{-}^{\alpha \bar \delta} h_{-}^{\gamma \bar \beta} \right) =0.
\end{eqnarray}

The action \rref{whatiswyllardaction} and the stability condition \rref{modifiedduystabilityinwhatis} represent the state of the art knowledge of the Born-Infeld action. As already pointed out in \cite{wyllard} the calculation leading to \rref{whatiswyllardaction} contains a large number of contributions but the final result is surprisingly simple. While this might be considered as a strong indication that the final result is indeed correct, we felt the need existed to have a second and independent calculation of the four derivative correction to the Born-Infeld action. This calculation is presented in chapter \ref{Beta-Function_Calculations_in_Boundary_Superspace} and is the main subject of this thesis.



 \chapter{Non-Linear Sigma-Models with Boundaries} \label{Non_Linear_Sigma-Models_with_Boundaries}

The ultimate goal of this thesis and the work done in the past few years is to extend the calculation presented in subsection \ref{WhatisUsingBetaFuntionCalculations} of the previous chapter by including higher loop contributions to the counterterm in order to find higher derivative corrections to the D-brane effective action. As we already remarked it seems quite involved to go further than a one loop calculation using the model described in that subsection.

This is where superspace methods come to the rescue. A general observation, albeit not a rule, is that $n$-loop calculations in superspace tend to have the same degree of difficulty as $(n-1)$-loop calculations in ordinary space.  Ergo, by the same amount of effort one can do calculations to a higher order in loops. Before our inquiries the hope existed, and turned out to be fulfilled, that this would also apply to our case. 

However, to be able to use the powerful technology that superspace methods bring along, we needed a superspace description that is explicitly supersymmetric on the boundary. Indeed as can be seen from the before-mentioned calculation \cite{ACNY} one needs to restrict the propagator to the boundary to preform the calculation of the beta function for $A^i$ and hence the boundary itself needs to have a superspace description.
%

%

%
%

Non-linear sigma-models in two dimensions with $N=(2,2)$ supersymmetries play a central role in the description of strings in non-trivial NS-NS backgrounds. In the absence of boundaries, a case relevant to Type II strings, their geometry has been intensively studied in the past, see for instance \cite{luis}--\cite{BSVV}. However, much less is known for the case with boundaries which is precisely the one of our interest.


In this chapter we study, motivated by their relevance in describing D-branes, non-linear sigma-models in two dimensions in the presence of boundaries. 
Partial results were known for some time, see for instance \cite{Ooguri:1996ck}--\cite{Lindstrom:2005zr}, however only recently a systematic study
was performed in \cite{stock1} and \cite{stock2}, resulting in the most general boundary conditions compatible
with $N=1$ supersymmetry. Subsequently, these results were extended to $N=2$ supersymmetry in \cite{zab}
(see also \cite{Lindstrom:2002vp} for some specific applications and \cite{Lindstrom:2002mc} for a different approach).

While impressive, the results of \cite{stock1} and \cite{stock2} remain somewhat surprising.
Not only are the derivations quite involved, but the presence of a Kalb-Ramond background seems to require a non-local superspace description
of the model. This already occurs in the very simple setting where open strings move in a trivial gravitational background but in a non-trivial
electro-magnetic background. It is clear that in order to study the open string effective action through the calculation of supergraphs, the motivation for the work presented in this chapter, a local superspace description is called for.

In this chapter we reanalyze the models studied in \cite{stock1} and \cite{stock2} and we resolve many of the
difficulties encountered there. We start by reconsidering a non-supersymmetric non-linear sigma-model and study
the most general boundary conditions. In the next section we extend this to models with supersymmetry.
Motivated by the methods used in \cite{oliver} and \cite{joanna} (in quite a different setting however), we use a superspace formulation which is manifestly
invariant under only one combination of the two bulk supersymmetries. In this way the analysis of boundary conditions
compatible with $N=1$ supersymmetry is greatly facilitated and one finds that, just as for the case without boundaries,
$N=0$ automatically implies $N=1$. In addition, no non-local terms are needed and the cases with or without Kalb-Ramond background are treated on the same footing. The price we pay for this is that we loose manifest bulk $d=2$ super Lorentz covariance.

Next we investigate under which conditions the $N=1$ supersymmetry gets promoted to an $N=2$ supersymmetry. As for the case without boundaries, one needs two separately integrable covariantly constant complex structures. The metric has to be hermitian with respect to both of them. However, the presence of boundaries
requires that one of them gets expressed in terms of the other one and the remainder of the geometric data. Finally, we study the $N=2$ superspace formulation.

The work presented in this chapter was originally published in \cite{susyboundary}.

\section{No Supersymmetry}
Before turning to the case with supersymmetry present, we reconsider the bosonic non-linear sigma-model and study
the most general boundary conditions in this case. As we will see, many of the properties unveiled here are also present in the case with supersymmetry but become less transparent there.

We take as our starting point the following action,
\begin{eqnarray}
{\cal S}=\int d \tau d \sigma \left( \frac{1}{2}\dot X{}^i\, G_{ij}\,\dot X{}^j-
\frac{1}{2} X{}^i{}'\, G_{ij}\, X{}^j{}'+X^i{}'\,\mathcal{F}_{ij}\,\dot X^j
\right),\label{bosac}
\end{eqnarray}
where derivatives with respect to $ \tau $ and $ \sigma $ are denoted by a dot and a prime respectively. This action, as we already know from section \ref{bosonicstringssection} in the previous chapter, describes the bosonic string moving in a curved background containing a gravitational field $G_{ij}$, a Kalb-Ramond field and gauge field combined in $\mathcal{F}_{ij} = B_{ij}  + 2 \pi \alpha' F_{ij}$.

Varying this action we get, apart from the well-known bulk contribution, a boundary term,
\begin{eqnarray}
\int d \tau\, \delta X^i \,G_{ij}\,\left(- X^j{}'+\mathcal{F}^j{}_k\,\dot X^k\right).\label{bbdy}
\end{eqnarray}
Of course one deals with two boundaries when describing open strings but the present discussion readily generalizes to this case. 

The boundary term vanishes if we either impose Neumann boundary conditions in all directions,
\begin{eqnarray}
X^i{}'-\mathcal{F}^i{}_j\,\dot X^j=0,
\end{eqnarray}
or Dirichlet boundary conditions in all directions,
\begin{eqnarray}
\delta X^i=0.
\end{eqnarray}
In order to introduce mixed boundary conditions we need
a $(1,1)$-tensor $ {\cal R}^i{}_j(X)$ satisfying,
\begin{eqnarray}
{\cal R}^i{}_k{\cal R}^k{}_j= \delta ^i{}_j.\label{RRisone}
\end{eqnarray}
A manifold which allows for such a tensor is called an almost product manifold with almost product structure ${\cal R}^i{}_j(X)$. Introduction of this tensor allows us to construct the following projection operators ${\cal P}_\pm$,
\begin{eqnarray}
{\cal P}_\pm^i{}_j\equiv \frac{1}{2}\left( \delta ^i{}_j\pm {\cal R}^i{}_j\right),\label{bproj}
\end{eqnarray}
which satisfy,
\begin{eqnarray}
\begin{split}
& {\cal P}_\pm^i{}_j {\cal P}_\mp^j{}_k = 0\,, \\
& {\cal P}_\pm^i{}_j {\cal P}_\pm^j{}_k = {\cal P}_\pm^i{}_k\,.
\end{split}
\end{eqnarray}
Using these it is possible to impose simultaneously Neumann,
\begin{eqnarray}
{\cal P}_+^i{}_j \left (X^j{}'-\mathcal{F}^j{}_k\,\dot X^k\right)=0,\label{bbn}
\end{eqnarray}
and Dirichlet boundary conditions,
\begin{eqnarray}
{\cal P}_-^i{}_j \delta X^j=0.\label{bbd}
\end{eqnarray}
In other words $ {\cal P}_+$ and $ {\cal P}_-$ project onto
Neumann and Dirichlet directions respectively.
The boundary conditions, eqs.\ (\ref{bbn}) and (\ref{bbd}), can also be rewritten as,
\begin{eqnarray} \label{rewri}
\begin{split}
X^i{}'&={\cal P}_-^i{}_jX^j{}'+ {\cal P}^i_+{}_j \mathcal{F}^j{}_k \dot X^k, \\
\delta X^i&= {\cal P}_+^i{}_j \delta X^j.
\end{split}
\end{eqnarray}
It is not hard to see that using these conditions the boundary term eq.\ (\ref{bbdy}) vanishes
provided the metric is invariant under the $(1,1)$ tensor,
\begin{eqnarray}
{\cal R}^k{}_i {\cal R}^l{}_j G_{kl} = G_{ij}.\label{metinv}
\end{eqnarray}
Using equation \rref{RRisone} this implies $ {\cal R}_{ij}= {\cal R}_{ji}$.

If we require time independence of the boundary conditions, in others words if $X^i(\tau,\sigma)$ satisfies the boundary conditions,
we require that $X^i(\tau+\delta \tau,\sigma)$ should do so too, we can put $\delta X^j=\dot{X}^j \delta \tau$ in (\ref{bbd}) and find,
\begin{eqnarray}
{\cal P}_-^i{}_j \dot{X}^j=0.\label{bbd2}
\end{eqnarray}
If we subsequently use that $[ \delta , \partial/ \partial\, \tau ]=0$ on the boundary,
\begin{eqnarray}
0 = [ \delta , \partial/ \partial\, \tau ] X^k = 2 {\cal P}_+^l{}_{[i} {\cal P}_+^m{}_{j]} {\cal P}^k_+{}_{l,m} \delta X^i \dot{X}^j,\label{deldel}
\end{eqnarray}
we find the condition,
\begin{eqnarray}
{\cal P}_+^l{}_{[i} {\cal P}_+^m{}_{j]} {\cal P}^k_+{}_{l,m}=0.\label{integrability}
\end{eqnarray}

The necessity of this condition can also be seen in the case where $\mathcal{F}$ is exact, 
\begin{equation}
\mathcal{F}_{ij}= 2 \pi \alpha' \left(
\partial_iA_j- \partial_j A_i \right).
\end{equation}
 Then using (\ref{bbd2}), we can rewrite the action (\ref{bosac}) as,
\begin{eqnarray}
{\cal S}=\int d \tau d \sigma \left( \frac{1}{2}\dot X{}^i\, G_{ij}\,\dot X{}^j-
\frac{1}{2} X{}^i{}'\, G_{ij}\, X{}^j{}'
\right)+ 2 \pi \alpha'  \int d \tau A_i {\cal P}_+^i{}_j\dot X^j.\label{bosac1}
\end{eqnarray}
If we now vary this action one indeed obtains the Neumann boundary conditions \rref{bbn} provided the integrability condition \rref{integrability} holds.

The integrabitity condition  \rref{integrability} also ensures that the commutator of two infinitesimal displacements in the Neumann direction remains in the Neumann direction,
\begin{equation}
[ \delta _1, \delta _2 ] X^i= {\cal P}_+^i{}_j \left( [ \delta _1, \delta _2 ] X^j \right) ,
\end{equation}
as can easily be seen from,
\begin{eqnarray} \label{comdeldel}
\begin{split}
{\cal P}_-^i{}_j \left( [ \delta _1, \delta _2 ] X^j \right) 
&={\cal P}_-^i{}_j \left( \delta _1X^k \delta _2X^j{}_{,k}- \delta _2X^k \delta _1X^j{}_{,k} \right), \\
&= {\cal P}^i_+{}_{l,m}{\cal P}_+^l{}_{[j} {\cal P}_+^m{}_{k]} \delta _2X^j \delta _1X^k =0.
\end{split}
\end{eqnarray}
In order to prove this one has to use,
 \begin{equation}
 {\cal P}_-^i{}_n{\cal P}^n_+{}_{l,m}{\cal P}_+^l{}_{[j} {\cal P}_+^m{}_{k]} = {\cal P}^i_+{}_{l,m}{\cal P}_+^l{}_{[j} {\cal P}_+^m{}_{k]}.
\end{equation}

Summarizing, we can have mixed Neumann \rref{bbn}, and Dirichlet  \rref{bbd}, boundary conditions,
provided there exists a $(1,1)$-tensor $ {\cal R}$ which satisfies,
\begin{eqnarray}
&&{\cal R}^i{}_k {\cal R}^k{}_j= \delta ^i{}_j, \label{c1} \\
&& {\cal R}^k{}_i {\cal R}^l{}_j G_{kl}=G_{ij}, \label{c2}\\
&& {\cal P}_+^l{}_{[i} {\cal P}_+^m{}_{j]} {\cal P}^k_+{}_{l,m}=0. \label{c3}
\end{eqnarray}
Equation (\ref{c1}) tells us that $ {\cal R}$ is an almost product structure for which the metric is preserved, (\ref{c2}). The last condition, (\ref{c3}), tells us that the projection operator $ {\cal P}_+$ is integrable. Note that this is weaker than requiring that $ {\cal R}$ is integrable. The latter would require that the Nijenhuistensor is zero,
\begin{eqnarray}
{\cal R}^i{}_l{\cal R}^l{}_{[j,k]}+{\cal R}^l{}_{[j}{\cal R}^i{}_{k],l}=0.
\end{eqnarray} 
This is equivalent to the integrability of both $ {\cal P}_+$ and $ {\cal P}_-$ as can be seen from their respective integrability conditions which are given by,
\begin{eqnarray}
 {\cal P}_+^l{}_{[j} {\cal P}_+^m{}_{k]} {\cal P}^i_+{}_{l,m} 
 = - \frac{1}{4} {\cal P}_-^i{}_m\left({\cal R}^m{}_l{\cal R}^l{}_{[j,k]}+{\cal R}^l{}_{[j}{\cal R}^m{}_{k],l}\right)
 = 0&  \text{ for } {\cal P}_+\,, \quad \quad \label{IntPPl} \\ 
  {\cal P}_-^l{}_{[j} {\cal P}_-^m{}_{k]} {\cal P}^i_-{}_{l,m} 
 = - \frac{1}{4} {\cal P}_+^i{}_m\left({\cal R}^m{}_l{\cal R}^l{}_{[j,k]}+{\cal R}^l{}_{[j}{\cal R}^m{}_{k],l}\right)  
 = 0&  \text{ for } {\cal P}_-\,. \quad \quad \label{IntPMi}
\end{eqnarray}

\section{$N=1$ Supersymmetry}


\subsection{General Superspace with Boundaries} \label{General_Superspace_with_Boundaries}

Before looking at the case of interest, the non-linear sigma model with boundaries, we will analyse what happens if we add boundaries to a general $N=(1,1)$ superspace as described in section \ref{generalsuperspacesection} of the previous chapter. There it was stated that for any superfield $\Sigma$ the following action,
\begin{equation} \label{generalsuperspaceactienN11} 
\mathcal{S}_{N=(1,1)} = \int d^2\sigma d\theta^+  d\theta^- \Sigma = \int d^2\sigma \left[ D_+ D_- \Sigma \right],
\end{equation} 
is invariant under a supersymmetry transformation generated by $\varepsilon^+ Q_+ + \varepsilon^- Q_-$, where the supercharges are given by,
\begin{equation}
Q_+ = \frac{\partial}{\partial \theta^+} + i \theta^+ \partial_{\pp} \, , \quad \quad
Q_- = \frac{\partial}{\partial \theta^-} + i \theta^- \partial_{=} .
\end{equation}
This is true provided there are no boundaries and can easily be proven by writing out this transformation explicitly.
With boundaries present, performing the supersymmetry transformation yields the following non-zero boundary term,
\begin{equation} \label{boundarytermsuperspacegeneral}
\int d^2\sigma \partial_{\sigma}  \left[ D_+ D_- \left( \frac{i}{2} \left(\varepsilon^{+} \theta^{+} - \varepsilon^{-} \theta^{-} \right) \Sigma \right) \right].
\end{equation}
This of course means the action \rref{generalsuperspaceactienN11} is not supersymmetric if we include boundaries. To get out of this impasse we want to find an action that is manifestly supersymmetric in the presence of boundaries and that is in the bulk equal to $\mathcal{S}_{N=(1,1)} $ in order to ensure it describes the same bulk physics. To do so we change coordinates to,
\begin{equation}
\theta  = \theta^+ + \theta^-, \quad \quad \tilde{\theta}  =  \theta^+ - \theta^-,
\end{equation}
and introduce corresponding supercharges and superderivatives,
\begin{equation} \label{superchargesandsuperderivativesgeneralsuspa}
\begin{split}
& D = \frac{1}{2} (D_{+}+D_{-}), \quad  \quad Q=\frac{1}{2}(Q_+ + Q_-), \\
& \tilde{D} = \frac{1}{2}(D_{+}-D_{-}), \quad  \quad \tilde{Q}=\frac{1}{2}(Q_+ - Q_-).
\end{split}
\end{equation}
These obey the following relations,
\begin{equation}
\begin{split}
&Q^2=\tilde{Q}^2 = +\frac{i}{4}\frac{\partial}{\partial \tau}, \quad \quad  \{Q,\tilde{Q}\}  = + \frac{i}{2}\frac{\partial}{\partial \sigma}, \\
&D^2=\tilde{D}^2 = -\frac{i}{4}\frac{\partial}{\partial \tau}, \quad \quad   \{D,\tilde{D}\}  = - \frac{i}{2}\frac{\partial}{\partial \sigma}, \\
 &\{Q,D\}  =  \{Q,\tilde{D}\}  =  \{\tilde{Q},D\}  =  \{\tilde{Q},\tilde{D}\}  = 0.
\end{split}
\end{equation}

We then consider the following action,
\begin{equation} \label{improvedactiongeneralsuspa}
\begin{split}
{\cal S}_{\text{N=1}} & = -2 \int d^{2} \sigma \left[ D \tilde{D} \Sigma \right] \\
         & = -2 \int d^{2} \sigma d \theta \left[ \tilde{D} \Sigma \right],
\end{split}
\end{equation}
where the square brackets in the first line denote $\theta=\tilde{\theta}=0$ and in the second line $\tilde{\theta}=0$.

Using the following relation,
\begin{equation}
 D_+D_-= -2 D \tilde{D} - \frac{i}{2} \partial_{\sigma},
\end{equation}
it is easily shown that this action is equivalent to $\mathcal{S}_{N=(1,1)} $ modulo boundary terms. This relation can also be used to check that it is invariant under the supersymmetry transformation $\varepsilon Q$.  It is precisely the new term $- \frac{i}{2} \partial_{\sigma}$ that cancels the boundary term \rref{boundarytermsuperspacegeneral} with $\varepsilon = \varepsilon^{+} = \varepsilon^{-}$. However this action is not invariant under $\tilde{\varepsilon} \tilde{Q}$.  

Another possibility is to interchange $D$ and $\tilde{D}$ in \rref{improvedactiongeneralsuspa}, resulting in an action invariant under $\tilde{\varepsilon} \tilde{Q}$ and not under $\varepsilon Q$.

Summarizing we have shown that adding a boundary to an $N=(1,1)$ superspace breaks a linear combination of the left and right moving supersymmetries and leaves an $N=1$ superspace. 

\subsection{The Superspace Formulation} \label{N1superspace}

After the general analysis of the previous section, we now take a bottom-up approach of a more specific case and ask if we can formulate a superspace description for the nonlinear sigma-model Lagrangian,
\begin{equation}
\begin{split}
{\cal L}_{\text{bulk}}& = 
2(G_{ij}+\mathcal{F}_{ij}) \partial_{\pp} \, X^i \partial_= X^j
+ 2i G_{ij} \psi^i_+\nabla_=^{(+)} \psi ^j_+ \\
& + 2i G_{ij} \psi ^i_-\nabla_{\pp}^{(-)} \psi ^j_-
 +R^{(-)}_{ijkl} \psi _-^i \psi _-^j \psi _+^k \psi _+^l \\
&+2 (F^i-i \eta \Gamma ^{\ i}_{(-)kl} \psi ^k_- \psi ^l_+ )G_{ij}
(F^i-i \eta \Gamma ^{\ i}_{(-)mn} \psi ^m_- \psi ^n_+ ), \label{lagragnianN1suspa}
\end{split}
\end{equation}
in such a way that it is manifestly supersymmetric at the boundary. This Lagrangian describes a superstring moving in a curved background containing a gravitational field $G_{ij}$, a Kalb-Ramond field and gauge field combined in $\mathcal{F}_{ij} = B_{ij}  + 2 \pi \alpha' F_{ij}$. 

As was suggested by the previous section we can only hope to find an $N=1$ superspace description. To do so we introduce a single real fermionic coordinate $ \theta $ and matching supercharge and covariant derivative,
\begin{equation} \label{qd2butnotquite}
Q = \frac{\partial}{\partial \theta} + \frac{i}{4} \theta \partial_{\tau} \quad \quad
D = \frac{\partial}{\partial \theta} - \frac{i}{4} \theta \partial_{\tau} ,
\end{equation}
satisfying the following relations
\begin{eqnarray}
Q^2=+\frac i 4 \frac{ \partial\ }{ \partial \tau }\,,\qquad D^2=-\frac i 4 \frac{ \partial\ }{ \partial \tau }\,, \qquad \{Q,D\}=0.\label{qd2}
\end{eqnarray}
Notice that the supercharge is constructed in such a way that it does not contain a sigma derivative, which insures the absence of boundary terms when performing a supersymmetry  transformation. Hence the superspace at hand is indeed, as asked, supersymmetric in the presence of boundaries. Moreover it can easily be seen that upon putting $\tilde{\theta}=0$, the supercharge $Q$ and the covariant derivative $D$, see \rref{superchargesandsuperderivativesgeneralsuspa}, we found in the previous section precisely reduce  to the ones introduced here. 


To be able to accomodate the full field-content of the model at hand we need at least two superfields. So we introduce bosonic $N=1$ superfields $X^i$ and fermionic superfields $\Psi^i$, $i \in \{1,\ldots , D\}$. From the point of view of the target manifold the former will be coordinates while the latter are vectors. The $X$ superfields contain the bulk scalar fields and half of the bulk fermionic degrees of freedom. The $ \Psi $ superfields contain the other half of the bulk fermionic fields and the auxiliary fields. In this section, we will stick to Neumann boundary conditions and postpone the analysis of more general boundary conditions to a later section. On the boundary, the fermionic degrees of freedom are halved and no auxiliary fields are needed anymore. In other words, the boundary conditions should be such that $ \Psi $ gets expressed as a function of $X$ so that only $X$ lives on the boundary.
In order to do so we assume that the $\sigma$ derivatives act only on the $X$ superfields. In this way only the variation of $X$ will result in a boundary term.

We start our analysis with the following action,
\begin{eqnarray}
{\cal S}=\int  d^{2} \sigma d \theta \sum_{n=1}^{10}{\cal L}_{(n)}, \label{Sd1}
\end{eqnarray}
where,
\begin{equation} \label{Ld1}
\begin{split}
{\cal L}_{(1)}&= {\cal O}^{(1)}_{ij}(X)DX^i \dot X^j ,\\
{\cal L}_{(2)}&= {\cal O}^{(2)}_{ij}(X) DX^i  X^j{}',\\
{\cal L}_{(3)}&= {\cal O}^{(3)}_{ij}(X) \Psi^i X^j{}' ,\\
{\cal L}_{(4)}&= {\cal O}^{(4)}_{ij}(X) D \Psi^i \Psi ^j,\\
{\cal L}_{(5)}&= {\cal O}^{(5)}_{ij}(X)D \Psi^i DX^j ,\\
{\cal L}_{(6)}&= {\cal O}^{(6)}_{ij}(X) \Psi^i \dot X^j,\\
{\cal L}_{(7)}&= {\cal T}^{(1)}_{ijk}(X) \Psi^i \Psi^j \Psi^k , \\
{\cal L}_{(8)}&= {\cal T}^{(2)}_{ijk}(X) \Psi^i \Psi^j DX^k,\\
{\cal L}_{(9)}&= {\cal T}^{(3)}_{ijk}(X) \Psi^i DX^j DX^k,\\
{\cal L}_{(10)}&= {\cal T}^{(4)}_{ijk}(X)DX^i DX^j DX^k,
\end{split}
\end{equation}
where both $ {\cal O}^{(n)}_{ij}(X)$ and $ {\cal T}^{(n)}_{ijk}(X)$ are a priori undetermined functions
of $X^a$. 

On dimensional grounds\footnote{The worldsheet coordinates $\tau$ and $\sigma$ have worldsheet mass dimension $-1$ as usual. From this we find the derivatives $\partial_{\tau}$ and $\partial_{\sigma}$ to have dimension $1$. $D$ and $\Psi$ then have to posses dimension $1/2$ and $X$ is dimensionless.}, it is not hard to see that this is the most general action we can write down
under the assumption that only $X'$ and not $ \Psi '$ appears. Any other term one can write reduces upon
partial integrating $D$ or $ \partial_{\tau}$ to the terms listed above. In the next few paragraphs we are going to simplify \rref{Ld1} as much as possible.

The only terms containing a $\sigma$ derivative are ${\cal L}_{(2)}$ and ${\cal L}_{(3)}$.  By varying $X$ for these, one immediately gets the boundary condition,
\begin{eqnarray}
{\cal O}^{(3)}_{ji}(X) \Psi^j = -{\cal O}^{(2)}_{ji}(X) DX^j .
\end{eqnarray}
In order that this identifies $\Psi $ in terms of $X$, we require that ${\cal O}^{(3)}$ is invertible.

If we preform the integral over $\theta$, we find that $[D \Psi] = D \Psi |_{ \theta =0}$ is an auxiliary field as asked. If we then require that the auxiliary field equations of motion can be solved for the auxiliary fields, we find that ${\cal O}^{(4)}$ should invertible as well.

If we carry out a partial integration of the covariant derivative $D$ in ${\cal L}_{(4)}$ it becomes apparent the anti-symmetric piece  of ${\cal O}^{(4)}$ can be absorbed in $ {\cal T}^{(2)}$. In accordance with this observation we take ${\cal O}^{(4)}_{ij}={\cal O}^{(4)}_{ji}$ from now on.

The following field redefinition,
\begin{equation}
 \Psi ^i\rightarrow \Psi ^i+{\cal N}^i{}_j(X) DX ^j,
\end{equation}
modifies almost all terms except ${\cal O}^{(3)}$ and ${\cal O}^{(4)}$. 
Of particular interest to us is the effect on ${\cal O}^{(5)}$ and ${\cal O}^{(6)}$,
\begin{equation} \label{fr1}
\begin{split}
{\cal O}^{(5)}_{ij}&\rightarrow {\cal O}^{(5)}_{ij}+{\cal O}^{(4)}_{ik} {\cal N}^k{}_j,\\
{\cal O}^{(6)}_{ij}&\rightarrow {\cal O}^{(6)}_{ij}-\frac i 4 {\cal N}^k{}_j{\cal O}^{(4)}_{ki} .
\end{split}
\end{equation}

After this field redefinition, we can completely eliminate $ {\cal L}_{(5)}$ through partial integration. This affects
${\cal T}^{(3)}$ and modifies ${\cal O}^{(6)}$ on top of (\ref{fr1}) to,
\begin{equation}
{\cal O}^{(6)}_{ij}\rightarrow {\cal O}^{(6)}_{ij}-\frac i 4 {\cal O}^{(5)}_{ij} - \frac i 2 {\cal N}^k{}_j{\cal O}^{(4)}_{ki} .\label{fr2}
\end{equation}
Using the fact that  ${\cal O}^{(4)}$ is invertible, as we already mentioned, it is easily observed from this equation that a suitable
choice for ${\cal N}$ can be found such that $ {\cal L}_{(6)}$ vanishes as well.

If we rewrite the $\tau$ derivative inside ${\cal L}_{(1)}$ using $\partial_\tau = 4i D^2$ and subsequently preform a  partial integration on one $D$ it is easily shown that the anti-symmetric part of $ {\cal O}^{(1)}$ can be absorbed in $ {\cal T}^{(4)}$. So from this point on we take ${\cal O}^{(1)}_{ij}={\cal O}^{(1)}_{ji}$.

We are still free to perform a field redefinition of the form,
\begin{equation}
\Psi ^i \rightarrow {\cal M}^i{}_j(X) \Psi ^j.
\end{equation}
If we do so $ {\cal O}^{(1)}$ and $ {\cal O}^{(2)}$ are unchanged, $ {\cal O}^{(5)}$ and $ {\cal O}^{(6)}$ remain zero and $ {\cal O}^{(3)}$
and $ {\cal O}^{(4)}$ transform as,
\begin{equation}
\begin{split}
{\cal O}^{(3)}_{ij} &\rightarrow {\cal M}^k{}_i{\cal O}^{(3)}_{kj},\\
{\cal O}^{(4)}_{ij} &\rightarrow {\cal M}^k{}_i{\cal O}^{(4)}_{kl}{\cal M}^l{}_j.
\end{split}
\end{equation}
As ${\cal O}^{(3)}$ is invertible, we can make a suitable choice for $ {\cal M}$ such that ${\cal O}^{(3)}_{ij}=-4 G_{ij}$, with $G_{ij}$ the
target space metric. With this, we exhausted the field redefinitions of $ \Psi $.

Concluding we found that without any loss of generality we can put,
\begin{eqnarray}
{\cal O}^{(1)}_{[ij]}={\cal O}^{(4)}_{[ij]}=0,\qquad {\cal O}^{(5)}_{ij}={\cal O}^{(6)}_{ij}=0,
\qquad {\cal O}^{(3)}_{ij}=-4G_{ij}.\label{res1}
\end{eqnarray}
 in our action \rref{Sd1} or its components (\ref{Ld1}).

We now proceed with the comparison of our general action \rref{Sd1}, with components \rref{Ld1}, on which we imposed the conditions \rref{res1} with the non-linear sigma-model lagrangian \rref{lagragnianN1suspa} from which we have eliminated the auxiliary fields $F^i$.

To do so we perform the $ \theta $ integral in (\ref{Sd1}) and eliminate the auxiliary fields $\left[D \Psi^{i}\right]$. We then go on  by identifying the leading bosonic terms
$\dot X\dot X$, $\dot X X'$ and $X'X'$ to the corresponding terms in \rref{lagragnianN1suspa} and find that the remaining freedom in the $ {\cal O}$ tensors gets fully fixed,
\begin{eqnarray}
{\cal O}^{(1)}_{ij}=2i G_{ij},\quad \quad  {\cal O}^{(2)}_{ij}=-4i \mathcal{F}_{ij},  \quad \quad {\cal O}^{(4)}_{ij}=8 G_{ij}.\label{res2}
\end{eqnarray}

Next we want to identify the fermions $\left[ DX^i \right]$ and $\left[ \Psi^i \right]$ with the bulk fermions $ \psi ^i_+$ and $ \psi ^i_-$
which appear in  \rref{lagragnianN1suspa}.  The leading terms quadratic in the fermions in the lagrangian are given by,
\begin{equation}
\begin{split}
&2i G_{ij}\,\Big[ i DX^i+ \Psi ^i\Big] \partial_= \Big[i DX^j+ \Psi ^j\Big] \nonumber \\
+&2i G_{ij}\,\Big[ i DX^i- \Psi ^i\Big] \partial_\pp \, \Big[i DX^j- \Psi ^j\Big],
\end{split}
\end{equation}
where we used the boundary condition which follows from the lagrangian,
\begin{eqnarray}
\left[ \Psi ^i \right]=i \mathcal{F}^i{}_j \left[ DX^j \right].
\end{eqnarray}
Comparing this to the corresponding terms in  \rref{lagragnianN1suspa}, we identify,
\begin{equation}
\begin{split}
\psi ^i_+&= \left[  i DX^i+ \Psi ^i \right], \\
\psi^i_-&= \eta \left[ i DX^i- \Psi ^i \right],\label{id1}
\end{split}
\end{equation}
where $\eta \in\{+1,-1\}$ allows one to differentiate between Ramond and Neveu-Schwarz boundary conditions. It will
not play an essential role in the remainder of this thesis.

Next, we determine the ${\cal T}$ tensors by comparing the rest of the terms with the ones appearing in  \rref{lagragnianN1suspa}.
A somewhat tedious but straightforward calculation\footnote{See the next section for a shortcut.} yields a unique solution,
\begin{eqnarray}
\label{res3}
{\cal T}^{(1)}_{ijk}=-\frac{8i}{3}T_{ijk}, \quad {\cal T}^{(2)}_{ijk}= 8 \Big\{[ji]k\Big\},
\quad {\cal T}^{(3)}_{ijk} = -8i T_{ijk}, \quad {\cal T}^{(4)} = 0.
\end{eqnarray}

With this we have the full model in $N=1$ superspace. Its lagrangian is explicitly given by,
\begin{equation}
\begin{split}
{\cal L}&=2i G_{ij} DX^i \dot X^j
-4i \mathcal{F}_{ij} DX^i  X^j{}'
-4 G_{ij} \Psi^i X^j{}'
+8 G_{ij} \nabla \Psi ^i \Psi ^j \\
&-\frac{8i}{3}T_{ijk}\Psi ^i\Psi^j \Psi^k
-8iT_{ijk}\Psi^i DX^j DX^k,\label{finac}
\end{split}
\end{equation}
where the covariant derivative $\nabla \Psi^i$ is given by,
\begin{eqnarray}
\nabla \Psi ^i=D \Psi ^i+\Big\{{}^i{}_{jk}\Big\} DX^k \Psi ^j .
\end{eqnarray}

\subsection{Alternative Derivation of the Superspace Formulation} \label{Alternative_Derivation_of_the_Superspace_Formulation}

As we already know, the nonlinear sigma-model action associated with the Lagrangian \rref{lagragnianN1suspa}  that was the starting point of the derivation presented in the previous section is equivalent with the following $N=(1,1)$ superspace action,
\begin{equation}
{\cal S}_{\text{bulk}}= 2\, \int d^2\sigma \left[ D_+ D_- \left((G_{ij}+\mathcal{F}_{ij}) D_+ \Phi^i D_- \Phi^j\right)\right],
\label{bulksupbackgroundfieldsalternativederivation}
\end{equation}
as can easily be shown by working out the integration over $\theta^\pm$ explicitly.

As already pointed out in section \ref{General_Superspace_with_Boundaries}, in a more general setting, actions of this form are not supersymmetric once there are boundaries present. Continuing along the lines set out in that section we change coordinates to $\theta$ and $\tilde{\theta}$ and introduce the following action,
\begin{equation}
\begin{split}
{\cal S} & =  -4 \int d \tau d \sigma \left[ D \tilde{D} (G_{ij}+\mathcal{F}_{ij}) D_+ \Phi^i D_- \Phi^j\right] \\
         & =  -4 \int d \tau d \sigma d \theta \left[ \tilde{D} (G_{ij}+\mathcal{F}_{ij}) D_+ \Phi^i D_- \Phi^j \right],
\label{improvedbulksup}
\end{split}
\end{equation}
which is explicitly invariant under the supersymmetry transformation generated by $Q$ on the boundary and, modulo
boundary terms, equivalent to \rref{bulksupbackgroundfieldsalternativederivation}. Moreover, if we make the following identification,
\begin{equation}
\begin{split}
\left[ \Phi^i \right] & =  X^i, \\
i \left[ \tilde{D} \Phi^i \right] & =  \Psi^i ,
\label{bulktoboundary}
\end{split}
\end{equation}
and work out the $\tilde{D}$ derivative explicitly it is straightforward to see that we recover the model \rref{finac} found at the end of the previous section.

As we already pointed out in section \ref{General_Superspace_with_Boundaries}, another possibility is the action (\ref{improvedbulksup}) with $D$ and $\tilde{D}$ interchanged.
Substituting (\ref{bulktoboundary}) in (\ref{id1}) we easily arrive at the following relation,
\begin{equation}
\begin{split}
\psi ^i_+&= i \left[ D\Phi^i + \tilde{D}\Phi^i \right], \\
\psi^i_-&=i \eta \left[ D\Phi^i -\tilde{D}\Phi^i \right],
\end{split}
\end{equation}
from which it is easily observed that interchanging $D$ and $\tilde{D}$ is equivalent to putting $\eta \rightarrow -\eta$.

Of course, the argument in this section does not replace the exhaustive analysis of the previous section,
because it is not a priori clear that the most general boundary model could be written in the form (\ref{improvedbulksup}). 

\subsection{The Boundary Conditions}

If we vary the action associated with \rref{finac} we obtain the following boundary term,
\begin{eqnarray}
-4\int d \tau d \theta  \left( \Psi^i G_{ij}+ i DX^i \mathcal{F}_{ij} \right) \delta X^j.
\end{eqnarray}
This vanishes if we take Neumann boundary conditions in all directions,
\begin{eqnarray}
\Psi ^i=i \mathcal{F}^i{}_jDX^j,
\end{eqnarray}
or Dirichlet boundary conditions in all directions,
\begin{eqnarray}
\delta X^j=0.
\end{eqnarray}

The more general case which involves both Dirichlet and Neumann boundary conditions requires the introduction
of an almost product structure $ {\cal R}^i{}_j(X)$ satisfying (\ref{RRisone}) and projection operators as defined in (\ref{bproj}). Using the latter we impose Neumann,
\begin{eqnarray}
{\cal P}_+^i{}_j \left (\Psi ^j-i\mathcal{F}^j{}_kDX^k \right)=0,\label{bn}
\end{eqnarray}
and Dirichlet boundary conditions,
\begin{eqnarray}
{\cal P}_-^i{}_j \delta X^j=0.\label{bd}
\end{eqnarray}
If we rewrite equations  (\ref{bd}) and (\ref{bn}), we obtain,
\begin{equation}
\begin{split}
\delta X^i&= {\cal P}^i_+{}_j \delta X^j, \\
\Psi ^i&= {\cal P}^i_-{}_j \Psi ^j
+i {\cal P}^i_+{}_j \mathcal{F}^j{}_k  DX^k.\label{bfin}
\end{split}
\end{equation}
{From} this one observes that, as was to be expected, $ \delta X$ is completely frozen in the
Neumann directions while $ \Psi $ gets a component in the Neumann directions when
there is a non-trivial $\mathcal{F}^i{}_j$.

Equation \rref{bd} implies,
\begin{eqnarray}
{\cal P}^i_-{}_j D X^j = 0 \quad \quad \text{and} \quad \quad {\cal P}^i_-{}_j \dot X^j=0.\label{bdex}
\end{eqnarray}
These equations require that certain compatibility conditions are satisfied which follow from
the second expression of (\ref{qd2}), $D^2=-i/4 \partial_{\tau}$. Indeed acting with $D$ on the first equation in (\ref{bdex}),
we get\footnote{Essentially we observe here that if $DX$ lies in a Neumann direction, then so does $D^2X$.},
\begin{eqnarray}
0=-\frac i 4 {\cal P}^i_-{}_j \dot X^j+{\cal P}^i_-{}_{l,m} {\cal P}^l_+{}_j {\cal P}^m_+{}_k D X^k D X^j,
\end{eqnarray}
where we used the first equation of (\ref{bfin}). This is indeed consistent with the second expression
of (\ref{bdex}) provided,
\begin{eqnarray}
{\cal P}_+^l{}_{[i} {\cal P}_+^m{}_{j]} {\cal P}^k_-{}_{l,m}=0,
\end{eqnarray}
holds. This can easily be seen to be equivalent with the integrability conditions for ${\cal P}_+$ as found in (\ref{integrability}).

Following the argument in the previous section one would easily be led to believe that the derivation of the integrability condition necessitates supersymmetry as it uses $D^2=-i/4 \partial_{\tau}$. This is not true as we can derive it, just like in the case without supersymmetry, (\ref{deldel}), from the condition $[\delta ,\partial_{\tau}]=0$ which clearly has nothing to do with supersymmetry.  A third way to derive it is to impose $[D, \delta ]=0$. 

Having the projection operators at our disposal we can rewrite the boundary term in the variation as,
\begin{equation}
\begin{split} \label{boundarytermusingprojectionoperators}
&\int d \tau d \theta  \left(\Psi^i+iDX^k \mathcal{F}_k{}^i\right) G_{ij}\delta X^j \\
& = \int d \tau d \theta \left( {\cal P}_+^i{}_k +{\cal P}_-^i{}_k \right) \left(\Psi^k+iDX^l \mathcal{F}_l{}^k\right) G_{ij} \delta X^j \\
& =\int d \tau d \theta \big( {\cal P}_+^i{}_k \left(\Psi^k+iDX^l \mathcal{F}_l{}^k\right) G_{ij} \delta X^j 
+ \left(\Psi^i+iDX^l \mathcal{F}_l{}^i\right) G_{ij} {\cal P}_-^j{}_k\delta X^k\big),
\end{split}
\end{equation}
where, in order to make the last step, we had to impose the invariance of the metric under the almost product structure,
\begin{eqnarray}
{\cal R}^k{}_i {\cal R}^l{}_j G_{kl} = G_{ij}.\label{metinvnisone}
\end{eqnarray}
From  \rref{boundarytermusingprojectionoperators} we see without effort that imposing the
Neumann, equation (\ref{bn}), and the Dirichlet, equation (\ref{bd}), boundary conditions, the boundary
term in the variation of the action indeed vanishes.

So we conclude that {\em any} $N=0$ non-linear sigma-model with given boundary conditions,
allows for an $N=1$ supersymmetric extension as given in
equation (\ref{finac}). The Neumann and Dirichlet boundary conditions, (\ref{bn}) and (\ref{bd}), require the
existence of an almost  product structure $ {\cal R}$ which satisfies (\ref{c1}-\ref{c3}) just as in the case without supersymmetry.

We now briefly compare our results to those obtained in \cite{stock1} and \cite{stock2}. In the present derivation,
whether or not an an antisymmetric background is present, does not play any role.
When the antisymmetric background vanishes, $\mathcal{F}_{ij}=0$, equations (\ref{c1}-\ref{c3}) precisely agree with
the conditions derived in \cite{stock1}.
However as supersymmetry is kept manifest, the derivation of these conditions are tremendously simplified. Contrary to
\cite{stock1}, we remained off-shell all the time.
A drawback compared to \cite{stock1}, is the loss of manifest $d=2$ bulk super Lorentz covariance in the present
formulation. 

For a non-trivial antisymmetric background, the comparison with the results
in \cite{stock2} is a bit more involved.
A first bonus compared to \cite{stock2} is that we here have a regular superspace formulation. Indeed non-local superspace terms are not needed here.
Combining equations  (\ref{id1}), (\ref{bdex}) and (\ref{bn}), we schematically obtain the following boundary condition for the fermions,
\begin{eqnarray}
\psi_-=\eta \, \frac{ {\cal R}-  \mathcal{F}_{++}}{1+ \mathcal{F}_{++}}\psi_+, \label{bcfermions}
\end{eqnarray}
where 
\begin{equation}
\mathcal{F}_{++}{}^i{}_j = {\cal P}_+^i{}_k \mathcal{F}^k{}_l {\cal P}_+^l{}_j.
\end{equation}

Lastly, we make the following identification,
\begin{equation}
 {\cal R} = r \quad \quad \text{and} \quad \quad (1+  \mathcal{F}_{++})^{-1}( {\cal R}-  \mathcal{F}_{++}) = R,
\end{equation}
where $r$ and $R$ are the (1,1)-tensors  defined in \cite{stock2}. It is then straightforward to show that equations (\ref{c1}-\ref{c3}) imply the conditions in
eq.\ (3.22) of \cite{stock2}.

\section{More Supersymmetry}

\subsection{Promoting the $N=1$ to an $N=2$ Supersymmetry}
\label{N2susy}
The action (\ref{finac}), is manifestly invariant under the supersymmetry transformation,
\begin{eqnarray}
\delta X^i= \varepsilon Q X^a,\quad \quad \quad \delta \Psi ^a = \varepsilon Q \Psi ^a,
\end{eqnarray}
where the supersymmetry generator $Q$ was defined in \rref{qd2butnotquite}. Just as in the case without boundaries it is interesting to investigate the conditions under which this action exhibits a second supersymmetry. 

To start we write down the most general transformation rules consistent with dimensions and statistics. These are given by,
\begin{equation}
\begin{split}
\delta X^i =&\,\hat\varepsilon {\cal J}^{\ i}_{(1)j}(X)DX^j+ \hat\varepsilon{\cal J}^{\ i}_{(2)j}(X) \Psi^j, \\
\delta \Psi^i =& \,\hat\varepsilon{\cal K}^{\ i}_{(1)j}(X)D \Psi^j+ \hat\varepsilon{\cal K}^{\ i}_{(2)j}(X) \dot X ^j +
\hat\varepsilon{\cal K}^{\ i}_{(3)j}(X)  X'{}^j \\
+& \hat\varepsilon{\cal L}_{(1)jk}^{\ i}(X) \Psi ^j \Psi ^k + \hat\varepsilon{\cal L}_{(2)jk}^{\ i}(X) \Psi ^j DX^k
+ \hat\varepsilon{\cal L}_{(3)jk}^{\ i}(X) DX ^j DX^k .\label{2ndsusy}
\end{split}
\end{equation}

If we demand off-shell closure of the supersymmerty-algebra we have to put,
\begin{equation}
{\cal K}_{(3)}=0,
\end{equation}
and we shall do so in the remainder of this subsection.
Moreover, it can be shown that upon making a suitable redefinition for $ {\cal K}_{(1)}$, ${\cal L}_{(1)}$, $ {\cal L}_{(2)}$ and $ {\cal L}_{(3)}$ the term $ \hat\varepsilon{\cal K}^{\ a}_{(3)b}(X) X^{'b} $ is proportional to the equation of motion for $ \Psi $. We refer the reader to  the next subsection for a more detailed  discussion.

Requiring the bulk terms in the variation of the action (\ref{finac}) to vanish under (\ref{2ndsusy}) yields the following conditions,
\begin{equation}
\begin{split}
{\cal J}_{(1)} & =  \frac{1}{2}(J +\bar{J}), \quad i {\cal J}_{(2)}=-4 {\cal K}_{(2)}= \frac{1}{2}(J-\bar{J}),\quad {\cal K}_{(1)}=-\frac{1}{2}(J+\bar{J}),   \\
{\cal L}^{i}_{(1)jk}&=  \frac{i}{2} \left(\partial_{[j} J^i{}_{k]}-\partial_{[j} \bar{J}^i{}_{k]}\right) - \frac{i}{2}(J+\bar J)^{mi}T_{mjk},\\
{\cal L}^{i}_{(2)jk}&=- \frac{1}{2} \left(\partial_{j} J^i{}_{k}+\partial_{j} \bar{J}^i{}_{k}+ 2(J+\bar J)^{mi}\{mjk\} \right) ,\\
{\cal L}^{i}_{(3)jk} &= -\frac{i}{2}(J+\bar J)^{mi}T_{mjk}, \label{2ndsusy2}
\end{split}
\end{equation}
while $J$ and $\bar{J}$ satisfy,
\begin{equation}
\begin{split}
G_{i(j} \; J^i{}_{k)}& = G_{i(j} \; \bar{J}^i{}_{k)}=0,  \\
\nabla^+_k J^i{}_j & = \nabla^-_k \bar{J}^i{}_j=0  .\label{cs1}
\end{split}
\end{equation}

Before investigating the vanishing of the boundary terms in the variation, we impose the supersymmetry algebra.
In particular the first supersymmetry has to commute with the second one, which is trivially realized because of $\{Q,D\}=0$.
Subsequently, we need that
\begin{eqnarray}
{[} \delta (\hat\varepsilon_1), \delta (\hat\varepsilon_2){]}X^i=\frac i 2 \hat\varepsilon_1\hat\varepsilon_2\dot X^i,\qquad
{[} \delta (\hat\varepsilon_1), \delta (\hat\varepsilon_2){]} \Psi ^i=\frac i 2 \hat\varepsilon_1\hat\varepsilon_2\dot \Psi ^i,
\label{closure}
\end{eqnarray}
holds on-shell. This is indeed true provided
\begin{eqnarray}
J^i{}_j J^j{}_k = \bar{J}^i{}_j \bar{J}^j{}_k = - \delta^i_k\, , \quad N^i{}_{jk}[J,J]=N^i{}_{jk}[\bar{J},\bar{J}]=0 \, ,\label{cs2}
\end{eqnarray}
with the Nijenhuistensor $N[A,B]$ given by,
\begin{eqnarray}
N^i{}_{jk}[A,B]=A^m{}_{[j}B^i{}_{k],m} + A^i{}_{m} B^m{}_{[j,k]} + B^m{}_{[j}A^i{}_{k],m} +B^i{}_{m} A^m{}_{[j,k]} \, .
\end{eqnarray}
These conditions \rref{cs2} tell us that $J$ and $\bar{J}$ are complex structures such that the metric is hermitian with respect to both. The last fact follows from the first line of \rref{cs1} while the second line tells us that $J$ and $\bar{J}$ are covariantly constant with respect to two different connections as defined in \rref{torsieconn}.

A noteworthy fact is that the algebra closes off-shell, even if ${[}J, \bar J{]} \neq 0$. This has to be contrasted with the case
without boundaries where the $N=(2,2)$ algebra closes off-shell modulo terms proportional to ${[}J, \bar J{]}$ times equations of motion. If we would modify the transformation rules with an equation of motion term  (\ref{eomterm}) as pursued in the next subsection, then one indeed obtains off-shell closure modulo terms proportional to ${[}J,\bar J{]}$. Doing so is completely equivalent to letting ${\cal K}_{(3)} $ be non-zero in the analysis performed above.

We now turn to the boundary term in the supersymmetry variation of the action. Using an obvious matrix like notation, one shows that
this term vanishes provided\footnote{The integrable projection operator ${\cal P}_+$ defines a {\em foliation}, {\em i.e.} a set of branes
which together fill the whole target space.  We could restrict to one (or two) of these branes and
call its (their total) worldvolume $\gamma$. If we require the endpoints of the open string to lie on
the submanifold $\gamma$, the boundary will always be a part of $\gamma$.
Conditions (\ref{ibdy}) and (\ref{compa}) then only hold on $\gamma$. We will not follow this approach here
and require these conditions on the whole of target space.},
\begin{equation}
\begin{split}
&{\cal P}_-(J-\bar J) {\cal P}_-=0, \\
&{\cal P}_+(J-\bar J) {\cal P}_+= {\cal P}_+{[}\mathcal{F}, J + \bar J{]} {\cal P}_++ {\cal P}_+\mathcal{F}(J-\bar J)\mathcal{F} {\cal P}_+,\label{ibdy}
\end{split}
\end{equation}
holds.

Invariance of the boundary conditions (\ref{bn}) and (\ref{bd}), under the $N=2$ supersymmetry transformations requires,
\begin{equation}
\begin{split}
&{\cal P}_-(J+\bar J) {\cal P}_+=- {\cal P}_-(J-\bar J)\mathcal{F} {\cal P}_+, \\
& {\cal P}_+ (J+\bar J) {\cal P}_-= {\cal P}_+\mathcal{F}(J-\bar J) {\cal P}_-.\label{compa}
\end{split}
\end{equation}
Using the antisymmetry of $J$ and $\mathcal{F}$ and the symmetry of $ {\cal R}$, it is clear that the second equation in (\ref{compa}) is the transposed of the first one.

It is surprising that conditions (\ref{ibdy}) and (\ref{compa}) are strictly algebraic.  Indeed, all derivative terms
disappear using the integrability conditions (\ref{integrability}).
Using these conditions, together with the previously obtained equations, we can express $\bar J$ in terms of $J$,
\begin{equation}
\begin{split}
\bar J &=(1+\mathcal{F}_{++})^{-1}(1-\mathcal{F}_{++})J_{++}(1+\mathcal{F}_{++})(1-\mathcal{F}_{++})^{-1}+J_{--} \\
&-(1+\mathcal{F}_{++})^{-1}(1-\mathcal{F}_{++})J_{+-}-
J_{-+}(1+\mathcal{F}_{++})(1-\mathcal{F}_{++})^{-1} \\
&={\cal M}J {\cal M}^{-1},\label{btype}
\end{split}
\end{equation}
with
\begin{eqnarray}
{\cal M}= \frac{ {\cal R}-\mathcal{F}_{++}}{1+\mathcal{F}_{++}},\qquad
{\cal M}^{-1}= \frac{ {\cal R}+\mathcal{F}_{++}}{1-\mathcal{F}_{++}}.\label{defvanm}
\end{eqnarray}
Note that eqs.\ (\ref{cs1}) and (\ref{cs2}) are invariant under $J\rightarrow J$ and $\bar J\rightarrow-\bar J$, while this
change in eqs.\ (\ref{ibdy}) and (\ref{compa}) turns eq.\ (\ref{btype}) into
\begin{equation}
\begin{split}
\bar J &=-(1+\mathcal{F}_{++})^{-1}(1-\mathcal{F}_{++})J_{++}(1+\mathcal{F}_{++})(1-\mathcal{F}_{++})^{-1}-J_{--}\\
&+(1+\mathcal{F}_{++})^{-1}(1-\mathcal{F}_{++})J_{+-}+
J_{-+}(1+\mathcal{F}_{++})(1-\mathcal{F}_{++})^{-1} \\
&=- {\cal M}J {\cal M}^{-1}.\label{atype}
\end{split}
\end{equation}
The latter, defined by equation (\ref{atype}), are called A-type boundary
conditions while the former, as defined by quation (\ref{btype}) are B-type
boundary conditions.

In the above, we introduced the notation,
\begin{equation}
A_{++}\equiv {\cal P}_+ A {\cal P}_+, \, \, \, \, A_{-+}\equiv {\cal P}_- A {\cal P_+},  \, \, \, \, A_{+-}\equiv {\cal P}_+ A {\cal P}_-, \, \, \, \, A_{--}\equiv {\cal P}_- A {\cal P_-}.
\end{equation}

Using the conditions involving $J$, it is quite
trivial to show that $\bar J$ is indeed an almost complex
structure under which the metric $G$ is hermitian. However, the
covariant constancy and integrability of $J$ does not imply that
$\bar J$ as given in equation (\ref{btype}) or  (\ref{atype}) is
covariantly constant or integrable. This imposes further
conditions on the allowed boundary conditions, geometry and
torsion!

In the case there is no torsion, $T=0$, we find from (\ref{cs1}) and (\ref{cs2}) that the
geometry is K\"ahler.  Without further conditions on the geometry, there can be only one independent
K\"ahler form so that $J \sim \bar{J}$ with either B-type or A-type boundary conditions.

For the B-type boundary conditions, as defined by (\ref{btype}), one
finds the well known results,
\begin{eqnarray}
[J, {\cal R}]=[J, \mathcal{F}_{++}]=0.
\end{eqnarray}
This implies that the D-brane worldvolume is a K\"ahler submanifold
with K\"ahler form $J_{++}$ and that $\mathcal{F}$ is a (1,1)-form with respect to $J_{++}$.

Turning to A-type boundary conditions, (\ref{atype}), one gets,
\begin{eqnarray}
\mathcal{F}_{++}J_{++}\mathcal{F}_{++}=J_{++},\qquad J_{--}=\mathcal{F}_{++}J_{+-}=J_{-+}\mathcal{F}_{++}=0.
\end{eqnarray}
The latter implies the existence of a second almost complex structure $\tilde J$,
\begin{eqnarray}
\tilde J\equiv \mathcal{F}_{++}J_{++}+J_{+-}+J_{-+},
\end{eqnarray}
which is integrable in the case of a space filling brane.
The following relation exists between the dimension of the brane and the rank of $\mathcal{F}$,
\begin{equation}
{\rm dim}({\rm brane}) = \frac{1}{2} (D + {\rm rank}(\mathcal{F})) \, .
\end{equation}
In the special case $\mathcal{F}=0$, also $J_{++}=0$ and the brane worldvolume becomes a lagrangian
submanifold.  For a more detailed treatment we refer to \cite{Becker:1995kb},
\cite{Bershadsky:1995qy} and \cite{zab}.

Concluding, we find that a second supersymmetry is allowed provided two almost complex structures, $J$ and $\bar J$, exist which are
separately integrable and covariantly constant, albeit with respect to two different connections. Till this point, this is exactly equal to the
situation without boundaries. However
when boundaries are present, it turns out that one of the two complex structures can be expressed in terms of the other one and the remainder
of the geometric data.

\subsection{Deriving the Second Supersymmetry from the Bulk Action} \label{Deriving_The_Second_Supersymmetry_from_the_Bulk_Action}

In the previous section we derived the second supersymmetry by writing down the most general transformation rules. Subsequently we demanded the variation of the action to vanish and the supersymmetry algebra to hold on-shell. We then proceeded by analyzing under which conditions, firstly, the boundary term in the variation vanished and secondly the boundary conditions where invariant. The calculation leading to the presented results is rather lengthy but straightforward. The analysis of section \ref{Alternative_Derivation_of_the_Superspace_Formulation} makes one wonder if we can derive the second supersymmetry for the boundary model from the second pair of supersymmetry transformations in the bulk. 

It can easily be shown that the action \rref{bulksupbackgroundfieldsalternativederivation} is invariant under a second supersymmetry of the following form,
\begin{equation}
\delta \Phi^i=\hat{\varepsilon}_+ J^i{}_j D_{+} \Phi^j + \hat{\varepsilon}_- \bar{J}^i{}_j D_{-} \Phi^j .
\label{bulk2ndsusy}
\end{equation}
where $J$ and $\bar{J}$ satisfy the contions \rref{cs1} and \rref{cs2} just like in the previous section. The condition that $J$ and $\bar{J}$ are complex structures \rref{cs1} is needed so that the variation of the action \rref{bulksupbackgroundfieldsalternativederivation} vanishes. On the other hand, the conditions that the metric is hermitian with respect to both and that they both are covariantly constant \rref{cs2} are necessary to make the supersymmetry algebra close on-shell.

We only expect a linear combination of these symmetries to survive on the boundary so we put
$\hat{\varepsilon}_+=\hat{\varepsilon}_-=\frac{1}{2}\hat\varepsilon$.  Using (\ref{bulktoboundary})
we find for the bottom and top components of (\ref{bulk2ndsusy}),
\begin{equation}
\begin{split}
\delta X^i =&\,\hat\varepsilon {\cal J}^{\ i}_{(1)j}(X)DX^j+ \hat\varepsilon{\cal J}^{\ i}_{(2)j}(X) \Psi^j, \\
\delta \Psi^i =& \,\hat\varepsilon{\cal K}^{\ i}_{(1)j}(X)D \Psi^j+ \hat\varepsilon{\cal K}^{\ i}_{(2)j}(X) \dot X ^j +
\hat\varepsilon{\cal K}^{\ i}_{(3)j}(X)  X'{}^j \\
+& \hat\varepsilon{\cal L}_{(1)jk}^{\ i}(X) \Psi ^j \Psi ^k + \hat\varepsilon{\cal L}_{(2)jk}^{\ i}(X) \Psi ^j DX^k
+ \hat\varepsilon{\cal L}_{(3)jk}^{\ i}(X) DX ^j DX^k .  \label{2ndsusyfull}
\end{split}
\end{equation}
but now with,
\begin{equation}
\begin{split}
{\cal J}_{(1)} & =  \frac{1}{2}(J +\bar{J}), \quad \quad i {\cal J}_{(2)}=-4 {\cal K}_{(2)}= \frac{1}{2}(J-\bar{J}),  \\
{\cal K}_{(1)} & =  \frac{1}{2}(J +\bar{J}), \quad \quad {\cal K}_{(3)} = -\frac{1}{4} (J+\bar{J}),  \\
{\cal L}^{i}_{(1)jk}&= -\frac{1}{2} \left(\partial_{[j} J^i{}_{k]}-\partial_{[j} \bar{J}^i{}_{k]}\right),   \\
{\cal L}^{i}_{(2)jk}&= \frac{i}{2} \left(\partial_{j} J^i{}_{k}+\partial_{j} \bar{J}^i{}_{k} \right), \\
{\cal L}^{i}_{(3)jk} &= 0. \label{2ndsusyfull2}
\end{split}
\end{equation}
By construction, the action \rref{finac} in invariant under this transformation.

We see that this does not quite reproduce the transformations \rref{2ndsusy2}. There is however more freedom in the $N=1$ boundary superspace than in the $N=(1,1)$ bulk superspace model. When requiring only on-shell closure one finds that one can add the following transformation to (\ref{2ndsusyfull}),
\begin{equation}
\begin{split}
\delta \Psi ^i = \hat{\varepsilon} K^i{}_j & \left( \frac{1}{2} D \Psi^j - \frac{1}{8} X^j{}'
- \frac{1}{2} G^{jm} \{mkl\} \Psi^k DX^l \right.  \\ & \quad \left. - \frac{i}{4} G^{jm} T_{mkl} \left(\Psi^k \Psi^l + DX^k DX^l \right) \right) \, ,
\label{eomterm}
\end{split}
\end{equation}
where $K$ is an arbitrary antisymmetric tensor, $G_{i(j}K^i{}_{k)}=0$. Note that $K$ multiplies the equation of motion for $\Psi$, which has precisely the right dimension, so that this transformation vanishes on-shell. Using the fact that $K$ is antisymmetric one can show in a straightforward way that the action \rref{finac} is also invariant under this added transformation.

Combining the transformations \rref{2ndsusyfull} with  \rref{2ndsusyfull2} and \rref{eomterm} we find the most general susy transformation,
\begin{equation}
\begin{split}
\delta X^i =&\,\hat\varepsilon {\cal J}^{\ i}_{(1)j}(X)DX^j+ \hat\varepsilon{\cal J}^{\ i}_{(2)j}(X) \Psi^j, \\
\delta \Psi^i =& \,\hat\varepsilon{\cal K}^{\ i}_{(1)j}(X)D \Psi^j+ \hat\varepsilon{\cal K}^{\ i}_{(2)j}(X) \dot X ^j +
\hat\varepsilon{\cal K}^{\ i}_{(3)j}(X)  X'{}^j \\
+& \hat\varepsilon{\cal L}_{(1)jk}^{\ i}(X) \Psi ^j \Psi ^k + \hat\varepsilon{\cal L}_{(2)jk}^{\ i}(X) \Psi ^j DX^k
+ \hat\varepsilon{\cal L}_{(3)jk}^{\ i}(X) DX ^j DX^k .  \label{2ndsusya}
\end{split}
\end{equation}
with,
\begin{equation}
\begin{split}
{\cal J}_{(1)} & =  \frac{1}{2}(J +\bar{J}), \quad \quad \quad \quad i {\cal J}_{(2)}=-4 {\cal K}_{(2)}= \frac{1}{2}(J-\bar{J}), \\
{\cal K}_{(1)}&=\frac{1}{2}(J+\bar{J}+K), \quad \quad {\cal K}_{(3)}= -\frac{1}{8}(2J+2\bar{J}+K), \\
{\cal L}^{i}_{(1)jk}&= - \frac{1}{2} \left(\partial_{[j} J^i{}_{k]}+\partial_{[j} \bar{J}^i{}_{k]}\right) + \frac{i}{4} G^{im}\; K^l{}_n \; T_{jkl},   \\
{\cal L}^{i}_{(2)jk}&= \frac{i}{2} \left(\partial_{j} J^i{}_{k}+\partial_{j} \bar{J}^i{}_{k}-i G^{im}\;K^l{}_m \;\big\{ljk\big\} \right), \\
{\cal L}^{i}_{(3)jk} &= \frac{i}{4} G^{im}\;K^l{}_m \; T_{jkl} \, .\label{2ndsusyb}
\end{split}
\end{equation}
The algebra closes off-shell if and only if,
\begin{eqnarray}
K=0 \quad \mbox{ and } \quad [J,\bar J]=0,\label{offshellclossol2}
\end{eqnarray}
or
\begin{eqnarray}
K= -2(J+\bar J)\label{offshellclossol1}.
\end{eqnarray}
The latter leads to the transformations found in the previous section, (\ref{2ndsusy}) with (\ref{2ndsusy2}). 

\subsection{Generalized Boundary Conditions}

Having at our disposal the complex structures $J$ and $\bar{J}$,
we can generalize eq.\ (\ref{id1}) to,
\begin{equation}
\begin{split}
\psi ^i_+&= \left(e^{\alpha J}\right)^i{}_j (i DX^j+ \Psi ^j), \\
\psi^i_-&= \eta \,\left( e^{\pm \alpha \bar{J}}\right)^i{}_j (i DX^j- \Psi ^j) ,\label{idnew}
\end{split}
\end{equation}
where $\alpha$ is an arbitrary angle. This amounts to applying an R-rotation to the
original $\psi_+$ and $\psi_-$. Using (\ref{cs1}) and (\ref{cs2}), one can show that
both possibilities are symmetries of the bulk action. However, only one of them survives on the boundary.

For A-type boundary conditions, (\ref{idnew}) with the minus sign  leaves the boundary action invariant while taking the plus 
sign leads to a new model, with the boundary condition, (\ref{bcfermions}),
replaced by,
\begin{eqnarray}
\psi_- & = & \eta \, e^{ - \alpha \bar{J}} \, \frac{ {\cal R}-  \mathcal{F}_{++}}{1+ \mathcal{F}_{++}} \, e^{ \alpha J} \psi_+ \nonumber \\
       & = &\eta \, \frac{ {\cal R}-  \mathcal{F}_{++}}{1+ \mathcal{F}_{++}} \, e^{  2 \alpha J} \psi_+,
\end{eqnarray}
where we used equation (\ref{atype}). 

For the B-type boundary conditions it's the other way around,  (\ref{idnew})  with the plus 
sign leaves the boundary action invariant while taking the minus sign leads to a new model, with the boundary condition, (\ref{bcfermions}),
replaced by,
\begin{eqnarray}
\psi_- & = & \eta \, e^{ \alpha \bar{J}} \, \frac{ {\cal R}-  \mathcal{F}_{++}}{1+ \mathcal{F}_{++}} \, e^{ \alpha J} \psi_+ \nonumber \\
       & = &\eta \, \frac{ {\cal R}-  \mathcal{F}_{++}}{1+ \mathcal{F}_{++}} \, e^{ 2 \alpha J} \psi_+,
\end{eqnarray}
where we used equation (\ref{btype}).

\subsection{$N=2$ Superspace}

In section \ref{generalsuperspacesection} of the previous chapter we saw that if we want to reformulate an action, bearing some supersymmetry for which the algebra closes only on-shell, in superspace we need to introduce extra degrees of freedom known as auxiliary fields. These auxiliary fields are precisely needed to make the supersymmetry algebra close off-shell. The fact that for the the model at hand the supersymmetry algebra, (\ref{2ndsusy}) and (\ref{2ndsusy2}), closes off-shell, hints towards
the existence of an $N=2$ superspace formulation without the need of introducing further auxiliary fields. 

Analyzing the field content of the model \rref{finac} it is easily seen that working with general $N=2$ superfields automatically introduces extra fields and so we need constraint equations to kill these. The structure of equations (\ref{2ndsusy}) and (\ref{2ndsusy2}) shows that these constraints will  generically be non-linear. These are very difficult to work with and we will limit ourselves to linear constraints. (This will turn out to be sufficient for our purposes in further chapters, it however also means that the most general description remains unknown.)


We denote the fermionic coordinates of $N=2$ superspace by $ \theta $ and $\bar \theta $ and introduce fermionic derivatives $D$ and $\bar D$ which satisfy,
\begin{eqnarray}
\{D,\bar D\}=-i \partial_\tau, \quad \quad D^2=\bar D^2=0.
\end{eqnarray}

We now want to introduce superfields, which upon integrating out the extra fermionic coordinate, reduce to the fields
introduced in the previous section. As already pointed out, we will restrict ourselves to the simplest case where only linear constraints are used.
In order to achieve this we introduce the $N=1$ derivative $\hat D$ which corresponds to the $D$ in the previous sections
and the ``extra'' derivative $\check D$ defined as follows,
\begin{eqnarray}
\hat D \equiv \frac 1 2 \left(D+\bar D\right),\qquad \check D \equiv \frac i 2 \left(D-\bar D\right).
\end{eqnarray}
Explicitly these and their corresponding supercharges are given by,
\begin{equation} \label{qdnohathat2butnotquite}
\begin{split}
\hat D = \frac{\partial}{\partial \theta} - \frac{i}{4} \theta \partial_{\tau}  \quad \quad
\hat Q = \frac{\partial}{\partial \theta} + \frac{i}{4} \theta \partial_{\tau} , \\
\check D = \frac{\partial}{\partial \bar \theta} - \frac{i}{4}\bar \theta \partial_{\tau}  \quad \quad
\check Q = \frac{\partial}{\partial \bar \theta} + \frac{i}{4}\bar \theta \partial_{\tau}. \\
\end{split}
\end{equation}
With this the explicit form of  $D$ and $\bar D$ are defined retroactively.  Looking at \rref{qdnohathat2butnotquite}, $\hat D$ and $\hat Q$ are of course easily recognised as the covariant derivatives and first supercharge from the previous sections \rref{qd2butnotquite}. $\check D$ is a new covariant derivative and the corresponding supercharge $\check Q$ is the generator of the second supersymmetry from the previous sections.  $\hat D$ and $\check D$ satisfy (amongst others, but these are the only ones we will need) the following relations,
\begin{eqnarray}
\hat D^2=\check D^2=-\frac i 4 \partial_\tau,\qquad \{\hat D,\check D\}=0.\label{nis2d}
\end{eqnarray}

We then introduce the $N=2$ superfields $X^i$ and $ \Psi ^i$. When passing from $N=2$ to $N=1$ superspace, we do not
want to introduce extra auxiliary degrees of freedom. In order to achieve this, the $\check D$-derivatives of the fields
should satisfy constraints. The most general linear constraints one can write down are,
\begin{equation}
\begin{split}
\check D X^i&= {\cal C}_{(1)j}^{\ i}\hat D X^j+ {\cal C}_{(2)j}^{\ i} \Psi ^j,\\
\check D \Psi ^i&= {\cal C}_{(3)j}^{\ i} \hat D \Psi ^j+ {\cal C}_{(4)j}^{\ i} \dot X ^j+ {\cal C}_{(5)j}^{\ i} X^j{}',\label{suspacon}
\end{split}
\end{equation}
where $ {\cal C}_{(n)}$, $n\in\{1,\cdots ,5\}$ are constant. Equation (\ref{nis2d}) implies the following integrability conditions,
\begin{equation}  
\begin{split}
&{\cal C}_{(1)}^2=-{\bf 1}+4i {\cal C}_{(2)} {\cal C}_{(4)},\quad  \quad \quad {\cal C}_{(3)}^2=-{\bf 1}+4i {\cal C}_{(4)} {\cal C}_{(2)}, \\
& {\cal C}_{(2)} {\cal C}_{(5)}= {\cal C}_{(5)} {\cal C}_{(2)}=0, \\
& {\cal C}_{(1)} {\cal C}_{(2)}= {\cal C}_{(2)} {\cal C}_{(3)},\quad {\cal C}_{(3)} {\cal C}_{(5)}= {\cal C}_{(5)} {\cal C}_{(1)},\quad {\cal C}_{(3)} {\cal C}_{(4)}= {\cal C}_{(4)} {\cal C}_{(1)}.\label{cint}
\end{split}
\end{equation}

These integrability conditions allow one to solve the constraints, (\ref{suspacon}), in terms of an
unconstrained, fermionic, dimension -1/2 superfield $\Lambda$, and an unconstrained, bosonic, dimension 0
superfield $Y$,
\begin{equation}
\begin{split} 
X&=(\check D- {\cal C}_{(1)}\hat D)\Lambda+ {\cal C}_{(2)}Y, \\
\Psi &=(\check D- {\cal C}_{(3)}\hat D)Y+ {\cal C}_{(4)}\dot\Lambda+ {\cal C}_{(5)}\Lambda'.
\end{split}
\end{equation}

Motivated by the results in section \ref{Deriving_The_Second_Supersymmetry_from_the_Bulk_Action}, we propose the following
parameterization for the tensors $ {\cal C}_{(n)}$,
\begin{equation}
\begin{split}  \label{C_Nparam}
& {\cal C}_{(1)}=\frac 1 2 (J+\bar J),\qquad \, \, \,  \,  {\cal C}_{(2)}=-\frac i 2 (J-\bar J),\qquad {\cal C}_{(3)}=\frac 1 2 (J+\bar J+K), \\
& {\cal C}_{(4)}=-\frac 1 8 (J-\bar J),\qquad {\cal C}_{(5)}=-\frac 1 8 (2J+2\bar J+K),
\end{split}
\end{equation}
where $J^2=\bar J^2=-{\bf 1}$.
In order that the integrability conditions (\ref{cint}) are satisfied, one needs,
\begin{eqnarray} \label{cintparam}
K^2=-\{J+\bar J,K\},\qquad 2[J,\bar J]=K(J-\bar J)=(\bar J-J)K.
\end{eqnarray}
This has two obvious solutions:
\begin{eqnarray}
K= -2(J+\bar J)\label{sol1},
\end{eqnarray}
or
\begin{eqnarray}
K=0 \quad \mbox{ and } \quad [J,\bar J]=0.\label{sol2}
\end{eqnarray}

Passing to $N=1$ superspace by integrating out the extra fermionic coordinate $\bar \theta$ it is easily seen that $\check D$ and $\check Q$ are identified,
\begin{eqnarray}
\check D X^i = \check Q X^{i}, \qquad \check D \Psi^{i} = \check Q \Psi^{i}.
\label{suspacon3}
\end{eqnarray}
Using this relation, (\ref{suspacon}) with (\ref{cint})  now define the second supersymmetry,
\begin{equation}
\begin{split}
\delta X^i =&\, \hat\varepsilon \check Q X^i= \hat\varepsilon {\cal C}_{(1)j}^{\ i}\hat D X^j+ \hat\varepsilon{\cal C}_{(2)j}^{\ i} \Psi ^j,\\
\delta \Psi^i  =&\,\hat\varepsilon \check Q \Psi ^i= \hat\varepsilon{\cal C}_{(3)j}^{\ i} \hat D \Psi ^j+ \hat\varepsilon{\cal C}_{(4)j}^{\ i} \dot X ^j+\hat\varepsilon {\cal C}_{(5)j}^{\ i} X^j{}'.\label{linearsecondsusy}
\end{split}
\end{equation}
Using the parametrisation \rref{C_Nparam} this  is, modulo non-linear terms, recognised as the supersymmetry
transformation rules (\ref{2ndsusya}) with (\ref{2ndsusyb}). The integrability conditions \rref{cint} or \rref{cintparam} which are realised by imposing \rref{sol1} or \rref{sol2} are seen to correspond to the off-shell closure conditions for the supersymmetry algebra \rref{offshellclossol2} and \rref{offshellclossol1}. 

Taking the first possibility, (\ref{sol1}), in both \rref{linearsecondsusy} and \rref{2ndsusya} one sees that if the two are to be equal then we can only cover the trivial case of flat space\footnote{More precise, it only covers a model in flat space with zero torsion which could be relevant if we want to analyse models with only gauge fields present. We did not pursue this possibility}  
So if we want to stick to linear constraints we need to opt for the second possibility, (\ref{sol2})! In that case, the two commuting integrable
structures $J$ and $\bar{J}$ are simultaneously diagonalizable. 

We choose complex coordinates so that,
\begin{equation}
\begin{split}
J^{\alpha}{}_{\beta} & =  \bar{J}^{\alpha}{}_{\beta} = i \delta^{\alpha}{}_{\beta},  \quad \quad \! \!
J^{\bar{\alpha}}{}_{\bar{\beta}} = \bar{J}^{\bar{\alpha}}{}_{\bar{\beta}} = - i \delta^{\bar{\alpha}}{}_{\bar{\beta}}, \quad \quad \! \!  \! \!
\quad \alpha,\beta \in\{1,\cdots m\},  \\
J^{\mu}{}_{\nu} & =  - \bar{J}^{\mu}{}_{\nu} = i \delta^{\mu}{}_{\nu},  \quad
J^{\bar{\mu}}{}_{\bar{\nu}} = - \bar{J}^{\bar{\mu}}{}_{\bar{\nu}} = - i \delta^{\bar{\mu}}{}_{\bar{\nu}}, \quad
\quad \mu,\nu \in\{1,\cdots n\},
\end{split}
\end{equation}
and all other components vanishing.
In these coordinates, where we denote the bosonic superfield now by $Z$, the constraint equations (\ref{suspacon})
with \rref{C_Nparam} now take the following form,
\begin{equation}
\begin{split}
&\check D Z^ \alpha =+i\,\hat D Z^ \alpha ,\quad \quad \quad \quad \, \, \, \, \check DZ^{\bar \alpha }=-i\,\hat D Z^{\bar \alpha }, \\
&\check D \Psi^ \alpha =+i\, \hat D\Psi^ \alpha -\frac i 2 Z^ \alpha {}',\quad
\check D \Psi^{\bar \alpha }=-i\,\hat D\Psi^{\bar \alpha }+\frac i 2 Z^{\bar \alpha }{}',
\label{csf}
\end{split} \quad \quad \alpha \in\{1,\cdots m\},
\end{equation}
or equivalently,
\begin{equation}
\bar D Z^ \alpha =D Z^{\bar \alpha }=0,\quad \bar D \Psi^ \alpha=\frac 1 2 Z^ \alpha {}',\quad
D \Psi^{\bar \alpha }= \frac 1 2 Z^{\bar \alpha }{}', \quad \quad \alpha \in\{1,\cdots m\}, \label{chiralsf}
\end{equation}
and
\begin{equation}
\begin{split}
&\check D Z^ \mu =+\Psi ^ \mu ,\quad \quad \! \check DZ^{\bar \mu }=-\Psi^{\bar \mu }, \\
& \check D\Psi^\mu=-\frac i 4 \dot Z^ \mu ,\quad \check D\Psi^{\bar\mu}=+\frac i 4 \dot Z^{\bar \mu }, \label{tcsf}
\end{split} \quad \quad \quad \quad \quad
\mu \in\{1,\cdots n\}.
\end{equation}
Equations (\ref{csf}) and (\ref{tcsf}) are the boundary analogs of the two-dimensional chiral and twisted
chiral superfields respectively.

We will only consider the case where only one type of superfields is present. Contrary to the case without
boundaries, this yields two different cases. Having only chiral (twisted chiral) superfields results in a
K\"ahler geometry with B(A)-type supersymmetry. Taking exclusively chiral superfields ($n=0$), we introduce
two potentials $K(Z,\bar Z)$ and $V(Z, \bar Z)$ and the action,
\begin{eqnarray}
\int d^2 \sigma d^2 \theta\, K(Z,\bar Z)_{ ,\alpha \bar \beta }\left(-2i D Z^ \alpha \bar D Z^ {\bar \beta }-8i
\Psi^ \alpha \Psi^{\bar \beta }\right)+
\int d \tau d^2 \theta\, V(Z,\bar Z).
\end{eqnarray}
Passing to $N=1$ superspace by integrating out $\bar \theta$ one gets the action (\ref{finac}) with,
\begin{equation}
\begin{split}
&G_{ \alpha \bar \beta }=K_{ ,\alpha \bar \beta },\quad \quad \quad \, \, \,  \mathcal{F}_{ \alpha \bar \beta }=-\frac 1 2 V_{, \alpha \bar \beta }, \\
&G_{ \alpha  \beta } = G_{\bar \alpha \bar \beta }=0,\quad \quad \mathcal{F}_{ \alpha \beta }= \mathcal{F}_{\bar \alpha \bar \beta }=0.
\end{split}
\end{equation}
Solving the constraints in terms of unconstrained superfields $ \Lambda $ and $Y$,
\begin{equation}
\begin{split}
&Z^ \alpha =\bar D \Lambda ^ \alpha ,\quad \quad \quad \quad \quad \quad \! \! Z^{\bar \alpha }= D \Lambda ^{\bar \alpha }, \\ 
&\Psi ^ \alpha = \bar D Y^ \alpha + \frac 1 2 \Lambda ^ \alpha {}',\quad \quad \Psi^{\bar \alpha }=DY^{\bar \alpha }+ \frac 1 2 \Lambda ^{\bar \alpha }{}',
\end{split}
\end{equation}
and varying the action with respect to the unconstrained superfields, we get the following boundary term,
\begin{equation}
\begin{split}
\int d \tau  d^2 \theta \, \Big( & \delta \Lambda ^ \alpha \big(-4i  G_{ \alpha \bar \beta }\Psi^{\bar \beta }+
V_{, \alpha \bar \beta }\bar DZ^{\bar \beta }\big)  \\
& \, \, +\delta \Lambda ^{\bar \alpha} \big(4i G_{ \bar\alpha  \beta }\Psi^{ \beta }+ V_{, \bar\alpha \beta }D Z^{\beta }\big)
\Big).
\end{split}
\end{equation}
This vanishes if we take Neumann boundary conditions in all directions,
\begin{equation}
 \Psi^{ \alpha }-\frac i 4 G^{ \alpha \bar \gamma }V_{, \bar \gamma \delta }\bar D Z^{ \delta } =
  \Psi^{\bar \alpha }+\frac i 4 G^{\bar \alpha  \gamma }V_{, \gamma \bar\delta }\bar D Z^{\bar \delta } =0,
\end{equation}
or Dirichlet boundary conditions in all directions,
\begin{equation}
 \delta \Lambda ^{ \alpha } = \delta \Lambda ^{\bar \alpha }=0 \quad  \Leftrightarrow \quad  \delta Z^{ \alpha } = \delta Z^{\bar \alpha }=0,
 \end{equation}
where we pointed out that  the  Dirichlet boundary conditions imposed on the unconstrained superfields $\Lambda$ are compatible with Dirichlet boundary conditions imposed on $Z$.

The more general case which involves both Dirichlet and Neumann boundary conditions again requires us to introduce an almost product structure $ {\cal R}$ which satisfies,
\begin{equation}
\begin{split}
&{\cal R}^ \alpha {}_{ \bar \beta } = {\cal R}^{\bar \alpha }{}_ \beta =0, \\
&{\cal R}_{ \alpha \bar \beta }\equiv G_{ \alpha \bar \gamma }{\cal R}^{\bar \gamma }{}_{\bar \beta } =
{\cal R}_{\bar \beta \alpha  }\equiv G_{\bar \beta  \gamma } {\cal R}^ \gamma {}_ \alpha .
\end{split}
\end{equation}
Using the almost product structure to construct projection operators $ {\cal P}_+$ and $ {\cal P}_-$, we find that
the boundary term in the variation indeed vanishes if we impose simultaneously Neumann,
\begin{equation}
 {\cal P}_+^{ \alpha }{}_{ \beta }\left( \Psi^{ \beta }-\frac i 4 G^{ \beta \bar \gamma }V_{, \bar \gamma \delta }\bar D Z^{ \delta }\right) =
{\cal P}_+^{\bar \alpha }{}_{ \bar\beta }\left( \Psi^{\bar \beta }+\frac i 4 G^{\bar \beta  \gamma }V_{, \gamma \bar\delta }\bar D Z^{\bar \delta }
\right)=0,
\end{equation}
and Dirichlet boundary conditions,
\begin{equation} \label{Nis2dirichletprojected}
{\cal P}_-^{ \alpha }{}_{ \beta } \delta \Lambda ^{ \beta } = {\cal P}_-^{ \bar\alpha }{}_{\bar \beta } \delta \Lambda ^{\bar \beta }=0.
\end{equation}

Demanding compatibility of the \rref{Nis2dirichletprojected} with the  Dirichlet boundary conditions imposed on $Z$,
\begin{equation} \label{Nis2dirichletprojectedonZ}
 {\cal P}_-^ \alpha {}_{ \beta } \delta Z^ \beta  =  {\cal P}_-^ {\bar \alpha} {}_{ \bar \beta }\delta Z^{\bar \beta}  =0 
\end{equation}
  requires,
\begin{equation}
{\cal P}^ \alpha _+{}_{ \delta , \bar \varepsilon } {\cal P}^ \delta _+{}_ \beta 
{\cal P}_+^{\bar \varepsilon }{}_{\bar \gamma }={\cal P}^{\bar \alpha} _+{}_{ \bar \delta , \varepsilon } 
{\cal P}^{\bar \delta} _+{}_ {\bar \beta} {\cal P}_+^{ \varepsilon }{}_{ \gamma }=0.
\end{equation}
Finally, equation \rref{Nis2dirichletprojectedonZ} implies,
\begin{equation}
D Z^ \alpha = {\cal P}^ \alpha _+{}_ \beta D Z^ \beta \quad \text{and} \quad \bar D \bar Z^ {\bar \alpha} = {\cal P}^{\bar \alpha} _+{}_ {\bar \beta} \bar D \bar Z^ {\bar \beta}.
\end{equation}
Using this and  $D^2=0$ and $\bar D^2=0$ respectively, we get,
\begin{eqnarray}
{\cal P}^ \alpha _+{}_{ [\delta , \varepsilon ]} {\cal P}^ \delta _+{}_ \beta 
{\cal P}_+^{ \varepsilon }{}_{ \gamma }={\cal P}^{\bar \alpha} _+{}_{ [\bar \delta , \bar \varepsilon] } 
{\cal P}^{\bar \delta} _+{}_ {\bar \beta} {\cal P}_+^{ \bar \varepsilon }{}_{\bar \gamma }=0.
\end{eqnarray}
The conditions obtained here are completely equivalent to those in eqs.\ (\ref{c1}-\ref{c3}) and (\ref{btype}) for a K\"ahler geometry.

We now briefly turn to the case where we take exclusively twisted chiral superfields ($m=0$). The action,
\begin{equation}
\begin{split}
{\cal S}=\int d^2 \sigma d^2 \theta \Big(&-8K_{,\mu\bar\nu}\Psi^{\mu}\hat D Z^{\bar\nu} +
8K_{,\bar\mu\nu}\Psi^{\bar\mu}\hat D Z^{\nu}
\\ & \quad \quad + 2 K_{,\mu} Z^{\mu'} - 2 K_{,\bar{\mu}} Z^{\bar{\mu}'} \Big),
\end{split}
\end{equation}
correctly reproduces the bulk theory, however it does not give the right boundary terms. In other words, the
$N=2$ superspace description of type A boundary conditions remains unknown.

\subsection{$N=2$ Superspace with Non-Linear Constraints}

In the previous subsection we have seen that when sticking to linear constraints we could not use one of the two solutions for the integrability conditions, \rref{sol1}, to describe general models in curved space.  We were also unable to describe the model with type A boundary conditions. Presumably we can solve these problems by imposing non-linear constraints instead of the linear constraints \rref{suspacon}. We did not perform the full analysis but merely want to give an indication of the path one should follow to do so.

On would take as a starting point the following constraints,
\begin{equation}
\begin{split}
\check D X^i =& {\cal J}^{\ i}_{(1)j}(X)DX^j+ {\cal J}^{\ i}_{(2)j}(X) \Psi^j, \\
\check D \Psi^i =& {\cal K}^{\ i}_{(1)j}(X)D \Psi^j+ {\cal K}^{\ i}_{(2)j}(X) \dot X ^j + {\cal K}^{\ i}_{(3)j}(X)  X'{}^j \\
+&  {\cal L}_{(1)jk}^{\ i}(X) \Psi ^j \Psi ^k +  {\cal L}_{(2)jk}^{\ i}(X) \Psi ^j DX^k + {\cal L}_{(3)jk}^{\ i}(X) DX ^j DX^k , \label{nonlinearconstraintslike2ndsusya}
\end{split}
\end{equation}
with parametrisation \rref{2ndsusyb} and integrability conditions \rref{offshellclossol2} and \rref{offshellclossol1}. These are of course again the off-shell closure conditions for the supersymmetry algebra.

Subsequently one has to try and solve the constraint equations in terms of unconstrained fields and write down an $N=2$ superspace action that upon integrating out the extra fermionic coordinate $\bar \theta$  reduces to the action (\ref{finac}).

 \chapter{Beta-Function Calculations in Boundary Superspace} \label{Beta-Function_Calculations_in_Boundary_Superspace}
\chaptermark{Beta-Function Calculations in Superspace}

We are now ready and fully armed to tackle the main subject of this thesis. In this chapter we will derive, starting from the general setup developed in the previous chapter, the beta-functions for an open string sigma-model in the presence of a $U(1)$ background field. As explained in chapter \ref{What_is_string_theory}, asking for these beta-functions to vanish yields, after some work, the Born-Infeld action. 

As we will show, the number of derivatives acting on the fieldstrengths is equal to twice the number of loops minus two. Hence, a one loop calculation gives us the Born-Infeld action while a two loop calculation results in the two derivative corrections to this action. The three loop calculation we will perform give us the four derivative corrections and provides a verification of the results found in \cite{wyllard}.

This chapter is organized as follows. In the first section we set up the sigma-model in $N=2$ boundary superspace. This is followed by an analysis of the calculation to be performed and a derivation of the necessary ingredients such as the Feynman rules and the superspace propagators. We also introduce a diagrammatic language which we will use throughout this chapter and show, using a toy model example, how to derive the propagators for constrained superfields.

Next we focus on the beta-functions at one and two loop order and find complete agreement with results found in the literature. The two loop calculation is performed in great detail to illustrate the techniques used. We then turn to the ``meat'' of this chapter and perform the three loop calculation. We find complete agreement with the proposal made by Niclas Wyllard \cite{wyllard} in its incarnation as found in \cite{koerber:thesis}. We conclude this chapter with some considerations about higher loops.

The work presented in this chapter was originally published in \cite{betastijnalexalexwalter}, some preliminary results were written down in \cite{towards}.

\section{Setup}

We will use  the model derived in the last section of the previous chapter which was originally published in \cite{susyboundary}. However, to make our lives slightly more comfortable we make the following modifications:
\begin{itemize}
\item First we perform a Wick rotation on the time coordinate of the worldsheet in order to work in Euclidean space on the worldsheet.
\item This is followed by a conformal mapping of the worldsheet to the upper half plane.
\item We then absorb a factor $2$ in $\Psi$ and remove an overall factor $2$ from the action.
\item Lastly we absorb a factor $i/2$ in the potential $V$.
\end{itemize}
After this, the modified model is thus described by a $d=2$, $N=2$ ``boundary'' superspace with bosonic coordinates $\tau \in [-\infty,\infty]$, $\sigma \in [0,\infty]$  and fermionic coordinates $ \theta, \bar \theta $. Through the whole chapter our metric has Euclidean signature and the boundary is located at $ \sigma =0$. The fermionic derivatives $D$ and $\bar D$ satisfy,
\begin{eqnarray}
\{D,\bar D\}= \partial_\tau,\quad \quad D^2=\bar D^2=0. \label{fremderivsetup}
\end{eqnarray}

We use constrained superfields $Z^ \alpha $, $Z^{\bar \alpha }$, $\Psi^ \alpha $ 
and $ \Psi^{\bar \alpha }$ which satisfy the following chirality conditions,
\begin{eqnarray} \label{betafuncchiralitycond}
\bar D Z^ \alpha=0, \quad \quad D Z^{\bar \alpha }=0,\quad \quad \bar D \Psi ^ \alpha = \partial _ \sigma  Z^ \alpha ,\quad \quad
D \Psi ^{\bar \alpha }= \partial _ \sigma Z^{\bar \alpha }. \label{chiralsfbeta}
\end{eqnarray}
These constraints can be solved in terms of unconstrained fermionic, $\Lambda$, and bosonic, $Y$, superfields,
\begin{equation}\label{solvecon}
\begin{split}
&Z^ \alpha =\bar D \Lambda ^ \alpha ,\quad \quad \quad \quad \quad \quad \! \! Z^{\bar \alpha }= D \Lambda ^{\bar \alpha }, \\ 
&\Psi ^ \alpha = \bar D Y^ \alpha + \partial _ \sigma  \Lambda ^ \alpha,\quad \quad \Psi^{\bar \alpha }=DY^{\bar \alpha }+ \partial _ \sigma  \Lambda ^{\bar \alpha }.
\end{split}
\end{equation}

We then consider the following action,
\begin{eqnarray}
{\cal S}= {\cal S}_0+ {\cal S}_1, \label{s} 
\end{eqnarray}
where the bulk action is given by,
\begin{eqnarray}
{\cal S}_0 =\int d^2 \sigma d^2 \theta\, \left(  D Z^ \alpha \bar D Z^ {\bar \alpha } +
\Psi^ \alpha \Psi^{\bar \alpha }\right), \label{s0} 
\end{eqnarray}
and the boundary action by,
\begin{eqnarray} \label{s1}
{\cal S}_1 
= - \int d \tau d^2 \theta\,  V(Z,\bar Z). 
\end{eqnarray}
The corresponding euclidean functional integral is then,
\begin{eqnarray}
{\cal Z} = \int {\cal D}Z{\cal D}\bar{Z}{\cal D}\Psi{\cal D}\bar{\Psi} e^{-{\cal S}[Z,\bar{Z},\Psi,\bar{\Psi}]},  \label{paths}
\end{eqnarray}

We impose Neumann boundary conditions in all directions,
\begin{eqnarray}
\Psi^{ \alpha }  = 0, \quad \quad \quad \quad
\Psi^{ \bar\alpha }  = 0,\qquad \qquad \quad \quad\quad \quad\mbox{at } \sigma =0.\label{bdycds}
\end{eqnarray}
If we hit this equation with a $\bar{D}$ respectively $D$ derivative we get the following conditions,
\begin{eqnarray}
\partial_{\sigma}Z^{ \alpha } = 0, \quad \quad \quad\quad 
\partial_{\sigma}Z^{ \bar\alpha }  = 0,\qquad \quad \quad\quad \quad\quad \quad \mbox{at } \sigma =0.\label{bdycdshitd}
\end{eqnarray}
Which we recognize as the ordinary Neumann boundary conditions for the bosonic string. Notice that there is no dependence on the potential $V$ in these boundary conditions  since we will threat ${\cal S}_1$ as an interaction term in what follows.

If we now compare the bosonic sector to the action \rref{whatisbetaaction1}, which describes a bosonic string coupled on the boundary to an electromagnetic field,
we clearly see that the complex components of the gauge field are given by,
\begin{eqnarray} \label{potentialsifoV}
A_{ \alpha} = - \frac{i}{2} V_{\alpha}, \quad \quad \quad \quad A_{\bar \alpha} =  \frac{i}{2} V_{\bar \alpha},
\end{eqnarray}
and hence those of the field strength by,
\begin{eqnarray} \label{fspot}
 F_{\alpha\bar{\beta}} = i V_{\alpha \bar \beta}, \quad \quad F_{\bar{\alpha}\beta} = -i V_{\bar \alpha  \beta}, \quad \quad F_{\alpha\beta} = F_{\bar{\alpha}\bar{\beta}} = 0.
\end{eqnarray}
The last equation tells us that the gauge bundle should be a holomorphic vector bundle (see subsection \ref{Stable_Holomorphic_Vector_Bundles}).

In the former we introduced the notation,
\begin{eqnarray}
 V_{\alpha\beta \ldots \bar{\alpha}\bar{\beta} \ldots} = \partial_{\alpha} \partial_{\beta} \ldots \partial_{\bar{\alpha}} \partial_{\bar{\beta}} \ldots V,
\end{eqnarray}
with,
\begin{eqnarray}
 \partial_{\alpha} = \frac{\partial}{\partial Z^{\alpha}}, \qquad \partial_{\bar{\alpha}} =  \frac{\partial}{\partial Z^{\bar{\alpha}}}.
\end{eqnarray}


%
%

\section{Superspace Perturbation Theory} \label{suspa_pert_the}

In order to begin perturbative calculations we use the background field method and take the superfields to be the sum of a classical piece, satisfying the equations of motion, and a quantum piece,
\begin{eqnarray}
Z^{\alpha} = Z_{cl}^{\alpha} + Z_{q}^{\alpha}, &\quad& Z^{\bar{\alpha}} = Z_{cl}^{\bar{\alpha}} + Z_{q}^{\bar{\alpha}}, 
\nonumber \\
\Psi^{\alpha} = \Psi_{cl}^{\alpha} + \Psi_{q}^{\alpha}, &\quad& \Psi^{\bar{\alpha}} = \Psi_{cl}^{\bar{\alpha}} + \Psi_{q}^{\bar{\alpha}}.
\end{eqnarray}
In performing this split we can drop the terms independent of the quantum fields from the action in all calculations since they give only an overall constant in the path integral.  The terms linear in the quantum fields are zero due to the fact that they are proportional to the equations of motion for the classical fields. Dropping the terms independent of the quantum fields and the terms linear in the quantum fields and omitting the index $q$ for the quantum fields, the bulk action (\ref{s0}) retains its form
while the boundary action (\ref{s1}) has the following expansion,
\begin{eqnarray} \label{interactiontermsaction}
{\cal S}_1  \!  \! \!  \! &= & \!  \! \!  \!  - \int d \tau d^2 \theta \bigg( \frac{1}{2} V_{\alpha\beta}Z^{\alpha} Z^{\beta}  + V_{\alpha \bar{\beta}} Z^{\alpha} Z^{\bar{\beta}} 
+ \frac{1}{2} V_{\bar{\alpha}\bar{\beta}}Z^{\bar{\alpha}} Z^{\bar{\beta}}
\nonumber \\
& +&  \!  \! \!  \!  \frac{1}{3!} V_{\alpha\beta\gamma} Z^{\alpha} Z^{\beta} Z^{\gamma}
+ \frac{1}{2} V_{\bar{\alpha}\beta\gamma}Z^{\bar{\alpha}} Z^{\beta} Z^{\gamma} 
+ \frac{1}{2} V_{\bar{\alpha}\bar{\beta}\gamma}Z^{\bar{\alpha}} Z^{\bar{\beta}} Z^{\gamma}
+ \frac{1}{3!} V_{\bar{\alpha}\bar{\beta}\bar{\gamma}}Z^{\bar{\alpha}} Z^{\bar{\beta}} Z^{\bar{\gamma}} \nonumber \\
&   + &  \!  \! \!  \! \mathcal{O}(Z^{4}) \bigg), \label{s1q}
\end{eqnarray}
where the potential $V$ and its derivatives now are a function of the classical fields.

From this we see that all interactions are characterized by the dimensionless potential $V$ which we expect to receive loop corrections, resulting in a 
bare potential $V_{\text{bare}}$ of the form,
\begin{eqnarray}
V_{\text{bare}}= V + V_{\text{c}} =V + \sum_{r\geq 1} V_{(r)} \lambda ^r,
\end{eqnarray}
where $V_{\text{c}}$ is the counterterm and,
\begin{equation}
 \lambda = (\ln \, \Lambda) /\pi \quad \quad \quad \text{and} \quad \quad \quad  \Lambda =M/m,
\end{equation}
with $M$ the UV cut-off and $m$ the IR regulator. Making the loop expansion explicit we get,
\begin{eqnarray}
V_{(r)}=\sum_{s\geq r} V_{(r,s)}\,\hbar^s. \label{loopexp}
\end{eqnarray}

The beta-function for the $V$ can then be obtained in the usual fashion by differentiating with respect to $\Lambda$,
\begin{eqnarray} \label{betafunctionforv}
\beta_{V} \equiv \Lambda \frac{d \, }{d \Lambda} V = - \frac{1}{\pi}V_{(1)},
\end{eqnarray}
and the renormalization group recursively expresses $V_{(r)}$, $r\geq 2$, in terms of $V_{(1)}$,
\begin{eqnarray}
V_{(r+1)}(V)=- \frac{\pi}{r+1}\, V_{(r)}(V+ \beta_{V}
)\Big|_{\mbox{\small part linear in } \beta_{V} }. \label{rg}
\end{eqnarray}
Indeed, imposing the bare potential to be independent of the cutoff $\Lambda$ we get,
\begin{eqnarray}
\Lambda \frac{d \, }{d \Lambda} V_{\text{bare}} = 0 = \Lambda \frac{d \, }{d \Lambda} V +  \sum_{r\geq 1} \left( \frac{\partial V_{(r)}}{\partial V} \Lambda \frac{d V}{d \Lambda}  \lambda^{r} +  \frac{r}{\pi} V_{(r)} \lambda^{r-1} \right).
\end{eqnarray}
Collecting the terms at each order in $\lambda$ and putting their coefficient to zero, we obtain precisely \rref{betafunctionforv} and \rref{rg}.

Using \rref{potentialsifoV} one can define bare gauge fields from the bare potentials and thus the beta-function for the gauge potentials are given by,
\begin{eqnarray}
\beta _ \alpha =\frac{i}{2\pi} \partial _ \alpha V_{(1)}, \qquad
\beta _ { \bar \alpha} =-\frac{i}{2\pi} \partial _ { \bar \alpha}
V_{(1)}.\label{bfundef}
\end{eqnarray}

\subsection{Propagators of Constrained Superfields, a Toy Model Example}\label{toy}

In the next subsection we will need to derive the propagator for constrained superfields which generally can get quite involved. In this subsection we will describe, using a toy-model to illustrate it, a procedure to derive the propagator for constrained superfields using unconstrained fields which greatly facilitates the procedure. The toy model described lives in $d=2$, $N=2$ superspace with fermionic derivatives $D$ and $\bar D$ satisfying, 
\begin{equation}
\{D,\bar D\}= \partial_\tau,\quad \quad D^2=\bar D^2=0, \label{fremderiv}
\end{equation}
and has the following action,
\begin{equation}
{\cal S} =\int d^2 \sigma d^2 \theta\, D Z \bar{D} \bar{Z} =  \int d^2 \sigma d^2 \theta\,   Z \left(-\partial_{\tau}\right) \bar{Z} 
, \label{toy_s}
\end{equation}
with $Z$ and $\bar{Z}$ chiral and anti-chiral superfields,
\begin{equation}
\bar D Z = D \bar Z=0. \label{toy_chiralsf}
\end{equation}

To derive the $\left\langle Z \bar Z \right\rangle$ propagator we add the following term to the action,
\begin{equation}
{\cal S}_{J} =\int d^2 \sigma d^2 \theta\, \left( - \bar J Z - J \bar Z \right), \label{toy_sj}
\end{equation}
containing unconstrained sources $J$ and $\bar J$. The euclidean generating functional is then given by,
\begin{equation}
{\cal Z}[J,\bar J]= \int {\cal D}Z{\cal D}\bar{Z} e^{-\left( {\cal S}[Z,\bar{Z}] + {\cal S}_{J}[Z,\bar{Z},J,\bar J] \right)},  \label{toy_genrfunc}
\end{equation}
so that we get,
\begin{equation}
\left\langle Z \bar Z \right\rangle = \frac{1}{{\cal Z}[0,0]} \frac{\delta \, }{\delta \bar J} \frac{\delta \, }{\delta J} {\cal Z}[J,\bar J] \Big|_{ J = \bar J =0}. \label{toy_prop}
\end{equation}
If we were naive we would proceed by completing the squares,
\begin{equation}
{\cal S} + {\cal S}_{J} =\int d^2 \sigma d^2 \theta\,  \left( Z + J \frac{1}{\partial_{\tau}}\right) \left(-\partial_{\tau}\right)\left( \bar{Z} + \frac{1}{\partial_{\tau}} \bar J   \right)  + J \frac{1}{\partial_{\tau}} \bar J,
\end{equation}
and then introduce shifted fields as follows,
\begin{equation}
 Z_{s} = Z + J \frac{1}{\partial_{\tau}}, \quad \quad \quad \quad  \bar Z_{s} = \bar{Z} + \frac{1}{\partial_{\tau}} \bar J.
\end{equation}
However, this would mean that the shifted fields would no longer obey the chirality conditions \rref{toy_chiralsf} since the shifts do not obey them. So we have to be more careful. Rewrite the action,
\begin{eqnarray}
  \!  \! \!  \!   \!  \! \!  \!   \!  \! \!  \!  {\cal S} + {\cal S}_{J}   \!  \! \!  \! 
&=&   \!  \! \!  \!  \int d^2 \sigma d^2 \theta\, \left( Z \left(-\partial_{\tau}\right) \bar{Z} -  Z \frac{D\bar D}{\partial_{\tau}}\bar J  - \bar Z \frac{\bar D D}{\partial_{\tau}} J\right)
\nonumber \\
&=&   \!  \! \!  \!  \int d^2 \sigma d^2 \theta\,  \left( Z + \frac{\bar D D}{\partial_{\tau}} J \frac{1}{\partial_{\tau}}\right) \left(-\partial_{\tau}\right) \left( \bar{Z} + \frac{D\bar D}{\partial_{\tau}^{2}}\bar J  \right)  +  J \frac{D\bar D}{\partial_{\tau}^{2}} \bar J ,
 \label{toy_sjr1}
\end{eqnarray}
and introduce shifted fields as follows,
\begin{equation}
 Z_{s} = Z + \frac{\bar D D}{\partial_{\tau}} J \frac{1}{\partial_{\tau}}, \quad \quad \quad \quad
 \bar Z_{s} = \bar{Z} +  \frac{D\bar D}{\partial_{\tau}^{2}}\bar J. \label{toy_shift2}
\end{equation}
Using \rref{fremderiv} we see now directly that the shifted fields obey the chirality constraints \rref{toy_chiralsf} so that we can rewrite,
\begin{eqnarray}
{\cal Z}[J,\bar J]   \!  \! \!  \!  &=&   \!  \! \!  \!  \int {\cal D}Z{\cal D}\bar{Z} e^{-\left( \int d^2 \sigma d^2 \theta\,  \left( Z + \frac{\bar D D}{\partial_{\tau}} J \frac{1}{\partial_{\tau}}\right)\left(-\partial_{\tau}\right) \left( \bar{Z} + \frac{D\bar D}{\partial_{\tau}^{2}}\bar J  \right)  +  J \frac{D\bar D}{\partial_{\tau}^{2}} \bar J \right)}  
\nonumber \\
&=&   \!  \! \!  \!  \int {\cal D}Z_{s}{\cal D}\bar{Z}_{s} e^{-\left( \int d^2 \sigma d^2 \theta\,  Z_{s}  \partial_{\tau}  \bar{Z}_{s}  +  J \frac{D\bar D}{\partial_{\tau}^{2}} \bar J \right)},  
\label{toy_genrfuncshifttoy}
\end{eqnarray}
and the propagator is given formally by,
\begin{equation}
\left\langle Z \bar Z \right\rangle = - \frac{ D \bar{D}}{\partial_{\tau}^{2}}. \label{toy_prop1}
\end{equation}
However, this procedure gets very involved when the constraints become more complicated. Therefore we suggest an alternative derivation.

The constraints, eq.~(\ref{toy_chiralsf}), can be solved in terms of unconstrained fermionic, $\Lambda$ and  $\bar \Lambda$, superfields,
\begin{equation}
Z =\bar D \Lambda,  \quad \quad \quad \quad \bar Z= D \bar \Lambda. \label{toy_solvecon}
\end{equation}
Using this we can rewrite the action \rref{toy_s},
\begin{eqnarray}
{\cal S}  
  \!  \! \!  \!  &=&   \!  \! \!  \!  \int d^2 \sigma d^2 \theta\,  \Lambda \left(-\partial_{\tau} \bar{D} D \right) \bar \Lambda, \label{toy_su}
\end{eqnarray}
and the sourceterm \rref{toy_sj}, 
\begin{eqnarray}
{\cal S}_{J}   \!  \! \!  \!  &=&   \!  \! \!  \!  \int d^2 \sigma d^2 \theta\, \left( - \Lambda \bar{D} \bar J - J D \bar{\Lambda} \right). \label{toy_suj_l}
\end{eqnarray}
We again proceed by completing the squares,
\begin{eqnarray}   \!  \! \!  \!   \!  \! \!  \!   \!  \! \!  \!   \!  \! \!  \! 
{\cal S} + {\cal S}_{J}   \!  \! \!  \!  &=&   \!  \! \!  \!  \int d^2 \sigma d^2 \theta\,  \left( \Lambda + J \frac{D}{ \partial_{\tau}^{2}}\right) \left(-\partial_{\tau} \bar{D} D \right) \left( \bar{\Lambda} + \frac{\bar{D}}{\partial_{\tau}^{2}} \bar J   \right)  +  J \frac{ D \bar{D}}{\partial_{\tau}^{2}} \bar J \nonumber \\
&=&   \!  \! \!  \!  \int d^2 \sigma d^2 \theta\,  \left( \Lambda +  \frac{D}{ \partial_{\tau}} J \frac{1}{ \partial_{\tau}}\right)\left(-\partial_{\tau} \bar{D} D \right) \left( \bar{\Lambda} + \frac{\bar{D}}{\partial_{\tau}^{2}} \bar J   \right)  +  J \frac{ D \bar{D}}{\partial_{\tau}^{2}} \bar J,  \label{toy_sjr2}
\end{eqnarray}
and then introduce shifted fields as follows,
\begin{equation}
 \Lambda_{s} =  \Lambda + \frac{D}{ \partial_{\tau}} J \frac{1}{ \partial_{\tau}}, \quad \quad \quad \quad
 \bar \Lambda_{s} =\bar{\Lambda} + \frac{\bar{D}}{\partial_{\tau}^{2}} \bar J,
\end{equation}
which results in shifted constrained fields exactly as in \rref{toy_shift2},
\begin{equation}
 Z_{s} = \bar{D} \Lambda_{s} 
 = Z +  \frac{\bar{D} D}{ \partial_{\tau}} J \frac{1}{ \partial_{\tau}}, \quad \quad \quad \quad
 \bar Z_{s} = D \bar \Lambda_{s} 
 =  \bar{Z} +  \frac{D \bar{D}}{\partial_{\tau}^{2}} \bar J.
\end{equation}
We see that the propagator we get from  \rref{toy_sjr2} is exactly the same as the one we get from \rref{toy_sjr1} namely,
\begin{equation}
\left\langle Z \bar Z \right\rangle = - \frac{ D \bar{D}}{\partial_{\tau}^{2}}. \label{toy_prop2}
\end{equation}


\subsection{Feynman Rules}

To derive the propagator we add the following term to the action,
\begin{equation}
{\cal S}_J=\int d^2 \sigma d^2 \theta\, \sum_\alpha\left( - J^{\bar \alpha }Z^{ \alpha } - J^{ \alpha }Z^{\bar \alpha } -
\Omega^{\bar \alpha }\Psi^{ \alpha } - \Omega^{ \alpha }\Psi^{\bar \alpha }
\right),\label{sqj} 
\end{equation}
containing unconstrained bosonic sources $J$ for $Z$ and unconstrained fermionic sources $\Omega$ for $\Psi$. From this we get the following generating functional,
\begin{equation}
{\cal Z}[J,\bar J,\Omega,\bar \Omega] = \int {\cal D}Z{\cal D}\bar{Z}{\cal D}\Psi{\cal D}\bar{\Psi} e^{-\left( {\cal S}
 + {\cal S}_{J} \right) }. \label{gen_func}
\end{equation}
Now we want to derive the propagator using the familiar method of completing the squares and performing a shift in the fields. However due to the fact that our quantum fields obey chirality conditions \rref{betafuncchiralitycond} the shifts would have to obey the same conditions thereby making the derivation quite subtle. Luckily we can adopt the trick developed in the previous subsection, using the unconstrained superfields as defined in \rref{solvecon}, which allows us to circumvent these subtleties. We then find the propagators,
\begin{equation}
\begin{split}
\langle Z^ \alpha (1) Z^{\bar \beta }(2)\rangle& = - G^{ \alpha \bar \beta } \frac{D_2\bar D_2}{\Box_2 } \delta ^{(4)}(1-2),\\
\langle \Psi ^ \alpha (1)\Psi ^{\bar \beta }(2) \rangle &= -G^{ \alpha \bar \beta } \left( 1-\bar D_2D_2 \frac{ \partial _{ \tau _2}}{\Box_2} \right)\delta ^{(4)}(1-2),\\
\langle Z ^ \alpha (1)\Psi ^{\bar \beta }(2) \rangle &= -G^{ \alpha \bar \beta }\bar D_2 \frac{ \partial _{ \sigma  _2}}{\Box_2} \delta ^{(4)}(1-2),\\
\langle \Psi ^ \alpha (1) Z ^{\bar \beta }(2) \rangle &=G^{ \alpha \bar \beta } D_2 \frac{ \partial _{ \sigma _2}}{\Box_2} \delta ^{(4)}(1-2) ,\label{fullprops}
\end{split}
\end{equation}
which satisfy the boundary conditions (\ref{bdycds}). By construction these propagators are compatible with the constraints \rref{betafuncchiralitycond}. 

In writing down the propagators we used and will use the following definitions  involving $ \delta $-functions,
\begin{equation}
\begin{split}
\delta ^{(2)}( \theta _1- \theta _2)&\equiv ( \bar \theta _1-\bar \theta _2) ( \theta _1- \theta _2), \\
\delta ^{(3)}(1-2)&\equiv \delta ( \tau _1- \tau _2) \delta ^{(2)}( \theta _1 - \theta _2), \\
\delta ^{(4)}(1-2)&\equiv \delta^{(2)} ( \sigma  _1- \sigma  _2) \delta ^{(2)}( \theta _1 - \theta _2).
\end{split}
\end{equation}
The following identities involving these will be extremely usefull,
\begin{equation}
\begin{split}
\delta ^{(2)}( \theta _1- \theta _2) \delta ^{(2)} ( \theta _2- \theta _1) & =0, \\
\delta ^{(2)}( \theta _1- \theta _2) D_a\left(\delta ^{(2)} ( \theta _2- \theta _1) f( \tau _2- \tau _1)\right) &=0, \\
\delta ^{(2)}( \theta _1- \theta _2)\bar D_a\left( \delta ^{(2)} ( \theta _2- \theta _1) f( \tau _2- \tau _1)\right)&=0, \\
\delta ^{(2)}( \theta _1- \theta _2) \left(D_a \bar D_a\delta ^{(2)} ( \theta _2- \theta _1) f( \tau _2- \tau _1)\right)&= \\
-\delta ^{(2)}( \theta _1- \theta _2)\left(\bar D_a  D_a\delta ^{(2)} ( \theta _2- \theta _1)f( \tau _2- \tau _1)\right)&=\delta ^{(2)}( \theta _1- \theta _2)f( \tau _2- \tau _1),
\end{split}
\end{equation}
where the subindex $a$ is $1$ or $2$ and we did not sum over repeated subindices $a$.

The vertices on the boundary are easily derived from the interaction terms in the action \rref{interactiontermsaction}. 
From this we observe that we only need the first of the propagators in equation (\ref{fullprops}) evaluated at the boundary which is given by,
\begin{equation}
\begin{split}
\ID{}^{ \alpha \bar \beta }(1-2)& \equiv \langle Z^ \alpha (1) Z^{\bar \beta }(2)\rangle\Big|_{\sigma_{1},\sigma_{2} \to 0}\\
&= G^{ \alpha \bar \beta } D_2\bar D_2 \left(\frac 1 \pi \int_m^M \frac{d\,p}{p}\,\cos\big(p\,( \tau _1- \tau _2)\big)\, \delta ^{(2)}( \theta _1- \theta _2)\right) \\
&=-G^{ \alpha \bar \beta } \bar D_1 D_1 \left(\frac 1 \pi \int_m^M \frac{d\,p}{p}\,\cos\big(p\,( \tau _1- \tau _2)\big)\, \delta ^{(2)}( \theta _1- \theta _2)\right),\label{propone}
\end{split}
\end{equation}
where we introduced an IR regulator $m$ and the UV cut-off $M$.

Now supergraph techniques can be used to compute quantum corrections to the theory at the boundary. The background always appears as a background field in the Feynman diagrams and conventional $D$-algebra has to be preformed in standard fashion.

\begin{figure}[h]
\begin{center}
\psfig{figure=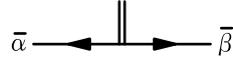}
\caption{A diagram containing vertices with only holomorphic or only anti-holomorphic indices will
be UV finite.
\label{fig:vertex}}
\end{center}
\end{figure}

In writing down the Feynman diagrams we use a notation inspired by the one introduced in  \cite{Grisaru2} which boils down to the following rules:
\begin{itemize}
\item The propagator is diagramatically represented as follows,
\begin{equation} \label{FreePropagatorFen}
 \psfig{figure=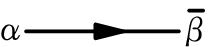,bbllx=0,bblly=3,bburx=59,bbury=11} \, \,  = \ID{}^{ \alpha \bar \beta }.
\end{equation}
The arrow always points from the holomorphic index to the anti-holomorphic index.
\item A vertex with $n$ holomorphic indices and $m$ anti-holmorphic indices is represented as,
\begin{equation}
 \psfig{figure=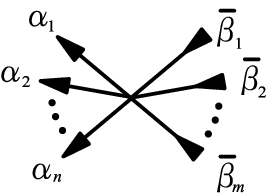,bbllx=0,bblly=24,bburx=77,bbury=53} \, \, = \frac{(n+m)!}{n! \, m!} V_{\alpha_{1}\alpha_{2}\ldots\alpha_{n}\bar\beta_{1}\bar\beta_{2}\ldots\bar\beta_{m}},
\end{equation} 

\vspace{.6cm}

\noindent where it should be clear that \psfig{figure=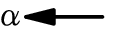,bbllx=0,bblly=3,bburx=30,bbury=11} and \psfig{figure=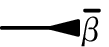,bbllx=0,bblly=3,bburx=29,bbury=11} represent free indices and should not be
confused with the propagator.

In the case of a $2$-point vertex we add a double line, to show explicitly the coupling to the background fields, at the interaction point because it would be invisible otherwise. See for example figure (\ref{fig:vertex}).
\end{itemize}

When calculating the beta-functions, we are solely interested in the UV divergences which we will treat using minimal subtraction.  The actual calculation is further simplified by several obsevations made in \cite{Grisaru2} which carry over to the present case. Using the rules of supergraph power counting, we easily see that all divergences are logarithmic at every loop order. As the interactions are fully characterized by a dimensionless potential $V$, the counterterm $V_{\text{c}}$ will be dimensionless as well. Since the superfields $Z_{cl},\bar{Z}_{cl}$ are dimensionless this implies that the counterterms will not involve any derivative ($[D]=[\bar{D}]=\frac{1}{2}$, $[\partial_{\tau}]=[\partial_{\sigma}]=1$) acting on the background fields. Thus the $D$-algebra can be preformed by integrating by parts the covariant spinor derivatives $D,\bar{D}$ only on the internal quantum lines of the diagrams. 
Another way of implementing this is by treating the interaction vertices as read of from the action \rref{interactiontermsaction} as effective constants. 

This implies the following drastic simplification on the Feymann rules used for the computation of UV divergent counterterms. We can neglect all terms in the boundary action  \rref{interactiontermsaction} which contain only quantum chiral or only quantum anti-chiral fields since any vertex of this type will produce a contribution with at least one $\bar{D}$ or $D$ acting on the external background fields. As argued before, these terms have too high a dimension to correspond to local divergent structures. In other words, vertices with
only holomorphic (or only anti-holomorphic) indices will never contribute to the UV divergences and we need not to consider them.

%
%

Consider for example the diagram in figure (\ref{fig:vertex}). It gives rise to a contribution of the form,
\begin{eqnarray}
\frac 1 2 \int d \tau _3 d^2 \theta _3 V_{ \gamma \delta  } \ID{}^{ \gamma \bar \alpha } (3-1)
\ID{}^{ \delta \bar \beta } (3-2),
\end{eqnarray}
which, upon partially integrating a fermionic derivative in one of the propagators,
can be seen to only contribute
to the UV finite part of the diagram. Exactly the same reasoning can be made with the effective
propagators appearing later on. 

A second important consequence is that the loop expansion is an expansion in the number of derivatives on the fieldstrengths. Indeed, a vertex where $n$ lines meet corresponds to $n-2$ derivatives acting on the fieldstrength (remember, $ F_{\alpha\bar\beta} = iV_{\alpha\bar\beta} $). Since no extra derivatives can be transferred to the potential $V$ in the calculation of UV divergences, we see that the total amount of derivatives on the fieldstrengths is given by,
\begin{equation}
(\#\text{derivatives}) = 2 (\#\text{propagators}) - 2 (\#\text{vertices}).
\end{equation}
This combined with the familiar relation,
\begin{equation} \label{topolooppropvert}
(\#\text{loops}) = (\#\text{propagators}) - (\#\text{vertices}) +1,
\end{equation}
results in,
\begin{equation}
(\#\text{derivatives}) = 2 (\#\text{loops}) - 2.
\end{equation}

\subsection{The Effective Propagator}

\begin{figure}[h]
\begin{center}
\psfig{figure=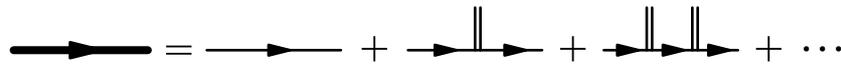}
\caption{The diagrammatic expansion of the effective propagator.
\label{fig:prop}}
\end{center}
\end{figure}

Using the previous observations, one calculates the relevant effective tree level propagator. It is diagramatically shown in figure (\ref{fig:prop}).

\noindent Both even and odd numbers of vertices contribute and the final expression reads as,
\begin{equation}
{\cal D}^{ \alpha \bar \beta }(1-2)= \ID_+^ { \alpha \bar \beta }(1-2)
+\ID_-^ { \alpha \bar \beta }(1-2),\label{propeff}
\end{equation}
with,
\begin{eqnarray}
\ID_\pm^ { \alpha \bar \beta }(1-2)= h_\pm^{ \alpha \bar \beta
}\ID_\pm(1-2)= h_\pm^{ \alpha \bar \beta } D_2 \bar D_2\left( \delta
^{(2)}( \theta _1- \theta _2) \Delta_\pm (1-2)\right), \label{Dpm}
\end{eqnarray}
with,
\begin{eqnarray}
\Delta _\pm( \tau )= \frac{1}{2 \pi }\int_m^M \frac{dp}{p}\, e^{\mp
i\,p\, \tau }. \label{t-space prop}
\end{eqnarray}
Note that $ \Delta _+( - \tau )= \Delta _-( \tau )$.

In the previous we used the open string metric(s), 
\begin{eqnarray}
h^\pm_{ij}\equiv G_{ij} \pm F_{ij},
\end{eqnarray}
where we obviously have that $h_{ij}^+=h_{ji}^-$.
The inverse, $h_\pm^{ij}$, is defined by,
\begin{eqnarray}
h^{ik}_+\,h^+_{kj}=h^{ik}_-\,h^-_{kj}= \delta ^i_j.
\end{eqnarray}
The holomorphicity  of the gauge bundle implies that,
\begin{eqnarray}
h_{ \alpha \beta }^\pm= h_{ \bar \alpha \bar \beta }^\pm=0.
\end{eqnarray}
It is also useful to define the following symmetric and anti-symmetric tensors,
\begin{eqnarray}
{\cal G}^{ij}\equiv \frac 1 2 \big(h_+^{ij}+h_-^{ij}\big), \quad \quad \quad \text{and} \quad \quad \quad {\cal B}^{ij}\equiv \frac{1}{2\,i}\big(h_+^{ij}-h_-^{ij}\big).
\end{eqnarray}

Since it is the effective propagator we will be using most, we will represent it, replacing the fat notation of figure (\ref{fig:prop}), diagrammatically as,
\begin{equation}
 \psfig{figure=propagator_59x11.eps,bbllx=0,bblly=3,bburx=59,bbury=11} \, \,  = {\cal D}^{ \alpha \bar \beta },
\end{equation}
thereby replacing the notation of the free propagator \rref{FreePropagatorFen}. If we need the latter we will explicitly mention it. Lastly we will be needing the the following two propagators,
\begin{equation}\label{diag.prop.propagator}
\psfig{figure=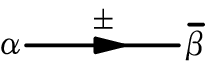,bbllx=0,bblly=4,bburx=59,bbury=16} = \ID_\pm^ { \alpha \bar \beta }.
\end{equation}

\subsection{Diagrammatic Notation} \label{Diagrammatic_Notation} 

In keeping track of all contributions to the bare potential and checking the renormalization group equations it proved to be extremely helpful to introduce a diagrammatic notation for the index structure of the different contributions. One of the most attractive features of this notation is that the diagram of the index structure is, as will become clear, exactly the same as the diagram of the full contribution it corresponds to. Of course, this same feature is also a possible cause for confusion. We hope however that the precise meaning of the diagram will always be clear from the context. When diagrams are used several times in the same formula, not necessarily always with the same meaning, the part that indicates the index structure only (without the momentum integral and symmetry factor), will appear in between brackets. (For an example of this, see equation (\ref{2loop:diag}).) More concretely,
we introduce the notations.

\begin{equation}
h_{\pm}^{\alpha\bar{\beta}} =
\psfig{figure=hplusmin_59x16.eps,bbllx=0,bblly=4,bburx=59,bbury=16} \, \, ,\label{diag.prop}
\end{equation}

\begin{equation}
{\cal G}^{\alpha \bar \beta} =
\psfig{figure=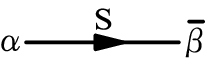,bbllx=0,bblly=4,bburx=59,bbury=16} \, \, ,
\end{equation}

\begin{equation}
\partial_\gamma F_{\alpha \bar \beta} =
\partial_\alpha F_{\gamma \bar \beta} =
\psfig{figure=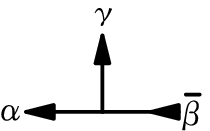,bbllx=0,bblly=15,bburx=58,bbury=37} \, \, ,
\end{equation}

\noindent and

\begin{equation}
\partial_{\bar \gamma} F_{\alpha \bar \beta} =
\partial_{\bar \beta} F_{\alpha \bar \gamma} =
\psfig{figure=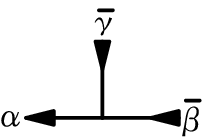,bbllx=0,bblly=15,bburx=58,bbury=37} \, \, .
\end{equation}

\vspace{.4cm}

\noindent Here it should be clear that
\psfig{figure=vertex_in_30x11.eps,bbllx=0,bblly=3,bburx=30,bbury=11}
and
\psfig{figure=vertex_out_29x11.eps,bbllx=0,bblly=3,bburx=29,bbury=11}
represent derivatives on the fieldstrength or potential and should not be
confused with the object defined in eq. (\ref{diag.prop})
corresponding to the propagator.

It's very easy to compute derivatives of different expressions by
using (for example),
\begin{eqnarray}
\partial_{\bar \gamma} h_\pm^{\alpha \bar \beta} = \pm\, h_\pm^{\alpha \bar
\delta} h_\pm^{\varepsilon \bar \beta}\, \partial_{\bar \gamma}
F_{\varepsilon \bar \delta}.
\end{eqnarray}
In diagrammatic form this becomes,
\begin{equation}
\partial_{\bar \gamma}\left( \psfig{figure=hplusmin_59x16.eps,bbllx=0,bblly=4,bburx=59,bbury=16} \right) =
\pm \, \,
\psfig{figure=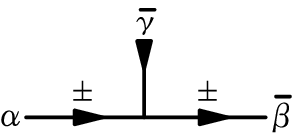,bbllx=0,bblly=14,bburx=84,bbury=36} \, \, .
\end{equation}

\vspace{.4cm}

\noindent The same formula evidently also holds for derivatives with
respect to holomorphic coordinates. 

\section{One Loop}

As a warming up exercise, we calculate the one loop contributions in this section and the two loop contributions in the next.
\begin{figure}[h]
\begin{center}
\psfig{figure=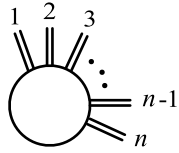}
\caption{A one loop diagram with $n$ vertices. The propagators are the free ones from equation (\ref{propone}).
\label{fig:1loop}}
\end{center}
\end{figure}

\noindent Performing the D-algebra, one finds that a one loop diagram with $2n$ vertices is given by,
\begin{equation}
\frac{1}{2n\,i}\int d \tau _1d^2 \theta _1d \tau _2 d^2 \theta _2\left(F^{2n}\right)_{ \alpha
\bar \beta }G^{ \alpha \bar \beta } \big(\Delta _+( \tau _1- \tau _2) - \Delta _-( \tau _1- \tau _2)\big)\delta ^{(3)}(1-2),
\end{equation}
where we used the relation between the fieldstrength and the potential (\ref{fspot}) and introduced the notation,
\begin{eqnarray}
(F^m)_{ \alpha \bar \beta }\equiv F_{ \alpha \bar \gamma _1}G^{ \bar \gamma _1 \delta _1}
 F_{ \delta _1 \bar \gamma _2}G^{ \bar \gamma _2 \delta _2}
 F_{ \delta _2 \bar \gamma _3}
 \cdots G^{\bar \gamma _{m-1} \delta _{m-1}}F_{ \delta _{m-1}\bar \beta }.
\end{eqnarray}
Because $\big( \Delta _+( \tau _1- \tau _2) - \Delta _-( \tau _1- \tau _2) \big)$ is an odd function, this vanishes and hence all one loop diagrams with an even amount of vertices are zero.

Turning to a one loop diagram with $2n+1$ vertices, one finds an ultra-violet divergence of the form,
\begin{equation}
- i\,  \frac{ \lambda }{2n+1} \int d \tau d^2\, \theta \,G^{ \alpha \bar \beta }\left(F^{2n+1}\right)_{ \alpha \bar \beta },
\end{equation}
where,
\begin{eqnarray}
\lambda \equiv \frac{1}{\pi}\ln \left( \frac{M}{m}\right).
\end{eqnarray}

When summing over all loops, we get the total divergent contribution at one loop,
\begin{equation}
- i\, \lambda \,   \int d \tau d^2 \theta\,  G^{ \alpha \bar \beta }
\left(\mbox{arcth}\,F\right)_{ \alpha \bar \beta }.
\end{equation}
Using minimal subtraction, this gives us the bare potential through order
$\hbar$,
\begin{equation}
V_{\text{bare}}= V+ \lambda\, V_{(1)} = V+ i\, \lambda   \, G^{ \alpha \bar \beta }
\left(\mbox{arcth}\,F\right)_{ \alpha \bar \beta },
\end{equation}
with,
\begin{equation}
V_{(1)}=i    G^{ \alpha \bar \beta } \left(\mbox{arcth}\,F\right)_{ \alpha \bar \beta }.
\end{equation}

Requiring the beta-funtion for $V$, as found in  \rref{betafunctionforv}, to vanish through this order,
\begin{equation} \label{betafunctionforvoneloop}
G^{ \alpha \bar \beta } \left(\mbox{arcth}\,F\right)_{ \alpha \bar \beta } = 0,
\end{equation}
we find the stability condition \rref{whatsisDUYdeformed}. As already mentioned in subsection \ref{Stable_Holomorphic_Vector_Bundles}, this stability condition together with the holomorphicity conditions \rref{fspot} is equivalent to the equations of motion for the Born-Infeld action.

Turning to the beta-functions for the gauge fields (\ref{bfundef}) we find that they vanish provided,
\begin{eqnarray}
 \partial_{\bar \beta} \left[ G^{ \alpha \bar \gamma }
\left(\mbox{arcth}\,F\right)_{ \alpha \bar \gamma } \right] = 0,\label{loeom}
\end{eqnarray}
which we easily recognize as the equations of motion for the Born-Infeld action in the case of holomorphic vector bundles. This provides a nice justification that asking the beta-function for $V$ to vanish is indeed sufficient.

In what follows we need the following identities,
\begin{equation} \label{identityderivV}
\partial_{\bar{\beta}} V_{(1)} = i {\cal G}^{\gamma \bar \delta} \partial_{\bar \beta} F_{\gamma \bar \delta} = i \, \,
\psfig{figure=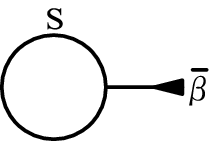,bbllx=0,bblly=15,bburx=60,bbury=39},
\end{equation}

\vspace{.4cm}

\noindent where we used the diagrammatic notation introduced in the previous section. Applying yet another derivative results in,
\begin{equation}
\begin{split}
\partial_{\alpha}\partial_{\bar{\beta}} V_{(1)} &= i \left[
\psfig{figure=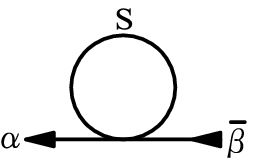,bbllx=0,bblly=22,bburx=71,bbury=47} \, \, + \right. \\ 
& \quad \quad \quad \quad \left. \frac{1}{2} \Bigg(
\psfig{figure=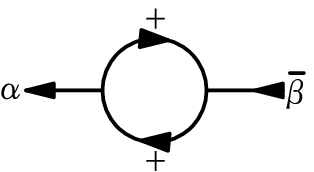,bbllx=0,bblly=22,bburx=88,bbury=47}
\, \, - \, \,
\psfig{figure=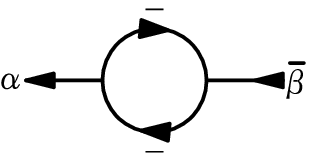,bbllx=0,bblly=22,bburx=88,bbury=47}
\Bigg)\right]. \label{ppartialV}
\end{split}
\end{equation}

\vspace{.2cm}

\section{Two Loops}

We now turn to the two loop contributions. Because the actual calculation is quite technical and would swamp the results we feel that it is necessary to split this paragraph into two pieces. In the first subsection we will simply summarize the results from the calculation and interpret them. The second will then be used to perform the actual calculation and elaborate, using the diagrammatical language we introduced, on how the magical addition found in the first section arises.

\subsection{Summary}

\begin{figure}[h]
\begin{center}
\psfig{figure=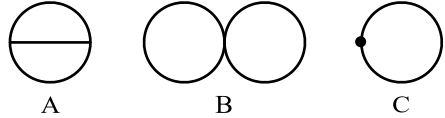}
\caption{The two 2-loop diagrams and the contribution arising from the one loop counter term.
The effective propagator as defined in equation (\ref{propeff}) is used.
\label{fig:2loop}}
\end{center}
\end{figure}

To start, we write down all three diagrams at two loops in figure (\ref{fig:2loop}).

\noindent Diagram A results in,
\begin{equation} \label{2loop:diagrA}
-\frac{ \lambda  ^2}{2} \int d \tau d^2 \theta \,V_{ \bar \beta _1\alpha _2 \alpha _3} V_{ \alpha _1 \bar \beta _2 \bar \beta_3}
\left( {\cal G}^{ \alpha_1 \bar \beta _1} {\cal G}^{ \alpha _2 \bar \beta _2} {\cal B}^{ \alpha _3 \bar \beta _3}+ {\cal B}^{\alpha_1 \bar \beta _1} {\cal G}^{ \alpha _2 \bar \beta _2} {\cal G}^{ \alpha _3 \bar \beta _3}  \right),
\end{equation}
while diagram B gives,
\begin{equation} \label{2loop:diagrB}
\frac{ \lambda  ^2}{2}\int d \tau d^2 \theta \, {\cal G}^{ \alpha \bar \beta } {\cal G}^{ \gamma \bar \delta } V_{ \alpha \bar \beta \gamma \bar \delta }. 
\end{equation}
Adding the two gives the following interesting (magical) result,
\begin{eqnarray}
-i\frac{ \lambda  ^2}{2} \int d \tau d^2 \theta \,{\cal G}^{ \alpha
\bar \beta } \partial _ \alpha
\partial _{ \bar \beta }\left(G^{ \gamma \bar \delta }
\left(\mbox{arcth}\,F\right)_{ \gamma \bar \delta  }\right).
\label{2loop:A+B}
\end{eqnarray}
Finally, the subdivergence (diagram C), which arises from the one
loop counter term, has the same structure and is given by,
\begin{eqnarray}
 i\;\lambda^2\int d \tau d^2 \theta \, {\cal G}^{ \alpha \bar \beta }
\partial _ \alpha
\partial _{ \bar \beta }\left(G^{ \gamma \bar \delta }
\left(\mbox{arcth}\,F\right)_{ \gamma \bar \delta  }\right).
\end{eqnarray}
Adding these contributions gives us the bare potential through two loops.
\begin{eqnarray}
V_{\text{bare}}=V+ \lambda\, V_{(1)}+ \lambda ^2\, V_{(2)},\label{Vbare2}
\end{eqnarray}
where,
\begin{eqnarray}
V_{(1)}&=&i    G^{ \alpha \bar \beta }
\left(\mbox{arcth}\,F\right)_{ \alpha \bar \beta } \nonumber\\
V_{(2)}&=& - \frac{i}{2}\, {\cal G}^{ \alpha \bar \beta } \partial _
\alpha \partial _{ \bar \beta }\left(G^{ \gamma \bar \delta }
\left(\mbox{arcth}\,F\right)_{ \gamma \bar \delta
}\right) \label{2loop:result}\\
&=& - \frac 1 2 \, \,
\psfig{figure=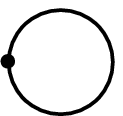,bbllx=0,bblly=13,bburx=33,bbury=31}
\, S\;. \nonumber
\end{eqnarray}

\vspace{.1cm}

\noindent In the last line we used the diagrammatic notation introduced previously. One easily verifies that the
two loop ${\cal O}( \lambda ^2)$ counterterm agrees with the
renormalization group result obtained from equation (\ref{rg}).

As the bare potential does not get an order $ \lambda  $ correction at two loops, the
beta-function will not be modified at this order and as a consequence, the Born-Infeld action
receives no two derivative corrections. This result agrees with \cite{andreevtseytlin}.

\subsection{The Explicit Two Loop Computation}\label{app 2l}

Now we turn to the actual calculation which we will perform very explicitly as an illustration of how the calculation, including the one at three loops, was done.

\begin{figure}[h]
\begin{center}
\psfig{figure=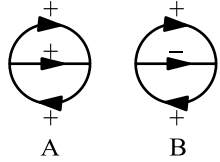} \caption{The basic
non-trivial diagrams at two loops. \label{fig:2loop_basic}}
\end{center}
\end{figure}

Let us first consider diagram A in figure (\ref{fig:2loop}). The effective propagator, as defined in \rref{propeff}, has a certain direction so that diagram A actually represents all possible inequivalent topologies with directed lines. Since the righthand side of \rref{propeff} consists of two terms, we have to consider all possible distinct diagrams with directed lines and all possible combinations of positive and negative frequency parts of the propagator. The latter can be found in equation \rref{Dpm}. It turns out that the only momentum integrals that have to be performed explicitly, are the ones belonging to the two diagrams in figure (\ref{fig:2loop_basic}). All other integrals are either zero (or finite) or are related to these ones by complex conjugation or partial integration. 

Let's start with diagram A of figure (\ref{fig:2loop_basic}). The Feynman rules for this diagram give,
\begin{equation}
\frac 1 2 \left\langle V_{ \bar \beta _1 \alpha _2 \alpha _3}(1) V_{ \alpha _1
\bar \beta _2 \bar \beta _3}(2) \ID_+^{ \alpha_1 \bar \beta _1}(2-1)
\ID_+^{ \alpha _2 \bar \beta _2}(1-2) \ID_+^{ \alpha _3 \bar \beta
_3}(1-2)\right\rangle_{1,2}\;,
\end{equation}
where $\langle \, \ldots \, \rangle_1$ means integrating over $\tau_1$, $\theta_1$ and $\bar
\theta_1$. Using \rref{Dpm} and performing the D-algebra leads to,
\begin{equation}
\frac 1 2 \left\langle V_{ \bar \beta _1 \alpha _2 \alpha _3} V_{\alpha _1 \bar \beta _2 \bar \beta _3} h_+^{ \alpha_1 \bar \beta _1} h_+^{ \alpha _2 \bar \beta _2} h_+^{ \alpha _3 \bar \beta _3} \, {\cal I}_A\right\rangle_1\;,
\end{equation}
with
\begin{equation}
{\cal I}_A = \int_{-\infty}^{+\infty} d\tau \; \Delta_+(\tau)\Delta_+(\tau) \partial_{\tau}\Delta_-(\tau)\;,\label{I}
\end{equation}
where we made the change of variable $\tau_2 \rightarrow \tau \equiv
\tau_1 - \tau_2$ and made use of $ \Delta _+( - \tau )= \Delta _-(
\tau )$.

Inserting the explicit form of the coordinate space propagators and,
\begin{eqnarray}
\partial_\tau \Delta_\pm (\tau) = \frac{1}{2\pi\tau} (e^{\mp iM\tau}- e^{\mp
im\tau})\;,
\end{eqnarray}
we arrive at the following expression,
\begin{equation} \label{I p-space}
\begin{split} 
{\cal I}_A &= \frac{1}{(2\pi)^3}\int_m^M \frac{dp}{p} \int_m^M \frac{dq}{q} \int \frac{d\tau}{\tau}\; \left(e^{-i(p+q-M)\tau}-e^{-i(p+q-m)\tau}\right) \\
&= -\frac{i\pi}{(2\pi)^3}\int_1^\Lambda \frac{dp}{p} \int_1^\Lambda \frac{dq}{q}\;[\epsilon(p+q-\Lambda)-\epsilon(p+q-1)]\;.
\end{split}
\end{equation}
In the last line we introduced the sign function,
\begin{equation}
  \epsilon(p)=1 \quad  \text{for}  \quad  p>0
  \quad \quad \quad \text{and}  \quad \quad \quad  
  \epsilon(p)=-1 \quad \text{for}  \quad  p<0. 
\end{equation}
Furthermore, all momenta are expressed in units of $m$ and again we used $\Lambda = M/m$.

The integral \rref{I p-space} can easily be written in terms of logarithms and dilogarithms,
\begin{equation}\label{finiteL}
{\cal I}_A = \frac{i}{(2\pi)^2}\left[ \ln(\Lambda-1)\ln\Lambda + \Li_2\left(\frac{1}{\Lambda}\right)- \Li_2\left(\frac{\Lambda-1}{\Lambda}\right)\right]\;,
\end{equation}
with,
\begin{equation}
\Li_2(x) = -\int_0^x dz\; \frac{\ln(1-z)}{z} = \sum_{k=1}^\infty \frac{x^k}{k^2}\;, \quad \quad \vert x \vert \leq 1\;.\label{poly2}
\end{equation}
The restriction to $\vert x \vert \leq 1$ is only required for the second equality.

Of course we are ultimately only interested in the UV divergent part of ${\cal I}_A$, so that we only need the behavior of  \rref{finiteL} for $\Lambda >> 1$,
\begin{eqnarray} \label{betaI_Ahighlambda}
{\cal I}_A = -\frac{i}{24}+\frac{i}{(2\pi)^2}\ln^2(\Lambda) + {\cal
O}\left( \frac{1}{\Lambda}\right) \, ,
\end{eqnarray}
where we used $\Li_2(1)=\zeta(2) = \pi^2/6$, as is clear from \rref{poly2}. The first term on the right hand side of \rref{betaI_Ahighlambda} is irrelevant when we only consider two loop contributions. However, at the three loop level (for instance for diagrams C and D of figure (\ref{fig:3loop})) this result will be multiplied by $\log(\Lambda)$, coming from the extra loop. This means that the first term will start playing a role as well, while the terms of order $1/\Lambda$ and higher will continue to be irrelevant for the UV behavior.

Putting everything together, we find the following contribution from diagram A of figure (\ref{fig:2loop_basic}) to \ref{2loop:diagrA},
\begin{equation}
i\;\frac{\lambda ^2}{8}\int d \tau d^2 \theta \, V_{ \bar \beta _1
\alpha _2 \alpha _3} V_{ \alpha _1 \bar \beta _2 \bar \beta _3}
h_+^{ \alpha_1 \bar \beta _1} h_+^{ \alpha _2 \bar \beta _2} h_+^{
\alpha _3 \bar \beta _3}\, .
\end{equation}
It is readily verified that this is indeed one of the terms appearing in \rref{2loop:diagrA}.

Turning now to diagram B of figure (\ref{fig:2loop_basic}), one finds that the equivalent of ${\cal I}_A$ for this diagram is,
\begin{equation}
{\cal I}_B = \int_{-\infty}^{+\infty} d\tau \; \Delta_+(\tau)\Delta_-(\tau) \partial_{\tau}\Delta_-(\tau)\, ,
\end{equation}
However, this integral at its turn can be related to ${\cal I}_A$ by partial integration and complex conjugation as follows,
\begin{equation}
{\cal I}_B = -\frac 1 2 \int_{-\infty}^{+\infty} d\tau \; \Delta_-(\tau)\Delta_-(\tau) \partial_{\tau}\Delta_+(\tau) = -\frac1 2 {\cal I}_A^* = \frac1 2 {\cal I}_A\, ,
\end{equation}
where in the last equality we used the fact that ${\cal I}_A$ is purely imaginary. 

Diagram B of figure (\ref{fig:2loop}) almost trivially leads to (\ref{2loop:diagrB}), so that it turns out that diagram A of figure (\ref{fig:2loop_basic}) leads to the only nontrivial integral we have to compute at two loops. The total contribution from diagrams A and B in figure (\ref{fig:2loop}) can very nicely and suggestively be written using the diagrammatic notation for the index structure explained in subsection \ref{Diagrammatic_Notation},
\begin{equation} \nonumber
\psfig{figure=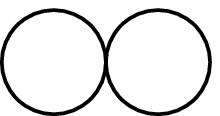,bbllx=0,bblly=13,bburx=61,bbury=31} \, \, + \, \,
\psfig{figure=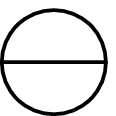,bbllx=0,bblly=13,bburx=31,bbury=31}  = - \frac i 2 \lambda^2
\left\langle  \psfig{figure=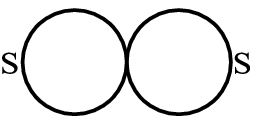,bbllx=0,bblly=13,bburx=72,bbury=31} \, \, + \frac 1 2
\Bigg(  \psfig{figure=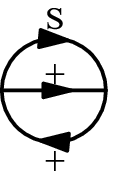,bbllx=0,bblly=21,bburx=31,bbury=48} \, \, - \, \,
\psfig{figure=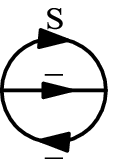,bbllx=0,bblly=21,bburx=31,bbury=48}  \Bigg)  \right\rangle
\end{equation}

\vspace{-0.3cm}

\begin{equation} \label{2loop:diag}
\end{equation}
Finally, the contribution from diagram C in figure (\ref{fig:2loop})
has to be taken into account. Using eq. (\ref{ppartialV}) one easily finds,
\begin{eqnarray}
\psfig{figure=2_loop_counterterm_33x31.eps,bbllx=0,bblly=13,bburx=33,bbury=31} &=& \lambda^2
\left\langle {\cal G}^{\alpha \bar \beta} \partial_{\alpha}\partial_{\bar{\beta}} V_{1} \right\rangle \nonumber\\
&=& i \lambda^2
\left\langle  \psfig{figure=2_loop_bril_SS_72x31.eps,bbllx=0,bblly=13,bburx=72,bbury=31} \, \, + \frac 1 2 \Bigg(  \psfig{figure=2_loop_setting_sun_spp_31x48.eps,bbllx=0,bblly=21,bburx=31,bbury=48} \, \, - \, \,  \psfig{figure=2_loop_setting_sun_smm_31x48.eps,bbllx=0,bblly=21,bburx=31,bbury=48}  \Bigg)  \right\rangle .
\end{eqnarray}

\vspace{.1cm}

\noindent Comparing this to (\ref{2loop:diag}) it is easy to understand why indeed we could write (\ref{2loop:A+B}) the way we did. Adding all two loop contributions, we indeed find (\ref{2loop:result}).

While the use of this diagrammatic notation might seem a bit unnecessary at two loops, it really becomes extremely useful (not to say necessary in every practical meaning of the word) at three loops.  Of course, even when using our diagrammatic notation, calculations will become quite lengthy at that stage and we will not show any of these explicitly there. Therefore we felt it was necessarily to illustrate it extensively on the two loop calculation.

\section{Three Loops}

Just as in the previous, it turns out to be very useful to split this section into two pieces. In the first subsection we limit ourselves to giving the results from the three loop calculation and interpreting them while in the second we will try to give an outline of the calculation together with some important intermediate results.

\subsection{Results}

\begin{figure}[h]
\begin{center}
\psfig{figure=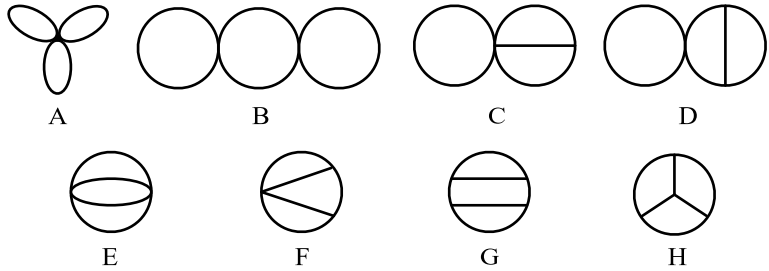}
\caption{All the 3-loop diagrams.
The effective propagator, eq. (\ref{propeff}), is used.
\label{fig:3loop}}
\end{center}
\end{figure}

The 3-loop diagrams are shown in figure \rref{fig:3loop}. As is explained in more detail in the next subsection, three loop diagrams give rise to terms proportional to $\lambda^3$
as well as terms linear in $\lambda$. It is a non-trivial check on the calculation that, with the contributions from the one and two loop counterterms taken into account, there are no terms quadratic in $\lambda$ present at this order. This must be the case, since there was no contribution to the beta-function at the two loop level. The $\lambda^3$ terms can be expressed quite concisely as,
\begin{eqnarray}
\begin{split}
V_{(3)}&= - \frac{1}{3}\, {\cal G}^{ \alpha \bar \beta }
\partial _ \alpha \partial _{ \bar \beta }V_{(2)} -\frac{i}{12}
\left( h_+^{\alpha \bar \beta}h_+^{\gamma \bar \delta} - h_-^{\alpha
\bar \beta}h_-^{\gamma \bar \delta}\right)
\partial_\alpha \partial_{\bar \delta} V_{(1)}
\partial_\gamma \partial_{\bar \beta} V_{(1)}\\
&= - \frac{1}{3}\,
\psfig{figure=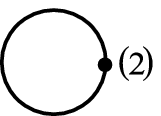,bbllx=0,bblly=12,bburx=44,bbury=31}
- \frac{i}{12} \left(
\psfig{figure=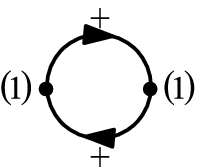,bbllx=0,bblly=20,bburx=56,bbury=46}
\, \, - \, \,
\psfig{figure=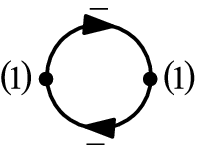,bbllx=0,bblly=20,bburx=56,bbury=46}
\right) \label{3loop:RG}
\end{split}
\end{eqnarray}
which, again, perfectly agrees with the renormalization group equations (\ref{rg})! This provides a very strong consistency check on our calculations. In the second line we again used the diagrammatic notation introduced before.  We also wish to remark that heavy use of this notation was made in checking equation \rref{3loop:RG}.

Adding the terms linear in $\lambda$ to the one loop result, we find up to terms containing four derivatives of the
fieldstrength,
\begin{equation}\label{result}
V_{(1)}=i    G^{ \alpha \bar \beta } \left(\mbox{arcth}\,\tilde
F\right)_{ \alpha \bar \beta } - \frac{1}{48} S_{ij\alpha \bar
\beta} S_{kl\gamma \bar \delta}\; h_+^{jk}h_+^{li}\left( {\cal
G}^{\alpha \bar \delta} {\cal B}^{\gamma \bar \beta} + {\cal
B}^{\alpha \bar \delta} {\cal G}^{\gamma \bar \beta} \right) + {\cal K}V_{(1,1)}.
\end{equation}
Here $\tilde F$ is the fieldstrength associated with the gauge potential $\tilde A$, related to the original $A$ by the field redefinition,
\begin{eqnarray}
\begin{split}
\tilde A_{\alpha} &= A_{\alpha} + \frac{1}{24}\partial_{\alpha}
\left( F_{\beta_1 \bar \gamma_2,\bar \gamma_3} F_{\beta_2 \bar
\gamma_1,\beta_3} h_+^{\beta_1 \bar \gamma_1}
h_+^{\beta_2 \bar \gamma_2} {\cal G}^{\beta_3 \bar \gamma_3} \right)\\
&= A_{\alpha} + \frac{1}{24}\partial_{\alpha}\,
\psfig{figure=2_loop_setting_sun_spp_31x48.eps,bbllx=0,bblly=21,bburx=31,bbury=48}\,;
\\ \\
\tilde A_{\bar \alpha} &= A_{\bar \alpha} +
\frac{1}{24}\partial_{\bar \alpha} \left( F_{\beta_1 \bar
\gamma_2,\bar \gamma_3} F_{\beta_2 \bar \gamma_1,\beta_3}
h_-^{\beta_1 \bar \gamma_1} h_-^{\beta_2 \bar \gamma_2} {\cal
G}^{\beta_3 \bar \gamma_3} \right)\\
&= A_{\bar \alpha} + \frac{1}{24}\partial_{\bar \alpha}\,
\psfig{figure=2_loop_setting_sun_smm_31x48.eps,bbllx=0,bblly=21,bburx=31,bbury=48}\,.
\label{field redef}
\end{split}
\end{eqnarray}
Again our diagrammatic language turned out to be extremely useful in working out the consequences of this field redefinition.
Notice that we can omit the tilde on the fields appearing in the second term in equation (\ref{result}), because the effect of this is higher order in $\hbar$. Recall that,
\begin{eqnarray} \label{ScurvBeta}
S_{ijkl}=\partial_i \partial_j F_{kl} +  h_+^{mn} \partial_i F_{km}\; \partial_j F_{ln} -  h_+^{mn} \partial_i F_{lm}\; \partial_j F_{kn} .
\end{eqnarray}
The use of Latin indices indicates a summation over real
coordinates, where the use of Greek indices, as usual, refers to a
summation over complexified holomorphic or anti-holomorphic
variables. Finally, in the last term of eq. (\ref{result}), ${\cal
K}$ is a derivative operator acting on the one loop part of
$V_{(1)}$ (see eq. (\ref{loopexp}) for notation). This term can be
written in a fairly transparent way by again making use of our
diagrammatic notation.
\begin{eqnarray}
{\cal K}V_{(1,1)} &=& \frac{1}{24}  \left(
\psfig{figure=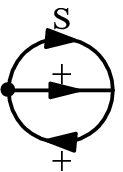,bbllx=0,bblly=21,bburx=33,bbury=48} \right. \, \, - \, \,
\psfig{figure=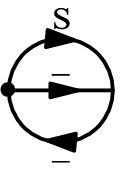,bbllx=0,bblly=21,bburx=33,bbury=48} \, \, + \, \,
\psfig{figure=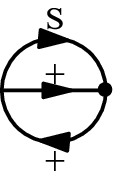,bbllx=0,bblly=21,bburx=33,bbury=48} \, \, - \, \,
\psfig{figure=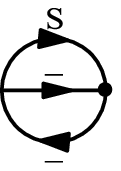,bbllx=0,bblly=21,bburx=33,bbury=48}\nonumber \\
&& \quad  \quad \, \,
\psfig{figure=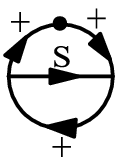,bbllx=0,bblly=21,bburx=33,bbury=48}  \, \, + \, \,
\psfig{figure=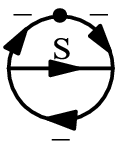,bbllx=0,bblly=21,bburx=33,bbury=48} \, \, + \, \,
\psfig{figure=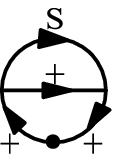,bbllx=0,bblly=21,bburx=33,bbury=48} \, \, + \, \,  \left.
\psfig{figure=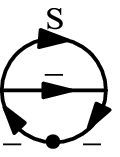,bbllx=0,bblly=21,bburx=33,bbury=48} \right) \\
&+& \frac{1}{48}  \left(
\psfig{figure=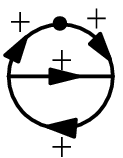,bbllx=0,bblly=21,bburx=33,bbury=48} \right.  \, \, - \, \,
\psfig{figure=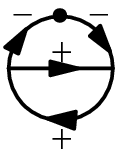,bbllx=0,bblly=21,bburx=33,bbury=48} \, \, - \, \,
\psfig{figure=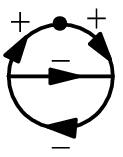,bbllx=0,bblly=21,bburx=33,bbury=48} \, \, + \, \,  \left.
\psfig{figure=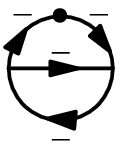,bbllx=0,bblly=21,bburx=33,bbury=48} \right)\nonumber
\end{eqnarray}

Since up to this order, we can put $V_{(1,1)}$ to zero and, in general, a field redefinition has no physical consequences, we arrive at the following correction to the stability condition for holomorphic vector bundles,
\begin{equation}
G^{ \alpha \bar \beta } \left(\text{arcth}\, F\right)_{ \alpha \bar
\beta } + \frac{1}{96} S_{ij\alpha \bar \beta} S_{kl\gamma \bar
\delta}\; h_+^{jk}h_+^{li}\left( h_{+}^{\alpha \bar \delta} h_{+}^{\gamma \bar \beta} - h_{-}^{\alpha \bar \delta} h_{-}^{\gamma \bar \beta} \right) =0.
\end{equation}
This is exactly the stability condition, as first written down in \cite{koerber:thesis}, which follows from the action presented in \cite{wyllard} (see subsection \ref{Higher_Derivative_Corrections}),
\begin{equation} \label{betafwyllardaction}
\begin{split}
{\cal S} =  \tau_9 \int d^{10}X\; \sqrt{h^{+}}\bigg[ 1 &+
\frac{1}{96} \Big( \frac 1 2 h_+^{ij}h_+^{kl}
S_{jk} S_{li} \\
&- h_+^{k_2 i_1}h_+^{i_2 k_1}h_+^{l_2j_1}h_+^{j_2 l_1}
S_{i_1 i_2 j_1j_2}S_{k_1k_2 l_1 l_2}\Big)\bigg],
\end{split}
\end{equation}
where,
\begin{equation}
S_{ij} = h_+^{kl} S_{klij}.
\end{equation}

\subsection{Calculational Outline} \label{beta_Calculational_Outline_3loops}

Before we start explaining the general procedure we used for calculating the three loop contributions, it's useful to understand some general features for any number of loops. It follows from the D-algebra that the number of $\tau$-derivatives\footnote{This will always equal the number of $\tau$
integrations (number of vertices minus the global position of the
diagram), so that the procedure will always make sense.}  appearing in the equivalent
of equation \rref{I} is given by,
\begin{equation}
(\#\tau\text{-derivatives}) = (\#\text{vertices}) -1.
\end{equation}
If we combine this with the topological relation \rref{topolooppropvert} we can conclude that the number of remaining momentum integrations will be,
\begin{equation} 
 (\#\text{propagators}) - (\#\tau\text{-derivatives}) = (\#\text{loops}),
\end{equation}
which, of course, makes very good sense. More importantly, the fact that in general there will be more than one derivative appearing in the equivalent of eq. (\ref{I}), will
lead to products of multiple sign functions in the equivalent of equation (\ref{I p-space}). One for every $\tau$-derivative to be precise.

Now, let us focus on the number of loops being three. One always ends up with having to do three momentum integrals of the type appearing in equation (\ref{I
p-space}) with, in general, a product of one, two or three sign functions. To this end, one separates the 3-dimensional domain of
integration into smaller parts, such that on each part the product
of the appearing sign functions has a definite value. From the
definition of the dilogarithm, equation (\ref{poly2}), and the
trilogarithm,
\begin{eqnarray}
\Li_3(x) = \int_0^x dz\; \frac{\Li_2(z)}{z} = \sum_{k=1}^\infty
\frac{x^k}{k^3}\;, \quad \quad \vert x \vert \leq 1,\label{poly3}
\end{eqnarray}
it is clear that one in general ends up with expressions involving
logarithms, dilogarithms and trilogarithms. Fortunately, since we
are only interested in the behavior of these expressions for
$\Lambda >> 1$, we can always convert them to expressions involving
only logarithms (up to terms of order $1/\Lambda$). This will be
accomplished by repeated use of identities like \cite{Lewin:81}:
\begin{eqnarray}
\Li_n(1) \! \! \! & \! =  \! & \! \! \! \zeta(n),\\
\Li_2(-x)+\Li_2\left(-\frac 1 x\right) \! \! \! &\! = \!  & \! \! \!  -\frac{\pi^2}{6}-\frac 1 2 \ln^2(x), \quad x>0, \label{inv2}\\
\Li_2(x)+\Li_2\left(\frac 1 x\right) \! \! \! &\! = \!  & \! \! \!  \frac{\pi^2}{3} -i\pi \ln x -\frac 1 2 \ln^2(x), \quad x>1,\\
\Li_3(-x)-\Li_3\left(-\frac 1 x\right) \! \! \! &\! = \!  & \! \! \!  -\frac{\pi^2}{6}\ln x -\frac 1 6 \ln^3(x), \quad x>0, \label{inv3}\\
\Li_3(x)-\Li_3\left(\frac 1 x\right) \! \! \! &\! = \!  & \! \! \!  \frac{\pi^2}{3} \ln x -i\frac{\pi}{2} \ln^2(x) -\frac 1 6 \ln^3(x), \quad x>1.
\end{eqnarray}
The most important consequence of the form of these equations is the appearance of terms linear in $\lambda \sim \ln \Lambda$, because
these lead to contributions to the beta-functions.

\begin{figure}[h]
\begin{center}
\psfig{figure=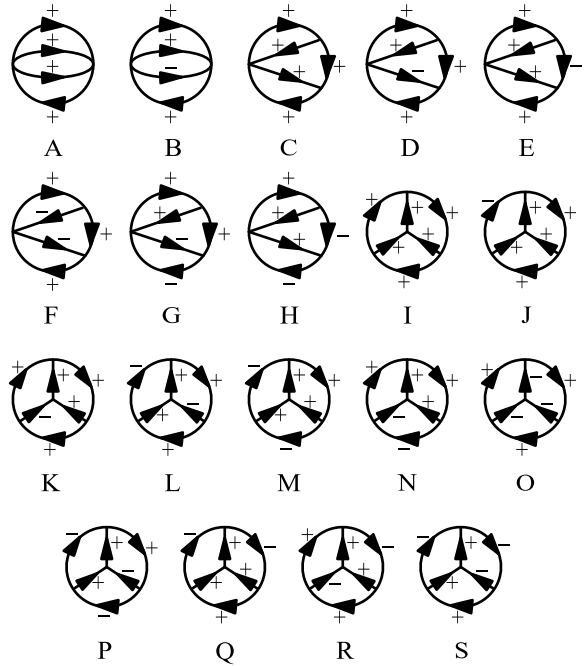} \caption{The basic
non-trivial diagrams at three loops. \label{fig:3loop_basic}}
\end{center}
\end{figure}

We saw in the previous section that, in the end we only needed to
perform one non-trivial integral to be able to calculate the entire
two loop contribution. At three loops, we would of course also like
to narrow down the number of necessary integrals to perform
explicitly as much as possible. The divergencies of diagrams A - D
of figure (\ref{fig:3loop}) can be computed from those at two loops
(and are, as we will show later-on, irrelevant for the beta-function).
To be able to compute the contributions from the other diagrams, E -
H of figure (\ref{fig:3loop}), it turns out that we still have to
perform 19 integrals in total. For completeness, we list the divergent part
of these integrals below.
\begin{eqnarray}
{\cal I}_A \! \! \! &\! = \!  & \! \! \!   \int d\tau \; \Delta_+(\tau)\Delta_+(\tau)
\Delta_+(\tau) \partial_{\tau}\Delta_-(\tau) = \frac{i}{8} \left(
\lambda^3 - \frac{\lambda}{2} \right) \quad \quad \quad \quad \quad \quad \quad \quad  \\
{\cal I}_B \! \! \! &\! = \!  & \! \! \!   \int d\tau \; \Delta_+(\tau)\Delta_+(\tau)
\Delta_-(\tau) \partial_{\tau}\Delta_-(\tau) = \frac{i}{12}
\lambda^3\\
{\cal I}_C \! \! \! &\! = \!  & \! \! \!   \int d\tau_1 d\tau_2 \; \Delta_+(1)\Delta_+(1+2)
\Delta_+(2) \partial_1\Delta_-(1) \partial_2\Delta_-(2) \nonumber \\  \! \! \! &\! = \!  & \! \! \! 
-\frac{1}{8} \left(
\lambda^3 - \frac{\lambda}{3} \right)\\
{\cal I}_D \! \! \! &\! = \!  & \! \! \!   \int d\tau_1 d\tau_2 \; \Delta_+(1)\Delta_+(1+2)
\Delta_-(2) \partial_1\Delta_-(1) \partial_2\Delta_-(2) \nonumber  \\  \! \! \! &\! = \!  & \! \! \! 
-\frac{1}{16} \left(
\lambda^3 - \frac{\lambda}{2} \right)\\
{\cal I}_E \! \! \! &\! = \!  & \! \! \!   \int d\tau_1 d\tau_2 \; \Delta_+(1)\Delta_-(1+2)
\Delta_+(2) \partial_1\Delta_-(1) \partial_2\Delta_-(2) \nonumber  \\  \! \! \! &\! = \!  & \! \! \! 
-\frac{1}{24} \left(
\lambda^3 - \frac{\lambda}{2} \right)\\
{\cal I}_F \! \! \! &\! = \!  & \! \! \!   \int d\tau_1 d\tau_2 \; \Delta_-(1)\Delta_+(1+2)
\Delta_-(2) \partial_1\Delta_-(1) \partial_2\Delta_-(2) \nonumber  \\  \! \! \! &\! = \!  & \! \! \! 
-\frac{1}{24}
\lambda^3 \\
{\cal I}_G \! \! \! &\! = \!  & \! \! \!   \int d\tau_1 d\tau_2 \; \Delta_+(1)\Delta_+(1+2)
\Delta_-(2) \partial_1\Delta_-(1) \partial_2\Delta_+(2) \nonumber  \\  \! \! \! &\! = \!  & \! \! \! 
\frac{1}{16} \left(
\lambda^3 - \frac{\lambda}{6} \right)\\
{\cal I}_H \! \! \! &\! = \!  & \! \! \!   \int d\tau_1 d\tau_2 \; \Delta_-(1)\Delta_+(1+2)
\Delta_-(2) \partial_1\Delta_-(1) \partial_2\Delta_+(2) \nonumber  \\  \! \! \! &\! = \!  & \! \! \! 
-\frac{1}{48} \left( \lambda^3 - \frac{\lambda}{2} \right)
\end{eqnarray}
\begin{eqnarray}
{\cal I}_I \! \! \! &\! = \!  & \! \! \! \int d\tau_1 d\tau_2 d\tau_3 \;
\Delta_+(1+2)\Delta_+(1+2+3) \Delta_+(2+3) \partial_1\Delta_-(1)
\partial_2\Delta_-(2) \partial_3\Delta_-(3)\nonumber\\
\! \! \! &\! = \!  & \! \! \! -\frac{i}{8} \left(
\lambda^3 - \frac{\lambda}{2} \right)\\
{\cal I}_J \! \! \! &\! = \!  & \! \! \! \int d\tau_1 d\tau_2 d\tau_3 \;
\Delta_+(1+2)\Delta_+(1+2+3) \Delta_-(2+3) \partial_1\Delta_-(1)
\partial_2\Delta_-(2) \partial_3\Delta_-(3)\nonumber\\
\! \! \! &\! = \!  & \! \! \! -\frac{i}{16} \left(
\lambda^3 - \frac{\lambda}{2} \right)\\
{\cal I}_K \! \! \! &\! = \!  & \! \! \! \int d\tau_1 d\tau_2 d\tau_3 \;
\Delta_+(1+2)\Delta_-(1+2+3) \Delta_+(2+3) \partial_1\Delta_-(1)
\partial_2\Delta_-(2) \partial_3\Delta_-(3)\nonumber\\
\! \! \! &\! = \!  & \! \! \! -\frac{i}{24} \left(
\lambda^3 - \lambda \right)\\
{\cal I}_L \! \! \! &\! = \!  & \! \! \! \int d\tau_1 d\tau_2 d\tau_3 \;
\Delta_-(1+2)\Delta_+(1+2+3) \Delta_-(2+3) \partial_1\Delta_-(1)
\partial_2\Delta_-(2) \partial_3\Delta_-(3)\nonumber\\
\! \! \! &\! = \!  & \! \! \! -\frac{i}{24} \left(
\lambda^3 - \frac{\lambda}{2} \right)\\
{\cal I}_M \! \! \! &\! = \!  & \! \! \! \int d\tau_1 d\tau_2 d\tau_3 \;
\Delta_+(1+2)\Delta_+(1+2+3) \Delta_-(2+3) \partial_1\Delta_-(1)
\partial_2\Delta_-(2) \partial_3\Delta_+(3)\nonumber\\
\! \! \! &\! = \!  & \! \! \! \frac{i}{48} \left(
\lambda^3 + \frac{\lambda}{2} \right) \\
{\cal I}_N \! \! \! &\! = \!  & \! \! \! \int d\tau_1 d\tau_2 d\tau_3 \;
\Delta_+(1+2)\Delta_-(1+2+3) \Delta_+(2+3) \partial_1\Delta_-(1)
\partial_2\Delta_-(2) \partial_3\Delta_+(3)\nonumber\\
\! \! \! &\! = \!  & \! \! \! \frac{i}{48} \left(
\lambda^3 - \frac{\lambda}{2} \right)\\
{\cal I}_O \! \! \! &\! = \!  & \! \! \! \int d\tau_1 d\tau_2 d\tau_3 \;
\Delta_+(1+2)\Delta_-(1+2+3) \Delta_-(2+3) \partial_1\Delta_-(1)
\partial_2\Delta_-(2) \partial_3\Delta_+(3)\nonumber\\
\! \! \! &\! = \!  & \! \! \! \frac{i}{24} \left(
\lambda^3 - \frac{\lambda}{2} \right)\\
{\cal I}_P \! \! \! &\! = \!  & \! \! \! \int d\tau_1 d\tau_2 d\tau_3 \;
\Delta_-(1+2)\Delta_+(1+2+3) \Delta_-(2+3) \partial_1\Delta_-(1)
\partial_2\Delta_-(2) \partial_3\Delta_+(3)\nonumber\\
\! \! \! &\! = \!  & \! \! \! 0\\
{\cal I}_Q \! \! \! &\! = \!  & \! \! \! \int d\tau_1 d\tau_2 d\tau_3 \;
\Delta_+(1+2)\Delta_+(1+2+3) \Delta_-(2+3) \partial_1\Delta_-(1)
\partial_2\Delta_+(2) \partial_3\Delta_-(3)\nonumber\\
\! \! \! &\! = \!  & \! \! \! 0\\
{\cal I}_R \! \! \! &\! = \!  & \! \! \! \int d\tau_1 d\tau_2 d\tau_3 \;
\Delta_+(1+2)\Delta_-(1+2+3) \Delta_+(2+3) \partial_1\Delta_-(1)
\partial_2\Delta_+(2) \partial_3\Delta_-(3)\nonumber\\
\! \! \! &\! = \!  & \! \! \! 0
\end{eqnarray}
\begin{eqnarray}
{\cal I}_S \! \! \! &\! = \!  & \! \! \! \int d\tau_1 d\tau_2 d\tau_3 \;
\Delta_-(1+2)\Delta_+(1+2+3) \Delta_-(2+3) \partial_1\Delta_-(1)
\partial_2\Delta_+(2) \partial_3\Delta_-(3)\nonumber\\
\! \! \! &\! = \!  & \! \! \! \frac{i}{48} \lambda \,.
\end{eqnarray}
The diagrams corresponding to the integrals are depicted in figure
(\ref{fig:3loop_basic}).

All of these integrals were performed in the same way as the two loop diagram of subsection (\ref{app 2l}). They all essentially involve integrals over products of sign functions. Take for instance ${\cal I}_M$. Using the definition of the $\tau$-space propagators (\ref{t-space prop}), we find more explicitly that
\begin{eqnarray}
\begin{split}
{\cal I}_M &= \frac{i}{2^6 \pi^3} \int_1^\Lambda \frac{dp}{p}
\int_1^\Lambda \frac{dq}{q} \int_1^\Lambda \frac{dk}{k} \big[
-\epsilon(p+q-\Lambda)
\epsilon(p+q-k-\Lambda) \epsilon(q-k+1) \\
& + \epsilon(p+q-\Lambda) \epsilon(p+q-k-1)
\epsilon(q-k+1)\\
& + \epsilon(p+q-\Lambda) \epsilon(p+q-k-\Lambda) -
\epsilon(p+q-\Lambda) \epsilon(p+q-k-1)\\
& + \epsilon(p+q-k-\Lambda) \epsilon(q-k+1) - \epsilon(p+q-k-1)
\epsilon(q-k+1)\\
& - \epsilon(p+q-k-\Lambda) + \epsilon(p+q-k-1)\big]\,.
\end{split}\label{I_3loop}
\end{eqnarray}
Clearly, this kind of computation is best done using a computer. We used the powerful and widely used computer algebra system Mathematica to perform the calculation of these intergrals.

\begin{figure}[h]
\begin{center}
\psfig{figure=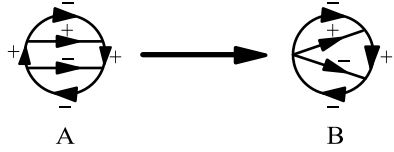} \caption{The integral of
diagram (A) can be related to the one of diagram (B).
\label{fig:ex_diagr}}
\end{center}
\end{figure}

The 19 integrals of which we have listed above can be seen as the building
blocks out of which every other non-trivial three loop integral can
be obtained quite easily. To illustrate this point, let us look at
the integral corresponding to diagram A of figure
(\ref{fig:ex_diagr}).

\noindent The reader may have noticed that this type of diagram did not appear
in the list of `building blocks' of figure (\ref{fig:3loop_basic}).
This is because an integral corresponding to this type of diagram
can always be related to one corresponding to a diagram of the type
of diagram B of figure (\ref{fig:ex_diagr}). This is easily
illustrated with the case at hand. The integral corresponding to
diagram A of figure (\ref{fig:ex_diagr}) equals,
\begin{eqnarray}
\begin{split}
{\cal I} &= \int d\tau_1 d\tau_2 d\tau_3 \; \Delta_-(1)\Delta_-(2)
\Delta_+(1+2+3) \partial_1\Delta_+(1)
\partial_2\Delta_+(2) \partial_3\Delta_-(3)\\
&+ \int d\tau_1 d\tau_2 d\tau_3 \; \Delta_+(1)\Delta_-(2)
\Delta_+(1+2+3) \partial_1\Delta_-(1)
\partial_2\Delta_+(2) \partial_3\Delta_-(3)\,.
\end{split}
\end{eqnarray}
Using the fact that,
\begin{eqnarray}
\int d\tau_3 \; \Delta_\pm(1+2+3) \partial_3\Delta_\mp(3) = \pm i
\Delta_\pm(1+2)\,,
\end{eqnarray}
we can perform the integral over $\tau_3$ and arrive at,
\begin{eqnarray}
\begin{split}
{\cal I} &= i\int d\tau_1 d\tau_2\; \Delta_-(1)\Delta_-(2)
\Delta_+(1+2) \partial_1\Delta_+(1)
\partial_2\Delta_+(2)\\
&+ i\int d\tau_1 d\tau_2 \; \Delta_+(1)\Delta_-(2) \Delta_+(1+2)
\partial_1\Delta_-(1)
\partial_2\Delta_+(2)\,,
\end{split}
\end{eqnarray}
which indeed corresponds to diagram B of figure
(\ref{fig:ex_diagr}). This diagram doesn't correspond to any of the
building blocks of figure (\ref{fig:3loop_basic}) either, but
comparison with the ones that do, shows that we can write,
\begin{eqnarray}
{\cal I} = i \left( {\cal I}_E + {\cal I}_G \right)\,.
\end{eqnarray}
This illustrates how other integrals one needs to perform can be
related to the `building blocks'.

 \section{Higer Loop Considerations}

\begin{figure}[h]
\begin{center}
\psfig{figure=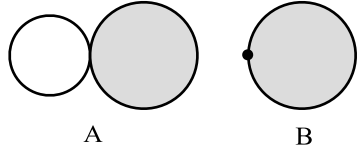}
\caption{A general diagram with external loop (A) and that same
diagram with the external loop replaced by the one loop counterterm
(B). \label{fig:blob}}
\end{center}
\end{figure}

An immediate question is whether the present program can be pushed to higher orders. Of course, already at four loops this procedure becomes extremely cumbersome. There might however be general arguments that lead to considerable simplifications. First of all, consider a diagram with an external loop, which will in general look like diagram A of figure (\ref{fig:blob}), where the bigger circle with shaded area can be any diagram\footnote{For some examples, see diagrams A - D of figure (\ref{fig:3loop})}. It is not very difficult to show that diagram B of figure (\ref{fig:blob}), resulting from replacing that external loop with the one loop counterterm, will contain a term which exactly cancels the original diagram. As a consequence, diagrams with (one or more) external loops cannot contribute to the beta-function. This is of course part of the bigger renormalization group picture. Diagrams which factorize will never contribute to the beta-function, because the divergences
encountered in the corresponding loop-integrals are already
accounted for at a lower-loop level.

\begin{figure}[h]
\begin{center}
\psfig{figure=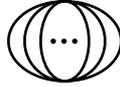}
\caption{An $n$-loop ``onion'' diagram. \label{fig:najuin}}
\end{center}
\end{figure}

More important simplifications might arise from an interesting observation made in \cite{wyllard}. There Niclas Wyllard noted that $S_{ijkl}$ which was introduced in (\ref{ScurvBeta}) can be viewed as the curvature tensor for a non-symmetric connection. Once this is better understood, this could lead to a method giving results to all order in the derivatives. Indeed the leading contribution to the beta-functions comes from the $n$-loop ``onion'' diagram shown in figure (\ref{fig:najuin}) which can be explicitly calculated in a reasonably straightforward way. The remainder of the beta-function should then follow as some sort of covariantization. This point of view is presently under investigation \cite{wip}. \\

From the calculational point of view it is important to note that the method presented in subsection \ref{app 2l} for two loops and \ref{beta_Calculational_Outline_3loops} for three loops can, in principle, easily be generalized to higher loops. First of all, superspace techniques can just as easily be applied to higher loop diagrams and the $\tau$-space integral one ends up with will always be a generalization of equations (\ref{I}) and (\ref{I_3loop}). The general procedure we deployed to handle these kinds of integrals can still be applied at higher loops. These integrals will always be expressible in terms of logarithms and polylogarithms. At $l$ loops, a general term will be of the form\footnote{Equation (\ref{generalterm}) only serves as a rough indication of the general form. More concretely, not all factors of ln need to have the same argument and there might be more polylogarithms involved in the same term. The important point is that all powers of ln and all orders of the polylogarithms involved add up to $l$.},
\begin{eqnarray}
\Li_n \left(r(\Lambda)\right)\;\ln^{l-n}\left(s(\Lambda)\right),
\qquad n \leq l,\label{generalterm}
\end{eqnarray}
where $r$ and $s$ are rational functions of $\Lambda$. The $n$-th
order polylogarithm $\Li_n$ is defined as,
\begin{equation}
\Li_n(x) = \int_0^x dz\; \frac{\Li_{n-1}(z)}{z} = \sum_{k=1}^\infty
\frac{x^k}{k^n}\;, \quad \quad \vert x \vert \leq 1,\label{polyn}
\end{equation}
and $\Li_2$ is defined in equation (\ref{poly2}). The last equality in (\ref{polyn}) is only valid for $\vert x\vert \leq 1$, as indicated, but the integral representation can be used as a
definition of $\Li_n$ on the whole (cut) complex plane. Since we are only interested in the behavior of equation (\ref{generalterm}) for $\Lambda>> 1$, we can use identities like \cite{Lewin:81}\footnote{For $n=2$ and 3, this expression reduces to (\ref{inv2}) and (\ref{inv3}), respectively.},
\begin{equation}
\Li_n(-x) + (-1)^n \Li_{n}\left(-\frac 1 x \right) = -\frac{1}{n!}
\ln^n(x) + 2 \sum_{k=1}^{[n/2]} \frac{\Li_{2k}(-1)}{(n-2k)!}\;
\ln^{n-2k}(x),
\end{equation}
where $[n/2]$ is the greatest integer contained in $n/2$ and,
\begin{eqnarray}
\Li_{n}(-1) = \left(2^{1-n}-1\right)\zeta(n),
\end{eqnarray}
to write the asymptotic behavior of (\ref{generalterm}) purely in terms of $\lambda \sim \ln(\Lambda)$. In the end we still arrive at a polynomial in $\lambda$, as desired.


\chapter{Closing Remarks} \label{Closing_Remarks}

\begin{quote}{\it ``I say, if your knees aren't green by the end of the day, you ought to seriously re-examine your life.''} 
\vspace*{-0,3cm}
\begin{flushright}{\scriptsize \sc Calvin (Calvin \& Hobbes by Bill Watterson)}\end{flushright}
\end{quote}

To conclude we summarize the results obtained in this thesis and point out some interesting paths for further research.

In chapter \ref{Non_Linear_Sigma-Models_with_Boundaries} we studied $d=2$ non-linear sigma-models in the presence of boundaries. In the absence of supersymmetry, we found that the boundary conditions require the existence of an almost product structure $ {\cal R}$ compatible with the metric and such that the projector $ {\cal P}_+=(1+ {\cal R})/2$ is integrable. Supersymmetrizing the model yields no further conditions. We obtained a manifest $N=1$ supersymmetric formulation of the model. Whether or not torsion is present did not essentially alter the discussion. Crucial in this was $ {\cal M}$ as found in \rref{bcfermions} (see also \rref{defvanm}) which relates the left movers to the right movers. In the case of constant magnetic background fields it was for the first time written down in \cite{Callan:1988wz}. 


Further restrictions where found when requiring more supersymmetry. Indeed, the existence of a second supersymmetry demands the presence of two complex structures $J$ and $\bar J$ both covariantly constant and such that the metric is hermitian with respect to both of them. Furthermore, one of the two should be expressed in terms of the other one, the metric, the Kalb-Ramond field and the almost product structure. 
No general manifest $N=2$ supersymmetric description, involving only linear superfield constraints, can be given. However, we showed that at least the type B K\"ahler models can be adequately described in $N=2$ superspace. It would be quite interesting to further investigate the $N=2$ superspace geometry, in particular for the A-type boundary conditions in the presence of non-trivial $U(1)$ backgrounds.

In chapter \ref{Beta-Function_Calculations_in_Boundary_Superspace} we calculated, leaning on the results of the previous chapter, the beta-functions through three loops for an open string sigma-model in the presence of $U(1)$ background. Requiring them to vanish is then reinterpreted as the string equations of motion for the background. Upon integration this yields the low energy effective action. Doing the calculation in $N=2$ boundary superspace significantly simplified the calculation. The one loop contribution gives the effective action to all orders in $ \alpha'$ in the limit of a constant fieldstrength. The result is the well known Born-Infeld action. The absence of a two loop contribution to the beta-function shows the absence of two derivative terms in the action. Finally the three loop contribution gives the
four derivative terms in the effective action to all orders in $ \alpha'$. Modulo a field redefinition we find complete agreement with the proposal made in \cite{wyllard}.

By doing the calculation in $N=2$ superspace, we get a nice geometric characterization of UV finiteness of the non-linear sigma-model: UV finiteness is guaranteed provided that
the background is a deformed stable holomorphic bundle.

A natural question is whether the methods deployed in chapter \ref{Beta-Function_Calculations_in_Boundary_Superspace} can be extended to higher loops. We argued that this is indeed possible although the calculations would become extremely lengthy and involved. A good understanding of the renormalization group might provide considerable simplifications. Another method might simply require to calculate the leading contribution, coming from the so called $n$-loop ``onion'' diagram, 
followed by some kind of covariantization to obtain the full beta-function.

Finally, one might wonder whether the present method extends to the non-abelian case, a subject we steered clear of in the whole thesis. When multiple branes coincide the gauge field living on the brane becomes non-abelian and the corresponding action will be a non-abelian generalization of the (abelian) Born-Infeld action.
 In that case the coupling to the gauge fields involves the introduction of a Wilson line. The path-ordering can be undone through the introduction of auxiliary fields, \cite{dorn}, and a first exploration was performed in \cite{klaus}. However, before the present analysis can be done for the non-abelian case, one needs to extend the superspace formulation we presented in chapter \ref{Non_Linear_Sigma-Models_with_Boundaries} such as to include Wilson loops and the auxiliary formulation of \cite{dorn}. This would certainly lead to significant information on non-abelian deformed stable holomorphic bundles. We leave this interesting question to future investigation.

\appendix


 \chapter{Conventions}
For the sake of quick reference we will summarize some of the common conventions used in this thesis.

Space-time coordinates will be labeled by greek indices, $\mu,\nu,\kappa,\lambda,\ldots$ and the Minkowski metric is written down in the mostly plus convention,
\begin{equation}\label{veldmetriek} 
\eta_{\mu\nu} = \diag \, (-,+,\ldots,+).
\end{equation}
Whenever we write roman indices $i,j,k,l,\ldots$ we are either working in the space part of space-time or we changed, by a Wick rotation, to Euclidean space-time.

On the worldsheet we use roman indices $a,b,c,d,\ldots$ and take our metric to have Minkowski signature, 
\begin{equation}
\eta_{ab} = \diag \, (-,+),
\end{equation}
with the excpection of chapter \ref{Beta-Function_Calculations_in_Boundary_Superspace} where we adopted Euclidean signature.

We also need complex coordinates related to the ordinary coordinates of a $2p$ dimensional subspace as follows,
\begin{equation}
\begin{split}
Z^\alpha &= \frac{1}{\sqrt 2}\left(X^{2\alpha -1}+iX^{2\alpha }\right) \\
\bar Z^{\bar\alpha} &= \frac{1}{\sqrt 2} \left(X^{2\alpha -1}-iX^{2\alpha }\right)
\end{split} \quad \quad  \quad  \text{for} \quad  \quad \quad \alpha \in \left\{1, \ldots ,p \right\}.
\end{equation}

The fieldstrength we use is given by,
\begin{equation}
F_{\mu\nu}  = \partial_\mu A_\nu-\partial_\nu A_\mu.
\end{equation}

We use square brackets, $[\ldots]$, for antisymmetrization of indices and round brackets, $(\ldots)$, for symmetrization. Our convention is that we sum over all permutations
and divide by the number of permutations.


 \chapter{Samenvatting}

\section{Inleiding}

Alle recente inspanningen die geleverd worden in theoretische natuurkunde kunnen grosso-modo worden opgedeeld in twee categorie\"en. De eerste spitst zich toe op het vinden van de onderliggende principes die fysische fenomenen drijven terwijl de tweede probeert om wat we observeren in de natuur te verklaren steunend op deze principes. Uiteraard zijn deze twee domeinen sterk afhankelijk van elkaar en is de grens tussen beiden vaak zeer vaag. De reden waarom we desondanks toch een onderschied willen maken is dat de queeste naar een theorie van alles in de eerste plaats een poging is om \'e\'en enkel onderliggend principe te vinden. Vertrekkende daarvan zou dan, in principe, alles verklaard kunnen worden. We zeggen in principe omdat dit niet noodzakelijk wil zeggen dat we in staat zullen zijn om alles te verklaren/berekenen/voorspellen/\ldots omdat sommige problemen naar alle waarschijnlijkheid te ingewikkeld zullen blijken. 

Op dit moment hebben we twee aanvullende theorie\"en die samen een uitstekende beschrijving geven van de natuur: het Standaard Model en Algemene Relativiteit. Het Standaard Model is een spontaan gebroken $SU(3) \times SU(2) \times U(1)$ ijktheorie en blijkt een kwalitatief en kwantitatief uitstekende beschrijving te geven van de elektrozwakke en de sterke kracht, zowel op lange als op  korte af\-standen. Daartegenover staat het feit dat het model afhankelijk is van zesentwintig vrije parameters die experimenteel moeten bepaald worden. De zwaartekracht is de enige kracht die niet opgenomen is in het Standaard Model. Alhoewel haar lange afstandsgedrag op uitstekende wijze beschreven wordt door Algemene Relativi\-teits\-theorie, is ze niet compatibel met de kwantummechanica. Niet alleen is algemene relativiteit een niet-renormaliseerbare kwantumveldentheorie, maar ook het bestaan van zwarte gaten geeft aanleiding tot fundamentele problemen en paradoxen. 

Uit de voorgaande paragraaf blijkt duidelijk dat er nog stappen te zetten zijn in de richting van \'e\'en enkele onderliggende theorie. In het ideale geval zou dat een theorie zijn die gravitatie met de andere krachten unificeert en die de zesentwintig vrije parameters in Standaard Model voorspelt. Een veelbelovende kandidaat is Snaartheorie in dewelke men de veronderstelling maakt dat elementaire deeltjes geen puntdeeltjes zijn maar kleine vibrerende snaartjes. Het resulterende model levert in het lage-energieregime dan een effectieve theorie op die essentieel uit een supersymmetrische versie van algemene relativiteit bestaat, gekoppeld aan een supersymmetrische ijktheorie. De stap naar snaartheorie geeft echter aanleiding tot diverse problemen. Het meest prominente probleem is misschien dat er niet \'e\'en maar vijf consistente snaartheorie\"en zijn. Deze zijn allen slechts goed gedefini\"eerd in tien dimensies, een tweede probleem.

Pogingen om het tweede probleem op te lossen, door compactificatie van de extra dimensies, leidde naar de ontdekking van D-branen wat een ware revolutie teweegbracht in snaartheorie. Een D$p$-braan is een $p$-dimensionaal dy\-namisch object, met een $p +1$-dimensionaal wereldvolume, dat gedefini\"eerd is doordat de uiteinden van open snaren erop kunnen eindigen. Via dualiteitstransformaties, waarin D-branen een cruciale rol spe\-len, kon men aantonen dat alle vijf snaartheorie\"en limietgevallen zijn van een onderliggend bouwwerk dat tegenwoordig onder de naam M-theorie bekend staat. 

Een heel interessant aspect van D-branen is het nauwe verband met ijktheorie\"en. Het $p+1$-dimensionaal wereldvolume wordt beschreven door een $p + 1$-dimensionale veldentheorie met $9 - p$ massaloze scalaire velden die de transversale positie van het D$p$-braan beschrijven en een $U(1)$ ijkveld dat te wijten is aan de open snaren, die als grondtoestand $U(1)$ ijkdeeltjes hebben, die erop eindigen. 

De effectieve actie van \'e\'en enkel D-braan, in de benadering van traag vari\"erende veldsterktes, is de Born-Infeld actie die in leidende orde niets meer is dan de actie voor Maxwell theorie. De in deze benadering gevonden actie is de volledige actie met alle termen die afgeleides op de $U(1)$ veldsterkte bevatten gelijk gesteld aan nul. Het eerste paper waarin afgeleide correcties bestudeerd werden was \cite{abelianbi4derivative}  waarin werd aangetoond dat de termen met twee afgeleiden nul zijn. Pas recent \cite{wyllard} werden de vier afgeleide termen berekend.

\section{Wat is Snaartheorie?}

In hoofdstuk 2 van deze thesis geven we een compacte samenvatting van snaartheorie met een sterke nadruk op de elementen die we in de volgende hoofstukken zullen nodig hebben. We beginnen met een korte beschrijving van het puntdeeltje maar stappen snel over op bosonische snaartheorie. Deze laatste gebruiken we om enkele belangrijke noties van snaartheorie in te voeren. Niet helemaal tevreden met enkele eigenchappen van bosonische snaartheorie richten we ons nadien op supersnaren. Onderweg introduceren we het concept van superruimtes wat relevant zal zijn voor hoofdstuk 3. We gaan verder met het beschrijven van niet-perturbatieve eigenschappen van snaartheorie door het introduceren, via $T$-dualiteit, van D-branen. Uiteindelijk concentreren we ons op verscheidene methoden om de Born-Infeld actie te berekenen.



\section{Niet-Lineaire Sigma-Modellen met Randen}

In hoofstuk 3 bestuderen we, gemotiveerd door hun relevantie voor het beschrijven van D-branen, niet-lineraire sigma-modellen met randen. Gedeeltelijke resultaten waren al een tijd bekend, zie bijvoorbeeld \cite{Ooguri:1996ck}--\cite{Lindstrom:2005zr}, maar het was pas recent, in \cite{stock1} en \cite{stock2}, dat een systematische studie werd uitgevoerd. Deze resulteerde in de meest algemene randvoorwaarden compatibel met $N=1$ supersymmetrie. Vervolgens werden deze resultaten uitgebreid naar $N=2$ supersymmetrie in \cite{zab} (zie ook \cite{Lindstrom:2002vp} voor enkele specifieke toepassingen en \cite{Lindstrom:2002mc} voor een andere aanpak).

De resulaten in  \cite{stock1} en \cite{stock2} zijn niet alleen indrukwekkend maar ook verrassend. Niet alleen zijn de afleidingen vrij ingewikkeld, de aanwezigheid van een niet triviaal Kalb-Ramond veld lijkt een niet lokale superruimte beschrijving te eisen. Dit gebeurt al in het simpelste geval waar open snaren bewegen in een triviale gravitationele achtergrond en een niet triviale electro-magnetische achtergrond. Het is duidelijk dat wanneer we de open string effectieve actie willen bestuderen door supergraaf berekeningen dat we op zijn minst een lokale actie nodig hebben.

We heranalyseren de modellen die bestudeerd werden in  \cite{stock1} en \cite{stock2} en lossen veel van de moeilijkheden die in deze papers naar voor komen op. We beginnen met het heranalyseren van niet-supersymmetrische niet-lineaire sigma-modellen en bestuderen de meest algemene randvoorwaarden. In de volgende sectie breiden we dit uit naar modellen met supersymmetrie. Gemotiveerd door de methodes gebruikt in \cite{oliver} en \cite{joanna} (echter in een zeer verschillende context) gebruiken we een superruimte formulatie die manifest invariant is onder slechts een van de twee bulk supersymmtrie\"en. Op deze manier simplificeren we de analyse van randvoorwaarden compatibel met $N=1$ supersymmetrie aanzienlijk en vinden we, net zoals in het geval zonder randen, dat $N=0$ automatisch $N=1$ impliceert. Meer nog, niet-locale termen zijn niet nodig en de gevallen met of zonder Kalb-Ramond veld worden op gelijke voet behandeld. De prijs die we hiervoor betalen is dat we manifeste bulk $d=2$ Lorentz covariantie verliezen.

Vervolgens onderzoeken we onder welke voorwaarden de $N=1$ supersymmetrie gepromoveerd kan worden naar een $N=2$ supersymmtrie. Net zoals in het geval zonder randen vinden we dat we twee apart integreerbare covariant constante complexe structuren nodig hebben en dat de metriek heremitisch moet zijn tenopzichte van beide. De aanwezigheid van randen verreist dat een van de twee kan uitgedrukt worden in functie van de andere en de rest van de geometrische data. Uiteindelijk bestuderen we de $N=2$ superruimte formulatie.

De resultaten neergeschreven in dit hoofdstuk zijn oorspronkelijk gepubliceerd in \cite{susyboundary}.

\section{Beta-Functie Berekeningen in Rand Superruimte}

Na de afleidingen in het voorgaande hoofdstuk zijn we klaar om in hoofdstuk 4 de hoofdbrok van deze thesis aan te pakken. We leiden, met de algemene opzet uit het vorige hoofdstuk als startpunt, de beta-functies af voor een open snaar sigma-model in de aanwezigheid van een $U(1)$ achtergrondveld. Vragen dat deze beta-functies nul zijn levert, na enige manipulatie, de Born-Infeld actie op.

Het aantal afgeleides werkend op de veldstrekte is gelijk aan twee maal het aantal lussen min twee. Hieruit volgt dat een \'e\'en lus brekening ons de Born-Infeld actie oplevert terwijl een twee lus berekening de twee afgeleide correcties op deze actie geeft. De drie lus berekening die we uitgevoerd hebben geeft ons de vier afgeleide correties en levert ons een onafhankelijke test op voor de reultaten in \cite{wyllard}.

Hoofdstuk 4 is als volgt opgedeeld. In de eerste sectie beschrijven we het gebruikte sigma-model in $N=2$ rand superruimte. Daarna maken we een analyse van de uit te voeren berekening en leiden we de nodige ingredienten, zoals de Feynmann regels en de superruimte propagatoren, af om deze te kunnen uitvoeren. We voeren ook een diagrammatische taal in die we doorheen het hele hoofdstuk zullen gebruiken en tonen, aan de hand van een toy model, hoe de propagators voor beperkte supervelden kunnen worden afgeleid.

Vervolgens focussen we ons op het berekenen van de beta-funties op een en twee lus orde en vinden complete overeenstemming met de literatuur. De twee lus be\-re\-ke\-ning voeren we zeer expliciet uit, teneinde de gebruikte technieken te illustreren. Nadien wenden we ons tot het belangrijkste stuk van deze thesis, het uitvoeren van de drie lus berekening en vinden complete overeenstemming met de resultaten bekomen door Niclas Wyllard \cite{wyllard} in de vorm zoals neergeschreven in \cite{koerber:thesis}. We sluiten dit hoofdstuk af met enkele beschouwing over hogere lus berekeningen.

De resultaten neergeschreven in dit hoofdstuk zijn oorspronkelijk gepubliceerd  in \cite{betastijnalexalexwalter}, enkele tussentijdse resultaten werden neergeschreven in \cite{towards}.


\end{document}